\documentclass[10pt,a4paper,twocolumn]{article}

\usepackage[utf8]{inputenc}
\usepackage{times}

\usepackage[a4paper,ignoreheadfoot,left=2cm,right=1.5cm,top=2cm,bottom=2.5cm]{geometry}

\usepackage[hyphens]{url} 
\usepackage{calc}
\usepackage{booktabs} 
\usepackage{xcolor}
\usepackage{chngcntr} 

\usepackage{graphicx}
\graphicspath{{figures/}}  

\usepackage[separate-uncertainty=true]{siunitx}
\DeclareSIUnit\cal{cal}
\DeclareSIUnit\mol{mol}
\DeclareSIUnit\eV{eV}
\DeclareSIUnit\atom{atom}
\DeclareSIUnit\cation{cation}

\newlength{\captionskip} 
\usepackage{amsmath}
\usepackage{amssymb}
\usepackage{bm} 
\usepackage{braket} 

\newcommand{\RR}{\ensuremath{\mathbb{R}}} 
\DeclareMathOperator*{\argmin}{arg\,min} 
\renewcommand{\vec}[1]{\ensuremath{\bm{#1}}} 
\newcommand{\mat}[1]{\ensuremath{\bm{#1}}} 

\usepackage{titletoc} 

\definecolor{linkcolor}{rgb}{0.15,0.15,0.15} 
\usepackage[
	pagebackref=true, 
	colorlinks=true, 
	allcolors=linkcolor,
]{hyperref}

\renewcommand*{\backrefalt}[4]{%
\ifcase #1 %
no citations%
\or
{\footnotesize\textcolor{linkcolor}{[p.~#2]}}%
\else
{\footnotesize\textcolor{linkcolor}{[pp.~#2]}}%
\fi
}

\usepackage[capitalise,noabbrev]{cleveref} 
\creflabelformat{equation}{#2\textup{#1}#3}
\crefname{subsection}{SI}{SI}

\usepackage[superscript]{cite}  
\usepackage[resetlabels]{multibib} 
\makeatletter
	\def\@mb@citenamelist{cite,citep,citet,citealp,citealt,citen,citenum}
\makeatother
\newcites{si}{References} 

\bibliographystylesi{reviewmod}

\newcommand{\eref}[1]{Reference~\citenum{#1}} 
\newcommand{\erefs}[1]{References~\citenum{#1}} 
\newcommand{\Erefs}[1]{References~\citenum{#1}} 

\usepackage{enumitem}
\renewlist{itemize}{itemize}{1}
\setlist[{itemize}]{label={\textbullet},align=left,leftmargin=2ex,itemindent=!,labelwidth=0pt,itemsep=0pt,parsep=5pt plus 2pt minus 3pt}

\newlist{romanenum}{enumerate}{2}
\setlist[romanenum,1]{label=\emph{(\roman*)},align=left,leftmargin=0pt,itemindent=!,labelwidth=0pt,itemsep=0pt,parsep=0.5\baselineskip,widest*=8}
\setlist[romanenum,2]{label=\roman{romanenumi}.\alph*,align=left,leftmargin=3.5ex,labelwidth=*,labelsep*=0pt,itemindent=0pt,itemsep=0pt}
\newlist{romanenum*}{enumerate*}{1}
\setlist[romanenum*,1]{label=(\alph*),afterlabel={{}},ref={(\roman{romanenumi}.\alph*)},itemjoin={{, }},itemjoin*={{, and }}}

\makeatletter
\renewcommand\maketitle{\par
	\begingroup
		\renewcommand\thefootnote{\@fnsymbol\c@footnote}%
		\def\@makefnmark{\rlap{\@textsuperscript{\normalfont\@thefnmark}}}%
		\long\def\@makefntext##1{\parindent 1em\noindent
			\hb@xt@1.8em{%
				\hss\@textsuperscript{\normalfont\@thefnmark}}##1}%
		\if@twocolumn
		\ifnum \col@number=\@ne
			\@maketitle
		\else
			\twocolumn[\@maketitle]%
		\fi
	\else
		\newpage
		\global\@topnum\z@   
		\@maketitle
	\fi
	\thispagestyle{plain}\@thanks
	\endgroup
	\setcounter{footnote}{0}%
}
\makeatother

\makeatletter
\def\@maketitle{%
  \newpage
  \null
  \vskip 2em%
  \flushleft
  \let \footnote \thanks
    {\LARGE\bf \@title \par}%
    \vskip 1.5em%
    {\large
      \lineskip .5em%
      \begin{tabular}[t]{@{}c}%
        \@author
      \end{tabular}\par}%
    \vskip 1em%
    {\large \@date}%
  \par
  \vskip 1.5em}
\makeatother

\makeatletter
\renewcommand*\l@section[2]{%
  \ifnum \c@tocdepth >\z@
    \addpenalty\@secpenalty
	\addvspace{0.4em \@plus\p@}%
    \setlength\@tempdima{1.5em}%
    \begingroup
      \parindent \z@ \rightskip \@pnumwidth
      \parfillskip -\@pnumwidth
      \leavevmode 
      \advance\leftskip\@tempdima
      \hskip -\leftskip
      #1\nobreak\hfil
      \nobreak\hb@xt@\@pnumwidth{\hss #2%
                                 \kern-\p@\kern\p@}\par
    \endgroup
  \fi}
\makeatother

\makeatletter
\renewcommand\subsection{\@startsection{subsection}{2}{\z@}%
	{-3.25ex\@plus -1ex \@minus -.2ex}%
	{1.5ex \@plus .2ex}%
	{\normalfont\large\bfseries}}
\makeatother

\newcommand{\dsgdb}{\texttt{qm9}}
\newcommand{\dsba}{\texttt{ba10}}
\newcommand{\dstco}{\texttt{nmd18}} 
\newcommand{\dstcou}{\texttt{nmd18u}}
\newcommand{\dstcor}{\texttt{nmd18r}}

\begin{document}

\newcommand{\titletext}{Representations of molecules and materials for interpolation\\ of quantum-mechanical simulations via machine learning}
\title{\titletext}
\author{%
	Marcel F. Langer\hspace{0.5ex}\footnotemark[1]\hspace{1ex}\footnotemark[2] \and 
	Alex Goe{\ss}mann\hspace{0.5ex}\footnotemark[1]\hspace{1ex}\footnotemark[3] \and
	Matthias Rupp\hspace{0.5ex}\footnotemark[1]\hspace{1ex}\footnotemark[4]\hspace{1ex}\footnotemark[5]
}
\date{%
	\setcounter{footnote}{0}%
	\footnotemark~~NOMAD Laboratory, Fritz Haber Institute of the Max Planck Society, Berlin, Germany;\quad 
	\footnotemark~~Machine Learning Group, Technische Universit{\"a}t Berlin, Germany;\quad 
	\footnotemark~~Institute of Mathematics, Technische Universit{\"a}t Berlin, Germany;\quad 
	\footnotemark~~Citrine Informatics, Redwood City, CA, USA\quad 
	\footnotemark~~Present address: Department of Computer and Information Science, University of Konstanz, Germany.\\[2ex] 
}

\twocolumn[\begin{@twocolumnfalse}

\vspace*{-30pt}
\maketitle

\subsubsection*{Abstract}
Computational study of molecules and materials from first principles is a cornerstone of physics, chemistry, and materials science, but limited by the cost of accurate and precise simulations.
In settings involving many simulations, machine learning can reduce these costs, often by orders of magnitude, by interpolating between reference simulations.
This requires \textit{representations} that describe any molecule or material and support interpolation.
We comprehensively review and discuss current representations and relations between them,
using a unified mathematical framework based on many-body functions, group averaging, and tensor products.
For selected state-of-the-art representations, we compare energy predictions for organic molecules, binary alloys, and Al-Ga-In sesquioxides in numerical experiments controlled for data distribution, regression method, and hyper-parameter optimization.

\bigskip
\medskip

\end{@twocolumnfalse}]

\raggedbottom

\noindent\begin{minipage}{\linewidth-4ex}
	\startcontents[main]
	\section*{\contentsname}
	\printcontents[main]{}{1}[2]{}
\end{minipage}

\noindent\hspace*{-4ex}\begin{minipage}{\linewidth+4ex}
\section*{Glossary}

\begin{tabular}{@{}p{0.19\linewidth-0.5\tabcolsep}@{\hspace{\tabcolsep}}p{0.81\linewidth-0.5\tabcolsep}@{}}
	\toprule
	Acronym & Meaning \\ 
	\midrule
	\multicolumn{2}{@{}l@{}}{\textit{Representations}} \\
	ACE    & atomic cluster expansion \\
	BoB    & bag of bonds \\
	BS     & bispectrum \\
	CM     & Coulomb matrix \\
	DECAF  & density-encoded canonically-aligned fingerprint \\
	FCHL   & Faber-Christensen-Huang-von Lilienfeld \\
	GM     & Gaussian moments \\
	HDAD   & histograms of distances, angles, and dihedral angles \\
	IDMBR  & inverse-distance many-body representation \\
	MBTR   & many-body tensor representation \\
	MOB    & molecular orbital basis \\
	MTP    & moment tensor potential \\
	NICE   & $N$-body iterative contraction of equivariants \\
	OMF    & overlap matrix fingerprint \\
	SF     & symmetry function \\
	SOAP   & smooth overlap of atomic positions \\	
	WST    & wavelet scattering transform \\
	\\
	\multicolumn{2}{@{}l@{}}{\textit{Methodology}} \\
	GPR    & Gaussian process regression \\
	HP     & hyperparameter (free parameter) \\
	KRR    & kernel ridge regression \\
	MAE    & mean absolute error \\
	ML     & machine learning \\
	QM     & quantum mechanics \\
	QM/ML  & ML model for accurate prediction of QM data \\
	RMSE   & root mean squared error \\
	system & poly-atomic system, e.g., a molecule or a crystal \\
	\\
	\multicolumn{2}{@{}l@{}}{\textit{Other}} \\
	SI     & supplementary information\\
	\bottomrule
\end{tabular}
\end{minipage}

\clearpage


\enlargethispage{-2\baselineskip}

\section{Introduction}\label{sec:introduction}

Quantitative modeling of atomic-scale phenomena is central for scientific insights and technological innovations in many areas of physics, chemistry, and materials science.
Solving the equations that govern quantum mechanics (QM), such as Schr{\"o}dinger's or Dirac's equation, allows accurate calculation of the properties of molecules, clusters, bulk crystals, surfaces, and other polyatomic systems.
For this, numerical simulations of the electronic structure of matter are used, with tremendous success in explaining observations and quantitative predictions. 

However, the high computational cost of these ab initio simulations (\cref{si:abinitiocosts}) often only allows investigating
from tens of thousands of small systems with a few dozen atoms to a few large systems with thousands of atoms, particularly for periodic structures.
In contrast, the number of possible molecules and materials grows combinatorially with the number of atoms: 
13~or fewer C, N, O, S, Cl atoms can form a billion possible molecules, \cite{br2009}
and for 5-component alloys, there are more than a billion possible compositions when choosing from 30 elements. (\cref{si:sizeccs})
This limits systematic computational study and exploration of molecular and materials spaces.
Similar considerations hold for ab initio dynamics simulations, which are typically restricted to systems with a few hundred atoms and sub-nanosecond timescales.

\begin{figure}
	\centering
	\includegraphics[width=\linewidth]{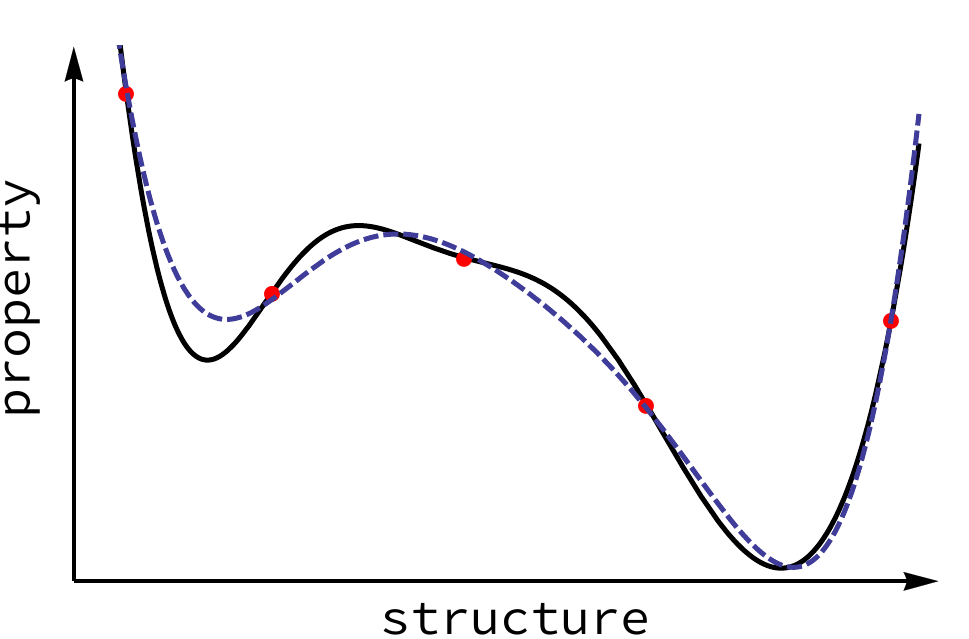}
	
	\vspace*{\captionskip}%
	\caption{\emph{Sketch illustrating the interpolation of quan\-tum-me\-chan\-i\-cal simulations by machine learning.}
		The horizontal axis represents chemical or materials space, the vertical axis the predicted property.
		Instead of conducting many computationally expensive ab initio simulations (solid line), machine learning (dashed line) interpolates between reference simulations (dots).\label{fig:qmmlsketch}}
	\vspace*{3\baselineskip}
\end{figure}

Such situations require many simulations of systems correlated in structure, implying a high degree of redundancy.
Machine learning \cite{g2015,jm2015} (ML) exploits this redundancy to interpolate between reference simulations \cite{rtml2012q,b2017q,cwc2018q,hl2020b} (\cref{fig:qmmlsketch}).
This ansatz replaces most ab initio simulations by ML predictions, based on a small set of reference simulations.
Effectively, it maps the problem of repeatedly solving a QM equation for many related systems onto a regression problem.
This approach has been demonstrated in benchmark settings \cite{bp2007q,bpkc2010q,rtml2012q} and applications \cite{b2017q,cdklc2018q,jlkkb2019q},
with reported speed-ups between zero to six orders of magnitude. \cite{kotm2016q,bkbc2018q,scaccr2019q,jkk2019q}
It is currently regarded as a highly promising avenue towards extending the scope of ab initio methods.

The most relevant aspect of ML models for interpolation of QM simulations (QM/ML models) after data quality (\cref{si:dataquality}) is the definition of suitable input features, that is, \emph{representations} of atomistic systems.
Representations define how systems relate to each other for regression and are the subject of this review.

\subsection{Scope and structure}

QM/ML models require a space in which interpolation takes place.
Such spaces can be defined explicitly, often as vector spaces, or implicitly, for example, via a kernel function in kernel-based machine learning. \cite{ss2002,hss2008}
\emph{This work reviews and compares explicit Hilbert-space representations of finite and periodic polyatomic systems for accurate interpolation of QM observables via ML},
focusing on representations that satisfy the requirements in \cref{sec:requirements} and energy predictions.

This excludes features that do not encode all input information, such as atomic numbers and coordinates,
for example, descriptors or fingerprints used in cheminformatics and materials informatics to interpolate between experimental outcomes, \cite{tc2009} 
and implicit representations learned by end-to-end deep neural networks \cite{gsrvd2017q,sacmt2017q,sksctm2017q,k2017bq,ssktm2018q,zhwsce2018q,tskylkr2018q,klt2018q,wgwbc2018q,um2019bq,ahk2019q,zhj2019q,kgg2020q,mgsn2020q,fwfw2020,qwamm2020q,czwe2020q,hsn2020q}
or defined via direct kernels between systems. \cite{hr1996q,um2017q,gzv2018q,km2020q,gzfd2020q} (\cref{si:scope,,si:rolerepr})

Characteristics and requirements of representations are discussed in \cref{sec:taxonomy,sec:requirements}.
\cref{sec:framework} describes a unified mathematical framework for representations.
Specific representations are delineated (benchmarked ones in \cref{sec:representations}, others in \cref{sec:other}), qualitatively compared (\cref{sec:analysis}), and empirically benchmarked (\cref{sec:benchmark}).
We conclude with an outlook on open problems and possible directions for future research in \cref{sec:outlook}.

\subsection{Related work}

Studies of QM/ML models often compare their performance estimates with those reported in the literature.
While such comparisons have value, they entertain considerable uncertainty due to different datasets, learning algorithms, including choice of hyperparameters (HPs, free parameters), sampling, validation procedures and reported quantities.
Accurate, reliable performance estimates require a systematic comparison that controls for the above factors, which we perform in this work.

Several recent studies systematically measured and compared prediction errors of representations (\cref{tab:relwork}).
We distinguish between studies that automatically (as opposed to manually) optimize
numerical HPs of representations, for example, the width of a normal distribution;
structural HPs of representations, for example, choice of basis functions;
and HPs of the regression method, for example, regularization strength.
\cref{si:relatedwork} discusses the individual studies in \cref{tab:relwork}.

\newcommand{\yes}{\checkmark}
\newcommand{\no}{$\times$}

\begin{table*}[tbh]
	\begin{minipage}[t]{0.6\linewidth}
		\begin{tabular}[t]{@{}l*{11}{c@{\hspace{\tabcolsep}}}c@{}} \toprule
			& \multicolumn{11}{c}{Reference} \\ \cmidrule(lr){2-12}
			& \citenum{fhhgsdvkrv2017q} 
			& \citenum{hjmfrgrf2020q} 
			& \citenum{zcldcbcstwo2020q} 
			& \citenum{sgc2019q}  
			& \citenum{nrbsmrcwh2019q} 
			& \citenum{strkghr2019q} 
			& \citenum{ook2020q} 
			& \citenum{kkclm2020q} 
			& \citenum{pdcfkdblg2020} 
			& \citenum{jmfhf2018q} 
			& \citenum{gfc2020}
			& here \\ \midrule
			Finite systems    & \yes & \yes & \no   & \yes & \no  & \yes & \yes & \yes & \yes & \yes  & \yes & \yes \\
			Periodic systems  & \no  & \yes & \yes  & \no  & \yes & \no  & \yes & \no  & \no  & \no   & \yes & \yes \\
			Other~properties  & \yes & \yes & \no   & \no  & \no  & \yes & \yes & \yes & \yes & \no   & \yes  & \no  \\
			Numerical HPs     & \no  & \yes & \yes  & \no  & \no  & \yes & \no  & \no  & \no  & \yes  & \no & \yes \\
			Structural HPs    & \no  & \no  & \no   & \no  & \no  & \no  & \no  & \no  & \no  & \no   & \no & \yes \\
			Regression HPs    & \yes & \yes & \yes  & \yes & \yes & \yes & \no  & \yes & \no  & \yes  & \no & \yes \\
			Timings           & \no  & \no  & \yes  & \yes & \no  & \no  & \no  & \no  & \no  & \no   & \no  & \yes \\
			\bottomrule
		\end{tabular}
	\end{minipage}%
	\begin{minipage}[t]{0.4\linewidth}%
		\caption{\emph{Related work.}
			See \cref{si:relatedwork} for details.
			\small\selectfont 
			Finite systems = study uses data\-sets of finite systems, such as molecules or clusters;
			Periodic systems = uses data\-sets of periodic systems, such as crystalline materials;
			Other properties = evaluate properties other than energy or its derivatives;
			numerical\,/\,structural\,/\,regression HPs = whether numerical hyperparameters of representations, structural hyperparameters of representations, or regression hyperparameters are optimized automatically.
			\label{tab:relwork}}
		\mbox{}
	\end{minipage}
\end{table*}


\section{Role and types of representations}\label{sec:taxonomy}

An $N$-atom system formally has $3N-6$ degrees of freedom.\linebreak[4]
Covering those with $M$ samples per dimension requires $M^{3N-6}$ reference calculations, which is infeasible except for the smallest systems.
How then is it possible to learn high-dimensional energy surfaces?

Part of the answer is that learning the whole energy surface is unnecessary, as configurations high in energy become exponentially unlikely---it is sufficient to learn low-energy regions.
Another reason is that the regression space's formal dimensionality is less important than the data distribution in this space. (\cref{si:strucdata})
Representations can have thousands of dimensions, but their effective dimensionality \cite{bbm2008} can be much lower if they are highly correlated.
The role of representations is, therefore, to map atomistic systems to spaces amenable to regression. 
These spaces, together with the data's distribution, determine the efficiency of learning.

We distinguish between \emph{local} representations that describe parts of an atomistic system, such as atoms in their environment, \cite{bp2007q,bkc2013q} and \emph{global} ones that describe the whole system.
For global representations, represented systems are either \emph{finite}, such as molecules and clusters, or \emph{periodic}, such as bulk crystals and surfaces.
(\cref{tab:classification})

\begin{table}[b!]
	\caption{\emph{Types of representations.}
		We distinguish between local (atoms in their environment) and global (holistic, whole system) representations, as well as between representations for finite (molecules, clusters) and periodic systems (bulk crystals, surfaces).
		Local representations have finite support, and thus do not need to distinguish between finite and periodic systems.
		See Glossary for abbreviations.
		\label{tab:classification}}

	\vspace{\captionskip}%
	\begin{tabular}{@{}p{7em}p{\linewidth-7em-2\tabcolsep}@{}}
		\toprule
		Category & Representation \\
		\midrule
		Local             & \raggedright ACE, BS, DECAF, FCHL, GM, MTP, NICE, OMF, SF, SOAP, WST \tabularnewline
		\\[-1.5ex]
		Global (finite)   & \raggedright BoB, CM, HDAD, IDMBR, MBTR, MOB, OMF, WST \tabularnewline
		\hphantom{Global} (periodic) & MBTR \\
		\bottomrule
	\end{tabular}
\end{table}

Local representations are directly suitable for local properties, such as forces, nuclear magnetic resonance shifts, or core-level excitations, \cite{rrl2015q} which depend only on a finite-size environment of an atom.
Extensive global properties (\cref{si:extensivity}) such as energies can be modeled with local representations via additive approximations, summing over atomic contributions (\cref{si:atomiccontr}).
Since local representations require only finite support, it does not matter whether the surrounding system is finite or periodic.
Global representations are suited for properties of the whole system, such as energy, band gap, or the polarizability tensor.
Since periodic systems are infinitely large, global representations usually need to be designed for or adapted to these.
Trade-offs between local and global representations are discussed in \cref{sec:analysis}.

Historically, interpolation has been used to reduce the effort of numerical solutions to quantum problems from the beginning.
Early works employing ML techniques such as Tikhonov regularization and reproducing kernel Hilbert spaces in the late 1980s and throughout the 1990s were limited to small systems. \cite{bbh1986q,dnu1991q,hhlr1992q,hhr1999q}
Representations for high-dimensional systems appeared a decade later, \cite{lhwgsr2006q,bp2007q,bpkc2010q}
underwent rapid development, and constitute an active area of research today.
\Cref{tab:refs} presents an overview.

\begin{table}[bhtp]
	\vspace*{\baselineskip}
	\caption{%
		\emph{Overview of representations.}
		For each representation (Repr.), year of publication (Year), original reference (Orig.), references for further methodological development (Dev.), and availability of implementations (Avail.) are shown.
		See Glossary for abbreviations.
		\label{tab:refs}
	}
	\vspace*{\captionskip}
	
	\begin{center}
	\begin{tabular}{@{\hspace{2pt}}llccc@{\hspace{2pt}}}
		\toprule
		& & \multicolumn{3}{c}{References} \\ \cmidrule(lr){3-5}
		Year & Repr. & Orig.                     & Dev.                                                                                                    & Avail. \\ 
		\midrule
		2007 & SF    & \citenum{bp2007q}         & \citenum{b2011eq,sir2017q,gsbbm2018q,rag2018q,auc2018q}                                                 & \citenum{availsf,hjmfrgrf2020q}   \\
		2010 & BS    & \citenum{bpkc2010q}       & \citenum{wt2018q,stt2019q,s2020bq}                                                                      & \citenum{availbs}                 \\
		2012 & CM    & \citenum{rtml2012q}       & \citenum{m2012eq,rtml2012bq,rdrl2015q,rrl2015q,bbhm2017q}                                               & \citenum{availmbtr,hjmfrgrf2020q} \\  
		2013 & SOAP  & \citenum{bkc2013q}        & \citenum{bpkc2010q,dbcc2016q,bdpbkcc2017q,cwc2018q,c2019dq,jkk2019q,kme2019q,stt2019q,s2020bq,kme2020q} & \citenum{availsoap,hjmfrgrf2020q} \\
		2013 & OMF   & \citenum{sgsmlg2013q}     & \citenum{zafsfrgsgwg2016q,pdcfkdblg2020}                                                                & ---                               \\
		2015 & BoB   & \citenum{hbrplmt2015q}    & \citenum{hl2016q}                                                                                       & \citenum{availqml}                \\
		2015 & WST   & \citenum{hpm2015}         & \citenum{hmp2017q,eehm2017q,bskqh2018q,eehmt2018q,hhrnh2019q,ssblkqh2020q}                              & \citenum{aaelrtzetal2020q}        \\
		2016 & MTP   & \citenum{s2016q}          & \citenum{ps2017q,gps2018q,ns2018q,s2019q}                                                               & \citenum{ngps2020}                \\
		2017 & MBTR  & \citenum{hr2017}          & \citenum{nrbsmrcwh2019q}                                                                                & \citenum{availmbtr,hjmfrgrf2020q} \\
		2017 & HDAD  & \citenum{fhhgsdvkrv2017q} & ---                                                                                                     & ---                               \\
		2018 & DECAF & \citenum{tzk2018q}        & ---                                                                                                     & \citenum{availdecaf}              \\
		2018 & FCHL  & \citenum{fchl2018q}       & \citenum{cbfgl2020q}                                                                                    & \citenum{availqml}                \\
		2018 & IDMBR & \citenum{ptm2018q}        & ---                                                                                                     & \citenum{availidmbr}              \\
		2018 & MOB   & \citenum{wcm2018q}        & \citenum{cwcm2019q,hsclm2020,lhdm2020}                                                                  & ---                               \\
		2019 & ACE   & \citenum{d2019bq}         & \citenum{dbcdeoo2020,d2020q}                                                                            & \citenum{availace}                \\
		2020 & NICE  & \citenum{npc2020q}        & ---                                                                                                     & \citenum{availnice}               \\
		2020 & GM    & \citenum{zk2020q}         & ---                                                                                                     & ---                               \\
		\bottomrule
	\end{tabular}
	\end{center}
\end{table}


\section{Requirements}\label{sec:requirements}

The figure of merit of ML models for fast, accurate interpolation of ab initio properties is sample efficiency:
The number of reference simulations required to reach a given target accuracy.
Imposing physical constraints on representations improves their sample efficiency by removing the need to learn these constraints from the training data.
The demands of speed, accuracy, and sample efficiency give rise to specific requirements,
some of which depend on the predicted property:

\begin{romanenum}
	\item \emph{Invariance} to transformations that preserve the predicted property, including \label{req:Invariance}%
		\begin{romanenum*}
			\item \label{req:InvIndexing} changes in atom indexing (input order, permutations of like atoms),
		\end{romanenum*}
		and often 
		\begin{romanenum*}[resume]
			\item \label{req:InvTranslation} translations
			\item \label{req:InvRotation} rotations 
			\item \label{req:InvReflection} reflections. 
		\end{romanenum*} 
		Predicting tensorial properties requires
		\begin{romanenum*}[resume]
			\item \label{req:Covariance} \emph{covariance} (equivariance) with rotations.
		\end{romanenum*} \cite{gsd2017q,gwcc2018q,cwc2018q,d2020q,ahk2019q,klt2018q,tskylkr2018q,htpak2019q}
		
		Dependence of the property on a global frame of reference, for example, due to the presence of a non-isotropic external field, can affect variance requirements.
	
	\item \emph{Uniqueness,} \label{req:Uniqueness}
		that is, variance against all transformations that change the predicted property:
		Two systems that differ in property should be mapped to different representations.
		
		Systems with equal representation that differ in property introduce errors: \cite{m2012eq,lrrk2015q,pwbocc2020q}
		Because the ML model cannot distinguish them, it predicts the same value for both, resulting in at least one erroneous prediction.
		Uniqueness is necessary and sufficient for reconstruction, up to invariant transformations, of an atomistic system from its representation. \cite{bkc2013q,kme2020q}
		
	\item%
		\begin{romanenum*} \item \label{req:Continuity} \emph{Continuity,} \end{romanenum*} 
		and ideally \begin{romanenum*}[resume] \item \ \emph{differentiability}, \end{romanenum*}
		with respect to atomic coordinates.
		
		Discontinuities work against the regularity assumptions in ML models, which try to find the least complex function compatible with the training data.
		Intuitively, continuous functions require less training data than functions with jumps.
		Differentiable representations enable differentiable ML models.
		If available, reference gradients can further constrain the interpolation function (``force matching''), improving sample efficiency. \cite{lhr2009q,bc2015q,csmt2018q}
		
	\item \emph{Computational efficiency} relative to the reference simulations. \label{req:Runtime}

        For an advantage over simulations alone (without ML), overall computational costs should be reduced by one or more orders of magnitude to justify the effort.
        The difference between running reference simulations and computing representations usually dominates costs. (\cref{si:compcosts})
		Therefore, the results of computationally sufficiently cheaper simulations, for example, from a lower level of theory, can be used to construct representations \cite{sgsmlg2013q,wcm2018q} or to predict properties at a higher level of theory (``$\Delta$-learning''). \cite{rdrl2015q,wcm2018q,sgc2019q}
		
	\item \emph{Structure}\label{req:Structure}
		of representations and the resulting data distribution should be suitable for regression. (\cref{si:strucdata,si:rolerepr})
		It is useful if feature vectors always have the same length.\cite{b2011eq,cgly2018q}.

		Representations often have a Hilbert space structure, featuring an inner product, completeness, projections, and other advantages.
		Besides the formal space defined by the representation, the structure of the subspace spanned by the data is critical. \cite{pwbocc2020q,gfc2020}
		This requirement is currently less well understood than \ref{req:Invariance}--\ref{req:Runtime} and evaluated mostly empirically (see \cref{sec:benchmark}).
		
	\item \emph{Generality},
		in the sense of being able to encode any atomistic system.

		While current representations handle finite and periodic systems,
		less work was done on charged systems, excited states, continuous spin systems, isotopes, and systems subjected to external fields.
\end{romanenum}

\noindent
\emph{Simplicity}, both conceptually and in terms of implementation, is, in our opinion, a desirable quality of representations, albeit hard to quantify.

\bigskip

\noindent
The above requirements preclude direct use of Cartesian coordinates, which violate requirement~\ref{req:Invariance},
and internal coordinates, which satisfy \ref{req:InvTranslation}--\ref{req:InvReflection} but are still system-specific, violating \ref{req:Structure} and possibly~\ref{req:InvIndexing} if not defined uniquely.
Descriptors and fingerprints from cheminformatics \cite{tc2009} and materials informatics violate \ref{req:Uniqueness} and \ref{req:Continuity}.

Simple representations such as the Coulomb matrix (\cref{sec:other}) either suffer from coarse-graining, violating~\ref{req:Uniqueness}, or from discontinuities, violating \ref{req:Continuity}.
In practice, representations do not satisfy all requirements exactly (\cref{sec:analysis}) but can achieve high predictive accuracy regardless;
for example, for some datasets, modeling a fraction of higher-order terms can be sufficiently unique already. \cite{jkvak2020q}
The optimal interaction orders to utilize in a representation also depend on the type and amount of data available. \cite{gzfd2020q}


\section{A unified framework}\label{sec:framework}

Based on recent work \cite{wmc2019q,cwc2018q,npc2020q} we describe concepts and notation towards a unified treatment of representations.
For this, we successively build up Hilbert spaces of atoms, $k$-atom tuples, local environments, and global structures,
using group averaging to ensure physical invariants and tensor products to retain desired information and construct invariant features.

\subsection{Representing atoms, environments, systems}

Information about a single atom, such as position and proton number, is represented as an abstract ket $\ket{\alpha}$ in a Hilbert space~$\mathcal{H}_{\alpha}$.
Relations between $k$~atoms, where their order can matter, are encoded as $k$-body functions $g_k : \mathcal{H}_{\alpha}^{\times k}\rightarrow \mathcal{H}_{g}$. (\cref{si:kbodyf})
These functions can be purely geometric, such as distances or angles, but could also be of (al)chemical or mixed nature.
Tuples of atoms and associated many-body properties are thus elementary tensors of a space $\mathcal{H} \equiv \mathcal{H}_{\alpha}^{\otimes k} \otimes \mathcal{H}_{g}$,
\begin{equation*}
	\ket{\mathcal{A}_{\alpha_1...\alpha_k}} \equiv \ket{\alpha_1} \otimes ... \otimes \ket{\alpha_k} \otimes g_k(\ket{\alpha_1},...,\ket{\alpha_k}).
\end{equation*}
A local environment of an atom $\ket{\alpha}$ is represented via the relations to its $k-1$ neighbors by keeping $\ket{\alpha}$ fixed:
\begin{equation*}
	\ket{\mathcal{A}_{\alpha}} \equiv \sum_{\alpha_1,\ldots,\alpha_{k-1}} \ket{\mathcal{A}_{\alpha,\alpha_1,\ldots,\alpha_{k-1}}} . \label{defLocal}
\end{equation*}
Weighting functions can reduce the influence of atoms far from $\ket{\alpha}$;
we include these in~$g_k$.
An atomistic systems as a whole is represented by summing over the local environments of all its atoms:
\begin{equation*}
	\ket{\mathcal{A}} = \sum_{\alpha_i} \ket{\mathcal{A}_{\alpha_i}} = \sum_{\alpha_1,\ldots,\alpha_{k}} \ket{\mathcal{A}_{\alpha_1,\ldots,\alpha_{k}}} . \label{defGlobal}
\end{equation*}
For periodic systems, this sum diverges, which requires either 
exploiting periodicity, for example, by working in reciprocal space, or 
employing strong weighting functions and keeping one index constrained to the unit cell. \cite{hr2017}

\subsection{Symmetries, tensor products, and projections}

Representations incorporate symmetry constraints (\cref{sec:requirements}) by using invariant many-body functions~$g_k$, such as distances or angles, 
or through explicit symmetrization via group averaging~\cite{wmc2019q}.
Explicit symmetrization transforms a tensor~$\ket{T}$ by integrating over a symmetry group $\mathcal{S}$ with right-invariant Haar measure~$dS$,
\begin{equation}
	\ket{T}_{\mathcal{S}} \equiv \int\limits_{\mathcal{S}} S\ket{T}dS ,   \label{equ:groupaveraging}
\end{equation}
where symmetry transformations $S\in \mathcal{S}$ act separately on each subspace of $\mathcal{H}$ or parts thereof.
For example, for rotational invariance, only the atomic positions in $\mathcal{H}_{\alpha}$ change.
Rotationally invariant features can be derived from tensor contractions, as any full contraction of contravariant with covariant tensors yields rotationally invariant scalars.\cite{zk2020q}

Sometimes group averaging can integrate out desired information encoded in $\ket{T}$. 
To counter this, one can perform {tensor products} of $\ket{T}$ with itself, effectively replacing $\mathcal{H}$ by $\mathcal{H}^{\otimes \nu}$.
Together, this results in a generalized transform
\begin{equation}
	\ket{T^\nu}_{\mathcal{S}} \equiv \int\limits_{\mathcal{S}} (S\ket{T})^{\otimes \nu}dS .
	\label{equ:generalizedtransform}
\end{equation}
To retain only part of the information in~$\mathcal{A}$, 
one can project onto orthogonal elements $\{\ket{h_l}\}_{l=1}^m$ of $\mathcal{H}$ 
via an associated projection operator $\mathcal{P} = \sum_l \ket{h_l}\bra{h_l}$. 
Inner products and induced distances between representations are then given by
\begin{equation}
	\label{equ:fwipdist}
	\braket{ \mathcal{A} | \mathcal{P} | \mathcal{A'} } 
	\;\text{and}\;
	d_{\mathcal{P}}(\ket{\mathcal{A}},\ket{\mathcal{A}'}) = ||\mathcal{P}\ket{\mathcal{A}}-\mathcal{P}\ket{\mathcal{A}'}||_{\mathcal{H}} .
\end{equation}


\section{Selected representations}\label{sec:representations}

We discuss three representations that fulfill the requirements in \cref{sec:requirements} and for which an implementation not tied to a specific regression algorithm and supporting finite and periodic systems was openly available.
These representations are empirically compared in \cref{sec:benchmark}.


\subsection{Symmetry functions}\label{sec:SF}

Symmetry functions\cite{bp2007q,b2011eq} (SFs) describe $k$-body relations between a central atom and the atoms in a local environment around it. (\cref{si:kbodyf,si:centralatomneighborhood})
They are typically based on distances (\emph{radial} SFs, $k=2$) and angles (\emph{angular} SFs, $k=3$).
Each SF encodes a local feature of an atomic environment, for example the number of H atoms at a given distance from a central C atom.

\begin{figure}[tbhp]
	\includegraphics[width=\linewidth]{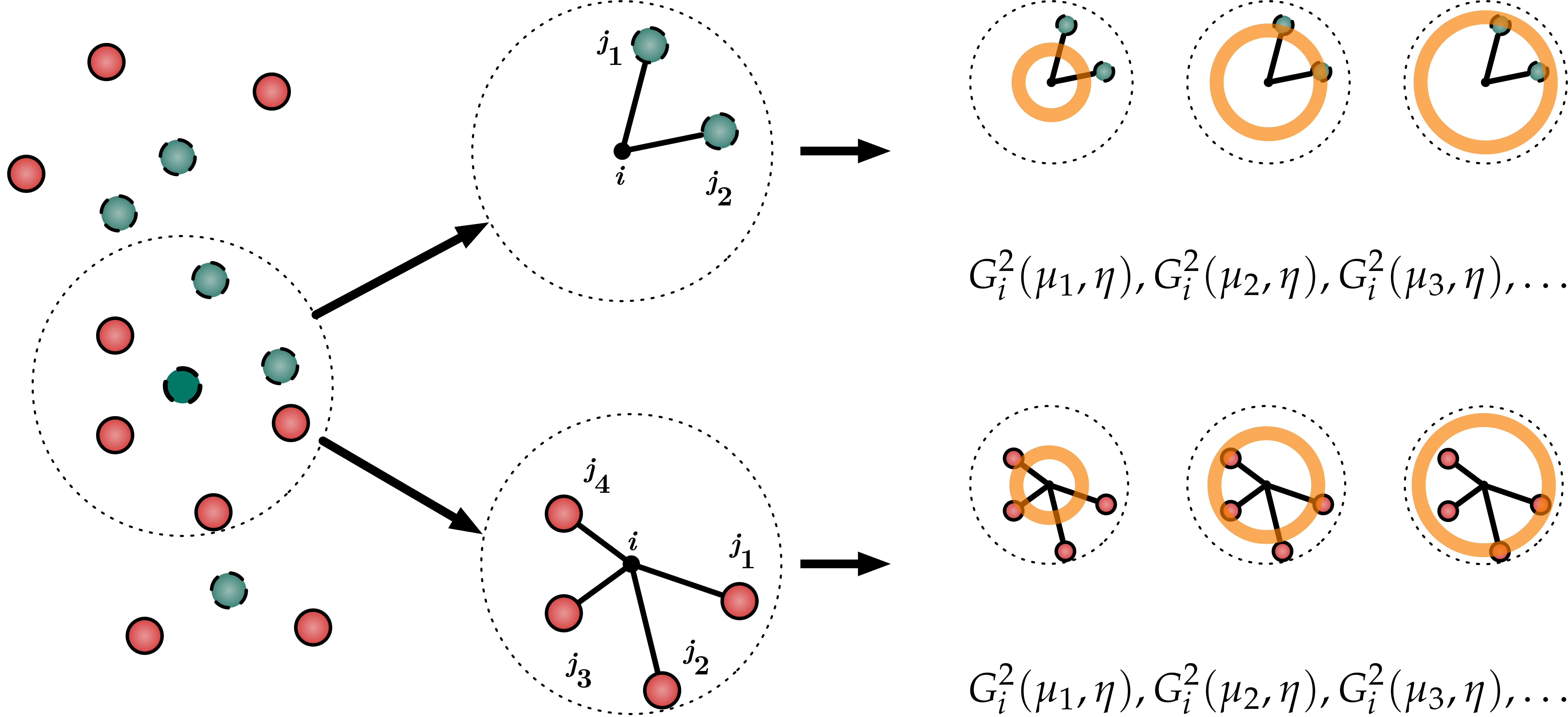}%
	
	\vspace*{\captionskip}
	\caption{%
		\emph{Symmetry functions.}
		Shown are radial functions $G_i^2(\mu,\eta)$ (\cref{equ:SFa}) for increasing values of~$\mu$.
		The local environment of a central atom is described by summing contributions from neighboring atoms separately by element.
	} \label{fig:SymmetryFunctions}
\end{figure}

For each SF and $k$-tuple of chemical elements, contributions are summed.
Sufficient resolution is achieved by varying the HPs of an SF.
For continuity (and differentiability), a cut-off function ensures that SFs decay to zero at the cut-off radius.
Two examples of SFs from \eref{b2011eq} (see \cref{tab:refs,si:hpsf} for further references and SFs) are
\begin{gather}
	G^2_i = \sum_{j} \exp \bigl(-\eta (d_{ij}-\mu)^2 \bigr) f_c(d_{ij}) \qquad\qquad \notag \\
	G^4_i = \,2^{1-\zeta} \sum_{j,k \neq i} (1 + \lambda \cos \theta_{ijk})^{\zeta} \;\cdot \qquad\qquad\quad \label{equ:SFa} \\
	        \exp{\bigl(-\eta (d_{ij}^2 + d_{ik}^2 + d_{jk}^2)\bigr)} 
	        f_c(d_{ij}) \, f_c(d_{ik}) \, f_c(d_{jk}) \notag
\end{gather}
where $\eta, \mu, \zeta, \lambda$ are numerical HPs controlling radial broadening, shift, angular resolution, and angular direction, respectively, $d_{ij}$ is a distance, $\theta_{ijk}$ is the angle between atoms $i$, $j$, $k$, and $f_c$ is a cut-off function.
\Cref{fig:SymmetryFunctions} illustrates the radial SFs in \cref{equ:SFa}.
The choice of which SFs to use is a structural HP.
Variants of SFs include partial radial distribution functions, \cite{sgbsmg2014q} SFs with improved angular resolution \cite{sir2017q} and reparametrizations for improved scaling with the number of chemical species \cite{gsbbm2018q,rag2018q,auc2018q}.

In terms of the unified notation, SFs use invariant functions~$g_k$ based on distances and angles, multiplied by a cut-off function, to describe local environments $\ket{\mathcal{A}_{\alpha}}$.
Projections~$\mathcal{P}$ onto tuples of atomic numbers then separate contributions from different combinations of chemical elements.

\subsection{Many-body tensor representation}\label{sec:MBTR}

\begin{figure}[thbp]
	\includegraphics[width=\linewidth]{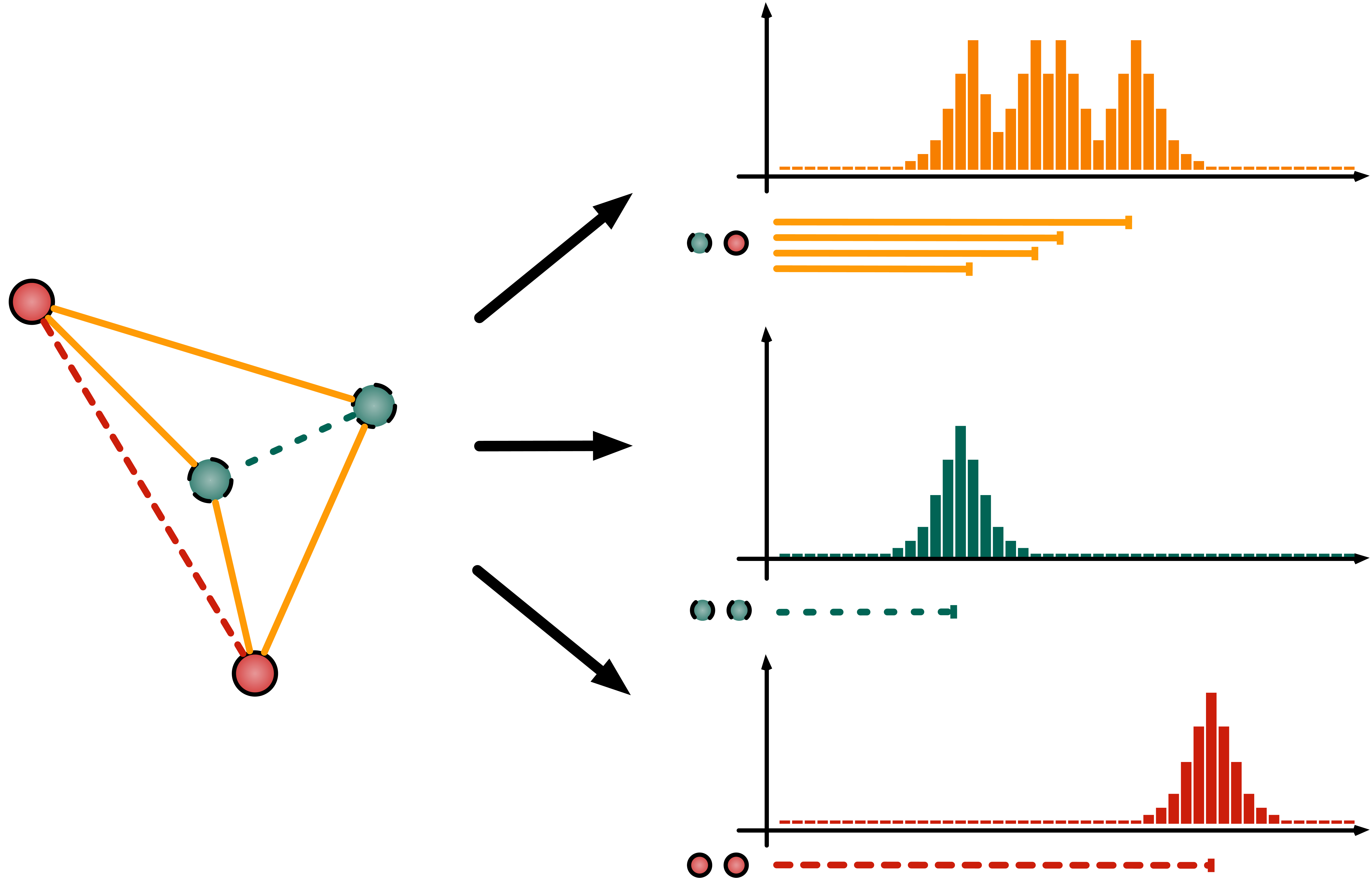}
	
	\vspace*{\captionskip}
	\caption{%
		\emph{Many-body tensor representation.}
		Shown are broadened distances (no weighting) arranged by element combination
	} \label{fig:ManyBodyTensor}
\end{figure}

The global many-body tensor representation~\cite{hr2017} (MBTR) consists of broadened distributions of $k$-body terms, arranged by element combination. 
For each $k$-body function and $k$-tuple of elements, all corresponding terms (for example, all distances between C and H atoms) are broadened and summed up (\cref{fig:ManyBodyTensor}). 
The resulting distributions describe the geometric features of an atomistic system:
\begin{equation}
	\label{equ:mbtr}
	f_k(x, z_1, \ldots, z_k) = \sum_{i_1,\ldots,i_k} w_k \; \mathcal{N}(x|g_k) \prod_{j=1}^k \delta_{z_j,Z_{i_j}} ,
\end{equation}
where $w_k$ is a weighting function that reduces the influence of tuples with atoms far from each other, and $g_k$ is a $k$-body function; both $w_k$ and $g_k$ depend on atoms $i_1, \ldots, i_k$.
$\mathcal{N}(x|\mu)$ denotes a normal distribution with mean~$\mu$, evaluated at~$x$.
The product of Kronecker~$\delta$-functions restricts to the given element combination $z_1,\ldots,z_k$.

Periodic systems can be treated by using strong weighting functions and constraining one index to the unit cell.
In practice, \cref{equ:mbtr} can be discretized.
Structural HPs include the choice of $w_k$ and $g_k$; numerical HPs include variance of normal distributions.
Requiring one atom in each tuple to be the central atom results in a local variant. \cite{localmbtr2019}

In terms of the unified notation, MBTR uses distribution-valued functions~$g_k$, including weighting, with distributions centered on $k$-body terms such as (inverse) distances or angles.
The outer-product structure of~$\ket{\mathcal{A}}$ corresponds to the product of $\delta$-functions in \cref{equ:mbtr}, which selects for specific $k$-tuples of chemical elements.


\subsection{Smooth overlap of atomic positions}\label{sec:SOAP}

\begin{figure}
	\includegraphics[width=\linewidth]{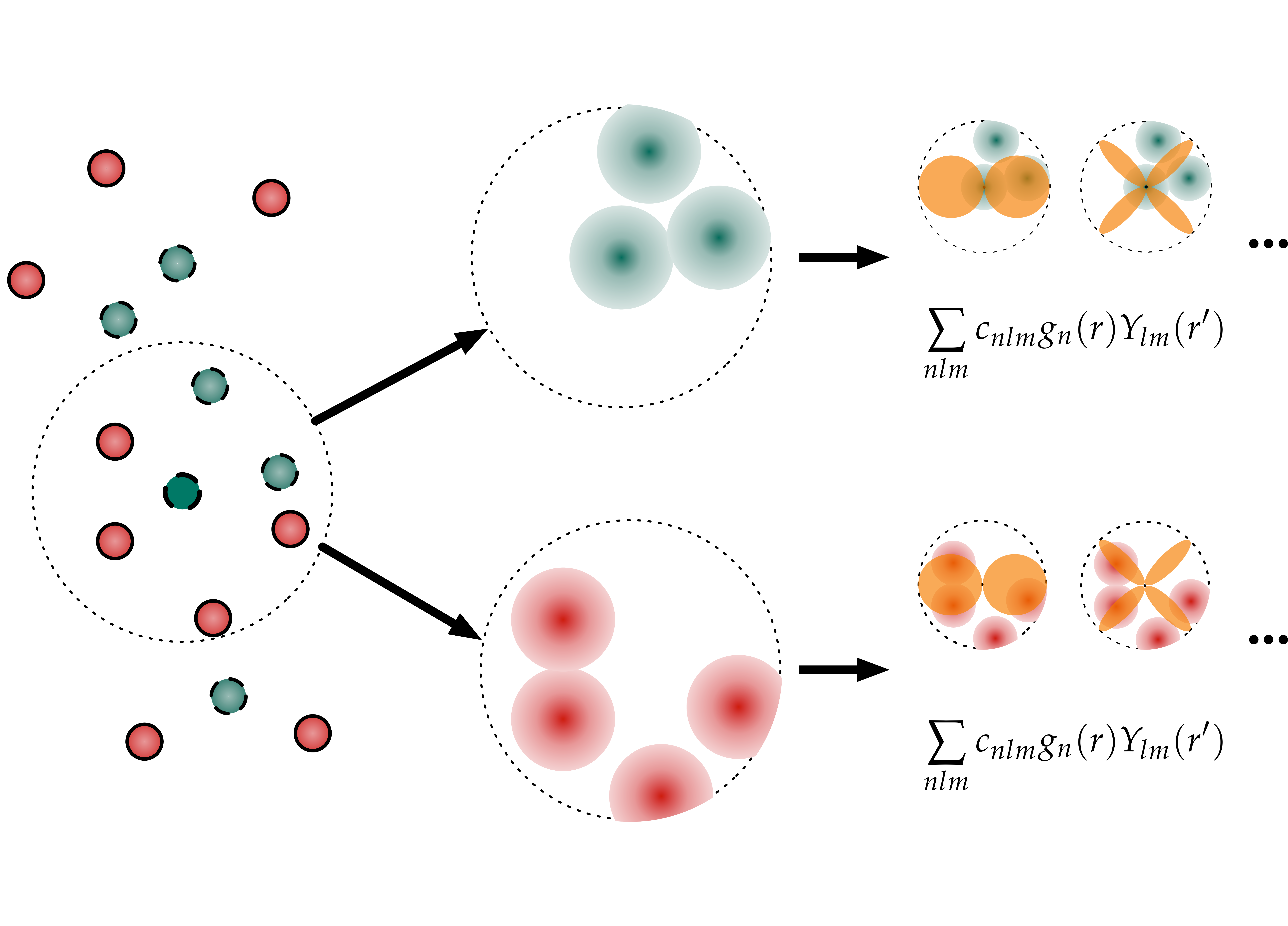}
	
	\vspace*{\captionskip}
	\caption{%
		\emph{Smooth overlap of atomic positions.}
		The local density around a central atom is modeled by atom-centered normal distributions and expanded into radial and spherical harmonics basis functions.
	} \label{fig:SmoothOverlapAtomicPositions}
\end{figure}

Smooth overlap of atomic positions \cite{bkc2013q} (SOAP) representations expand a central atoms' local neighborhood density, approximated by Gaussian functions located at atom positions, in orthogonal radial and spherical harmonics basis functions (\cref{fig:SmoothOverlapAtomicPositions}):
\begin{equation}
	\label{equ:soapexpansion}
	\rho(\vec r) = \sum_{n,l,m} c_{n,l,m} \, g_n(\vec{r}) \, Y_{l,m}(\vec{r}) ,
\end{equation}
where $c_{n,l,m}$ are expansion coefficients, $g_n$ are radial, and $Y_{l,m}$ are (angular) spherical harmonics basis functions.
From the coefficients, rotationally invariant quantities can be constructed, such as the {power spectrum} 
\begin{equation}
	\label{equ:powerspectrum}
	p_{n,n',\ell} = \sum_m c_{n,\ell,m} c^*_{n',\ell,m}  ,
\end{equation}
which is equivalent to a radial and angular distribution function, \cite{jkk2019q} and therefore captures up to three-body interactions.
Numerical HPs are the maximal number of radial and angular basis functions, the broadening width, and the cut-off radius.

An alternative to the power spectrum is the \emph{bispectrum} \cite{bpkc2010q} (BS), a set of invariants that couples multiple angular momentum and radial channels. 
The Spectral Neighbor Analysis Potential (SNAP) includes quadratic terms in the BS components. \cite{wt2018q}
Extensions of the SOAP framework include recursion relations for faster evaluation \cite{c2019dq} and alternative radial basis functions~$g_n$, such as third- and higher-order polynomials, \cite{c2019dq} Gaussian functions, \cite{hjmfrgrf2020q} and spherical Bessel functions of the first kind. \cite{kme2019q,kme2020q}

In terms of the unified notation, SOAP uses vector-valued~$g_k$ to compute the basis set coefficients in \cref{equ:soapexpansion}.
Analytic group-averaging (symmetry integration) then results in invariant features such as the power spectrum ($\nu=2$, \cref{equ:generalizedtransform}) or bispectrum ($\nu=3$).


\section{Other representations}\label{sec:other}

Many other representations were proposed.

The Coulomb matrix \cite{rtml2012q} (CM) globally describes a system via inverse distances between atoms but does not contain higher-order terms.
It is fast to compute, easy to implement, and in the commonly used sorted version (see footnote reference~25 in \eref{rtml2012q}), allows reconstruction of an atomistic system via a least-squares problem.
However, its direct use of atomic numbers to encode elements is problematic, and it suffers either from discontinuities in the sorted version or from information loss in the diagonalized version as its eigenspectrum is not unique \cite{m2012eq,rtml2012bq}.
A local variant exists. \cite{bbhm2017q}

The bag-of-bonds \cite{hbrplmt2015q} (BoB) representation uses the same inverse distance terms as the CM but arranges them by element pair instead of by atom pair.
The ``BA-rep\-res\-ent\-a\-tion'' \cite{hl2016q} extends this to higher-order interactions by using bags of dressed atoms, distances, angles, and torsions.
The inverse-distance many-body representation \cite{ptm2018q} (IDMBR) employs higher powers of inverse distances and separation by element combinations.

Histograms of distances, angles, and dihedral angles \cite{fhhgsdvkrv2017q} (HDAD) are histograms of geometric features organized by element combination.
This global representation is similar to MBTR but typically uses fewer bins, without broadening or explicit weighting. 

The Faber-Christensen-Huang-von~Lilienfeld representation \cite{fchl2018q,cbfgl2020q} (FCHL)
describes atomic environments with normal distributions over 
row and column in the periodic table ($k=1$), interatomic distances ($k=2$),
and angles ($k=3$), scaled by power laws. 
In the FCHL18 variant, \cite{fchl2018q} the full continuous distributions are used, requiring an integral kernel for regression.
Among other optimizations, FCHL19 \cite{cbfgl2020q} discretizes these distributions, 
similar to the approach taken by SFs, and can be used with standard vector kernels.

Wavelet scattering transforms \cite{hpm2015,hmp2017q,eehm2017q,bskqh2018q,eehmt2018q,hhrnh2019q,ssblkqh2020q} (WST) use a convolutional wavelet frame representation to describe variations of (local) atomic density at different scales and orientations.
Integrating non-linear functions of the wavelet coefficients yields invariant features, where second- and higher-order features couple two or more length scales.
Variations use different wavelets (Morlet, \cite{hpm2015,hmp2017q} solid harmonic, or atomic orbital \cite{eehm2017q,eehmt2018q,bskqh2018q,ssblkqh2020q}) and radial basis functions (exponential \cite{eehm2017q}, Laguerre polynomials \cite{bskqh2018q,ssblkqh2020q}).

Moment-tensor potentials \cite{s2016q} (MTP) describe local atomic environments using a spanning set of efficiently computable, rotationally and permutationally invariant polynomials derived from tensor contractions.
Related representations include
Gaussian moments \cite{zk2020q} (GM), based on contractions of tensors from (linear combinations of) Gaussian-type atomic orbitals;
the $N$-body iterative contraction of equivariants (NICE) framework, \cite{npc2020q} which uses recursion relations to compute higher-order terms efficiently;
and atomic cluster expansion \cite{d2019bq,dbcdeoo2020,d2020q} (ACE), which employs a basis of isometry- and permutation-invariant polynomials from trigonometric functions and spherical harmonics.

Overlap-matrix fingerprints \cite{sgsmlg2013q,zafsfrgsgwg2016q,pdcfkdblg2020} (OMF) and related approaches \cite{zhj2019q,qwamm2020q} employ the sorted eigenvalues (and derived quantities) of overlap matrices based on Gaussian-type orbitals as representation.
Eigenvalue crossings can cause derivative discontinuities, requiring post-processing \cite{pdcfkdblg2020} to ensure continuity.
Using a molecular orbital basis (MOB \cite{wcm2018q,cwcm2019q} and related approaches \cite{czwe2020q}) adds the cost of computing the basis, for example, localized molecular orbitals via a Hartree-Fock self-consistent field calculation.
Other matrices can be used, such as Fock, Coulomb, and exchange matrices, or even the Hessian, for example, from a computationally cheaper reference method.
%
Density-encoded canonically-aligned fingerprints \cite{tzk2018q} (DECAF) represent the local density in a canonical, invariant coordinate frame found by solving an optimization problem related to kernel principal component analysis.

Tensor properties require covariance (equivariance).
Proposed solutions include 
local coordinates from eigendecompositions, \cite{rrl2015q} which exhibit discontinuities when eigenvalues cross, 
related local coordinate systems, \cite{tzk2018q}
and internal vectors \cite{lkv2015q} (IV), based on inner products of summed neighbor vectors at different scales,
as well as covariant extensions of SOAP \cite{gwcc2018q,cwc2018q} and ACE \cite{d2020q}.


\enlargethispage*{2\baselineskip}\vspace*{-0.5\baselineskip}

\section{Analysis}\label{sec:analysis}

We discuss
relationships between specific representations,
to which degree they satisfy the requirements in \cref{sec:requirements},
trade-offs between local and global representations, 
and relationships to other models and modeling techniques, including systematic selection and generation of features.

\subsection{Relationships between representations}

All representations in \cref{sec:representations} and most representations in \cref{sec:other} are related through the concepts in \cref{sec:framework}.
We distinguish two primary strategies to deal with invariances,
the use of invariant $k$-body functions (BoB, CM, FCHL, HDAD, IDMBR, MBTR, SF)
and explicit symmetrization (ACE, BS, GM, MOB, MTP, NICE, OMF, SOAP, WST).
A similar distinction can be made for kernels. \cite{gzv2018q}
Some representations share specific connections:

An evenly-spaced grid of SFs can be seen as a histogram of distances, angles, or higher-order terms, similar to MBTR and HDAD. 
This suggests a local MBTR or HDAD variant by restricting summation to atomic environments, \cite{localmbtr2019} and a global variant of SFs by summing over the whole system.
A difference is that MBTR explicitly broadens $k$-body terms, whereas SFs implicitly broaden them via the exponential functions in \cref{equ:SFa}.
The original formulation of SFs represents each chemically distinct central atom separately,
whereas MBTR represents each (unique) tuple of $k$~elements in separate tensor components.
Both approaches correspond to using Kronecker~$\delta$ functions on element types.

ACE, BS, GM, MTP, NICE, and SOAP share the idea of generating tensors that are then systematically contracted to obtain rotationally invariant features.
These tensors should form an orthonormal basis, or at least a spanning set, for atomic environments.
Formally, expressing a local neighborhood density in a suitable basis before generating derived features avoids asymptotic scaling with the number of neighboring atoms, \cite{d2019bq}
although HPs, and thus runtime, still depend on it.
Within a representation, recursive relationships can exist between many-body terms of different orders. \cite{c2019dq,npc2020q,dbcdeoo2020}
\Erefs{d2019bq,d2020q,dbcdeoo2020} discuss technical details of the relationships between ACE and SFs, BS, SNAP, SOAP, MTP.

\subsection{Requirements}

Some representations, in particular early ones such as the CM, do not fulfill all requirements in \cref{sec:requirements}.
Most representations fulfill some requirements only in the limit,
that is, absent practical constraints such as truncation of infinite sums, short cut-off radii, and restriction to low-order interaction terms.
The degree of fulfillment often depends on HPs,
such as truncation order, the length of a cut-off radius, or the highest interaction order~$k$ used.
Effects can be antagonistic;
for example, in \cref{equ:soapexpansion}, both \ref{req:Uniqueness}~uniqueness and \ref{req:Runtime}~computational effort increase with $n,l,m$. \cite{bkc2013q}
In addition, not all invariances of a property might be known or require additional effort to model, for example, symmetries \cite{gsd2017q}.

Mathematical proof or systematic empirical verification that a representation satisfies a requirement or related property are sometimes provided:
The symmetrized invariant moment polynomials of MTPs form a spanning set for all permutationally and rotationally invariant polynomials; \cite{s2016q} basis sets can also be constructed. \cite{dbcdeoo2020}
For SOAP, systematic reconstruction experiments demonstrate the dependence of uniqueness on parametrization. \cite{bkc2013q}

While \ref{req:Uniqueness}~uniqueness guarantees that reconstruction of a system up to invariances is possible in principle, accuracy and complexity of this task vary with representation and parametrization.
For example, reconstruction is a simple least-squares problem for the global CM as it comprises the whole distance matrix $\mat{D}_{ij} = ||\vec{r_i}-\vec{r_j}||_2$,
whereas for local representations, (global) reconstruction is more involved.

If a local representation comprises only up to 4-body terms then there are degenerate environments that it cannot distinguish, \cite{pwbocc2020q} but that can differ in property.
Combining representations of different environments in a system can break the degeneracy.
However, by distorting feature space \ref{req:Structure}~structure, these degeneracies degrade learning efficiency and limit achievable prediction errors, even if the training set contains no degenerate systems. \cite{pwbocc2020q}
It is currently unknown whether degenerate environments exist for representations with terms of order $k>4$.
The degree to which a representation is unique can be numerically investigated through the eigendecomposition of a sensitivity matrix based on a representation's derivatives with respect to atom coordinates. \cite{pdcfkdblg2020}

\subsection{Global versus local representations}

Local representations can be used to model global properties by assuming that these decompose into atomic contributions.
In terms of prediction errors, this tends to work well for energies. (\cref{si:extensivity})
Learning with atomic contributions adds technical complexity to the regression model and is equivalent to pairwise-sum kernels on whole systems, (\cref{si:atomiccontr})
with favorable computational scaling for large systems (see \cref{si:compcosts,si:computetimes,tab:compcosts}).
Other approaches to creating global kernels from local ones exist. \cite{dbcc2016q}

Conversely, using global representations for local properties can require modifying the representation to incorporate locality and directionality of the property. \cite{rrl2015q,hjmfrgrf2020q} 
A general recipe for constructing local representations from global ones is to require interactions to include the central atom, starting from $k=2$. \cite{localmbtr2019}

\subsection{Relationships to other models and techniques}

Two modeling aspects directly related to representations are which subset of the features to use and the construction of derived features.
Both modulate feature space dimensionality and \ref{req:Structure}~{structure}.
Adding products of 2-body and 3-body terms as features, for example,
can improve performance, \cite{jkvak2020q} as these features relate to higher-order terms, (\cref{si:kbodyf})
but can also degrade performance if the features are unrelated to the predicted property,
or if there is insufficient data to infer the relationship.
Feature selection tailors a representation to a dataset by selecting a small subset of features that still predict the target property accurately enough.
Optimal choices of features depend on the data's size and distribution.

In this work, we focus exclusively on representations.
In kernel regression, however, kernels can be defined directly between two systems, without an explicit intermediate representation.
For example, $n$-body kernels between atomic environments can be systematically constructed from a non-invariant Gaussian kernel using Haar integration, or using invariant $k$-body functions (\cref{si:kbodyf}), yielding kernels of varying body-order and degrees of freedom. \cite{gzv2018q,gzfd2020q}
Similarly, while neural networks can use representations as inputs, their architecture can also be designed to learn implicit representations from the raw data (end-to-end learning).
In all cases, the Requirements in \cref{sec:requirements} apply.


\section{Empirical comparison}\label{sec:benchmark}

We benchmark prediction errors for the representations from \cref{sec:representations} on three benchmark datasets.
Since our focus is exclusively on the representations, we control for other factors, in particular for 
data distribution, regression method, and HP optimization.

\subsection{Data}

The \dsgdb{} consensus benchmarking dataset\cite{rdrl2014q,availqm9dft10b} comprises 133\,885 organic molecules composed of H, C, N, O, F with up to 9 non-H atoms. (\cref{si:dsgdb})
Ground state geometries and properties are given at the DFT/B3LYP/6-31G(2df,p) level of theory.
We predict $U_0$, the atomization energy at 0\,K.

The \dsba{} dataset \cite{availqm9dft10b,nrbsmrcwh2019q} (\cref{si:dsba})
contains the ten binary alloys AgCu, AlFe, AlMg, AlNi, AlTi, CoNi, CuFe, CuNi, FeV, and NbNi.
For each alloy system, it comprises all structures with up to 8 atoms for face-centered cubic (FCC), body-centered cubic (BCC), and hexagonal close-packed (HCP) crystal types, 15\,950 structures in total.
Formation energies of unrelaxed structures are given at the DFT/PBE level of theory.

The \dstco{} challenge \cite{kaggle} dataset \cite{sgylbhglzs2019} (\cref{si:dstco}) contains 3\,000 ternary (Al$_x$-Ga$_y$-In$_z$)$_2$O$_3$ oxides, $x+y+z=1$, of potential interest as transparent conducting oxides.
Formation and band-gap energies of relaxed structures are provided at the DFT/PBE level of theory.
The dataset contains both relaxed (\dstcor, used here) and approximate (\dstcou) structures as input.
In the challenge, energies of relaxed structures were predicted from approximate structures.

Together, these datasets cover finite and periodic systems, organic and inorganic chemistry, and ground state as well as off-equilibrium structures.
See \cref{si:dsgdb,si:dsba,si:dstco} for details.

\subsection{Method}

We estimate prediction errors as a function of training set size (learning curves, \cref{si:learncurves,si:subsets}).
To ensure that subsets are representative, we control for distribution of elemental composition, size, and energy. (\cref{si:sampling})
This reduces the variance of performance estimates and ensures the validity of the independent-and-identically-distributed data assumption inherent in ML.
All predictions are on data never seen during training.

We use kernel ridge regression \cite{r2015fq} (KRR; predictions are equivalent to those of Gaussian process regression, \cite{rw2006} GPR) with a Gaussian kernel as an ML model. (\cref{si:krr})
KRR is a widely-used non-parametric non-linear regression method.
There are two regression HPs, the length scale of the Gaussian kernel and the amount of regularization. (\cref{si:hpregression})
In this work, training is exclusively on energies; in particular, derivatives are not used.
All HPs, that is, regression HPs, numerical HPs (e.g., a weight in a weighting function), and structural HPs (e.g., which weighting function to use), are optimized with a consistent and fully automatic scheme based on sequential model-based optimization and tree-structured Parzen estimators. \cite{bbbk2011,byc2013} (\cref{si:hpopt})
This setup treats all representations on equal footing.
See \cref{si:hpregression,si:hpsf,si:hpmbtr,si:hpsoap} for details on the optimized HPs.

\subsection{Results}

\begin{figure*}[hpbt]
	\begin{minipage}[t]{0.5\linewidth-1ex}\centering
		\includegraphics[width=\linewidth]{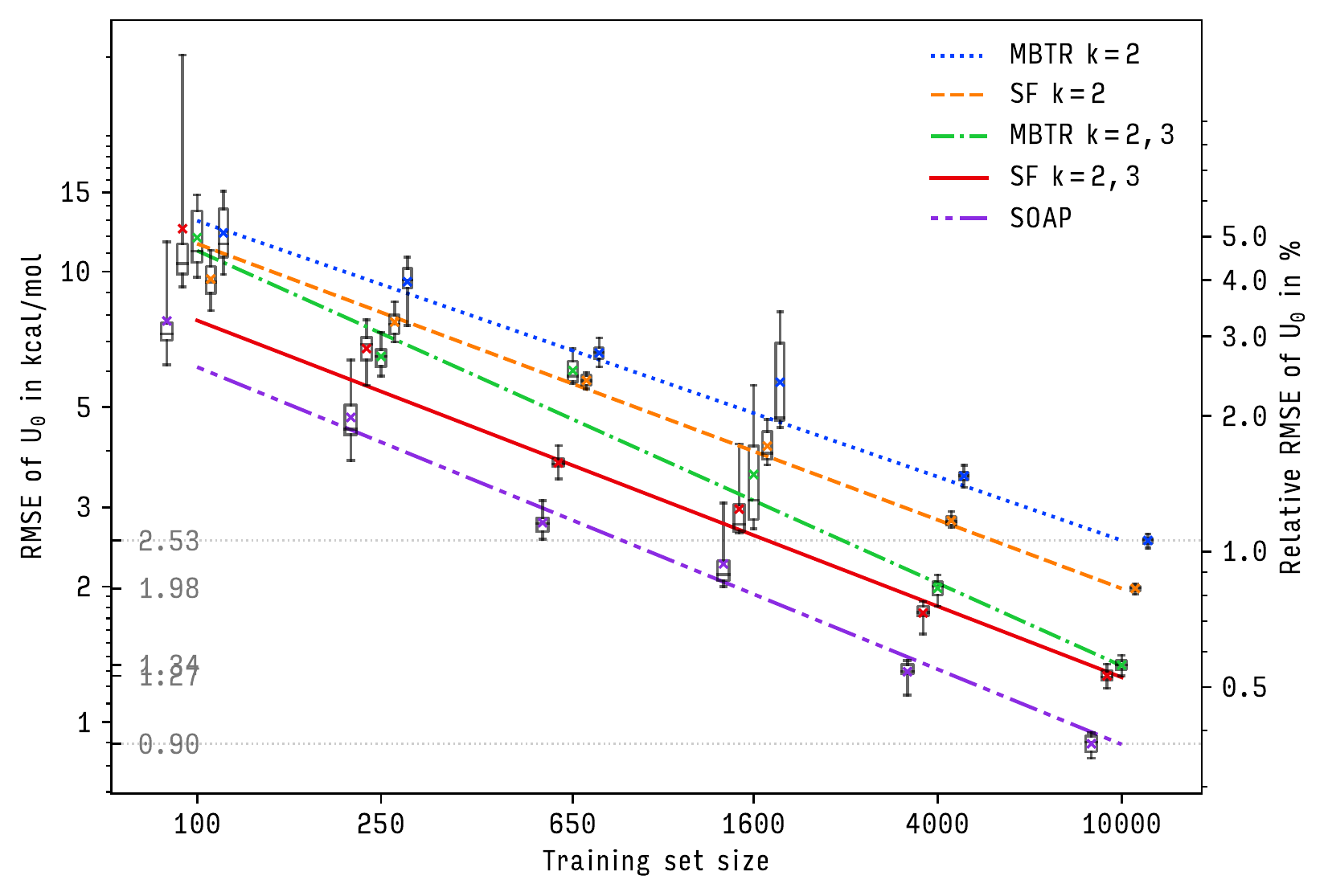}
		
		Dataset \dsgdb.
	\end{minipage}%
	\hfill%
	\begin{minipage}[t]{0.5\linewidth-1ex}\centering
		\includegraphics[width=\linewidth]{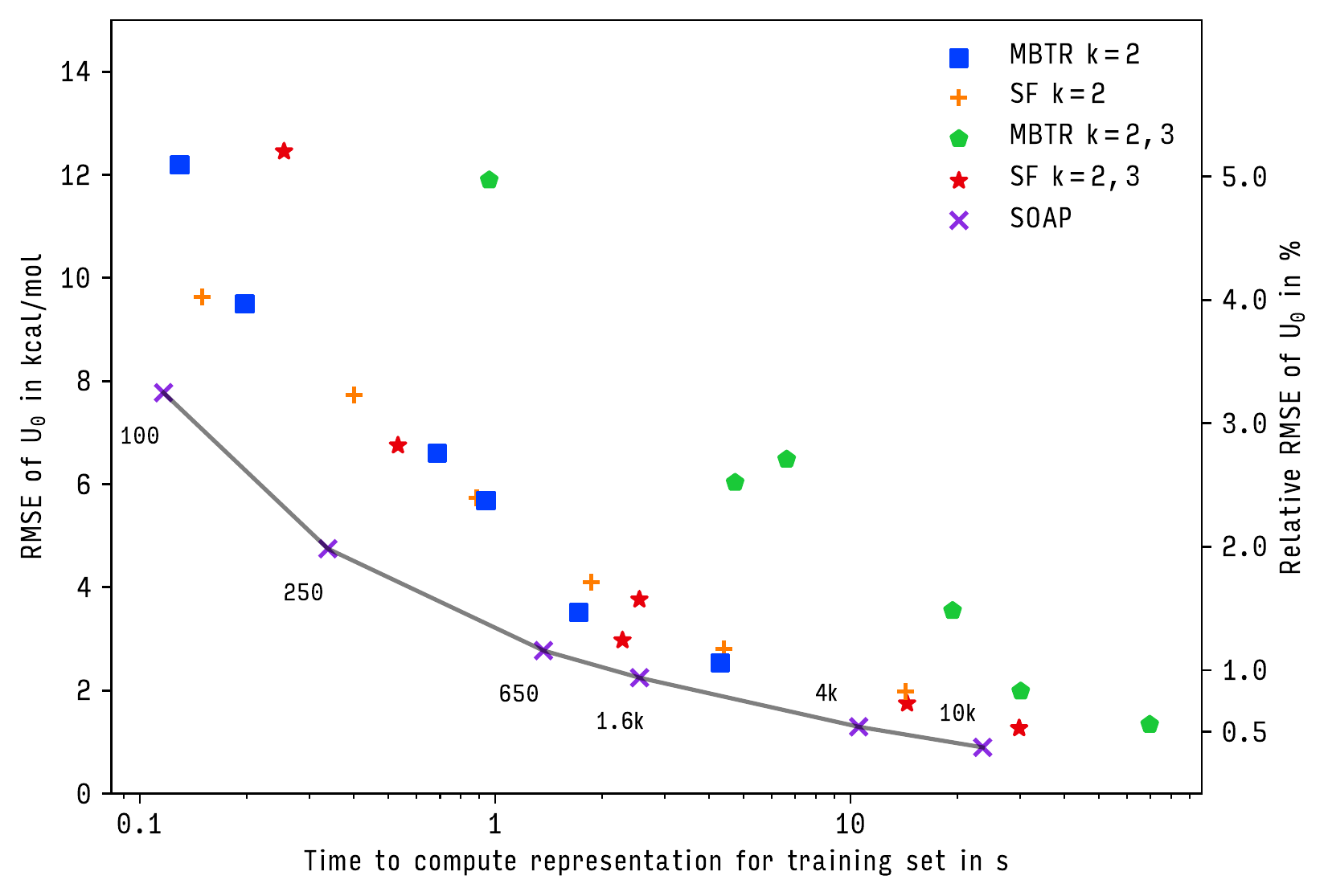}
		
		Dataset \dsgdb.
	\end{minipage}
	
	\medskip
	
	\begin{minipage}[t]{0.5\linewidth-1ex}\centering
		\includegraphics[width=\linewidth]{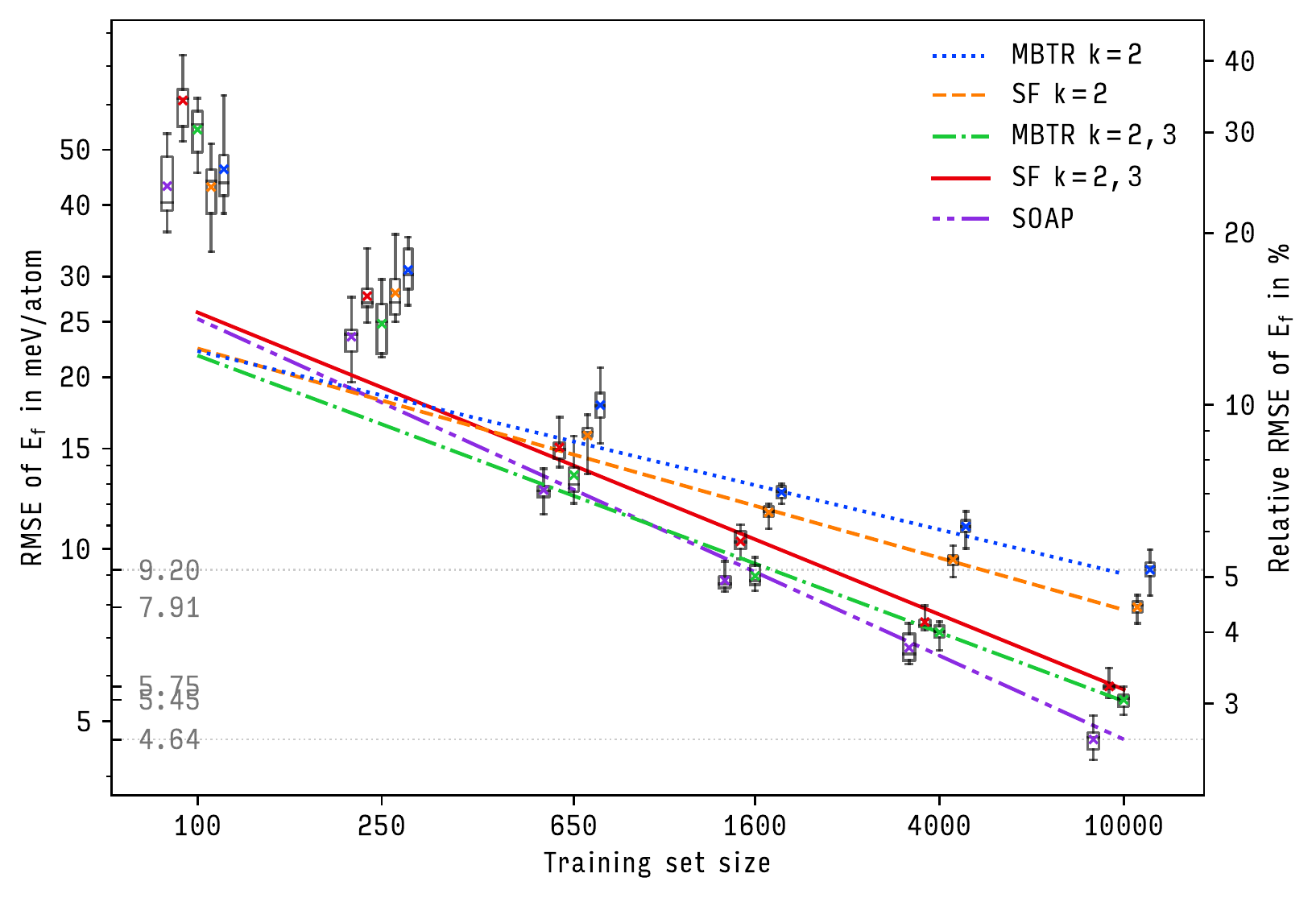}
		
		Dataset \dsba.
	\end{minipage}%
	\hfill%
	\begin{minipage}[t]{0.5\linewidth-1ex}\centering
		\includegraphics[width=\linewidth]{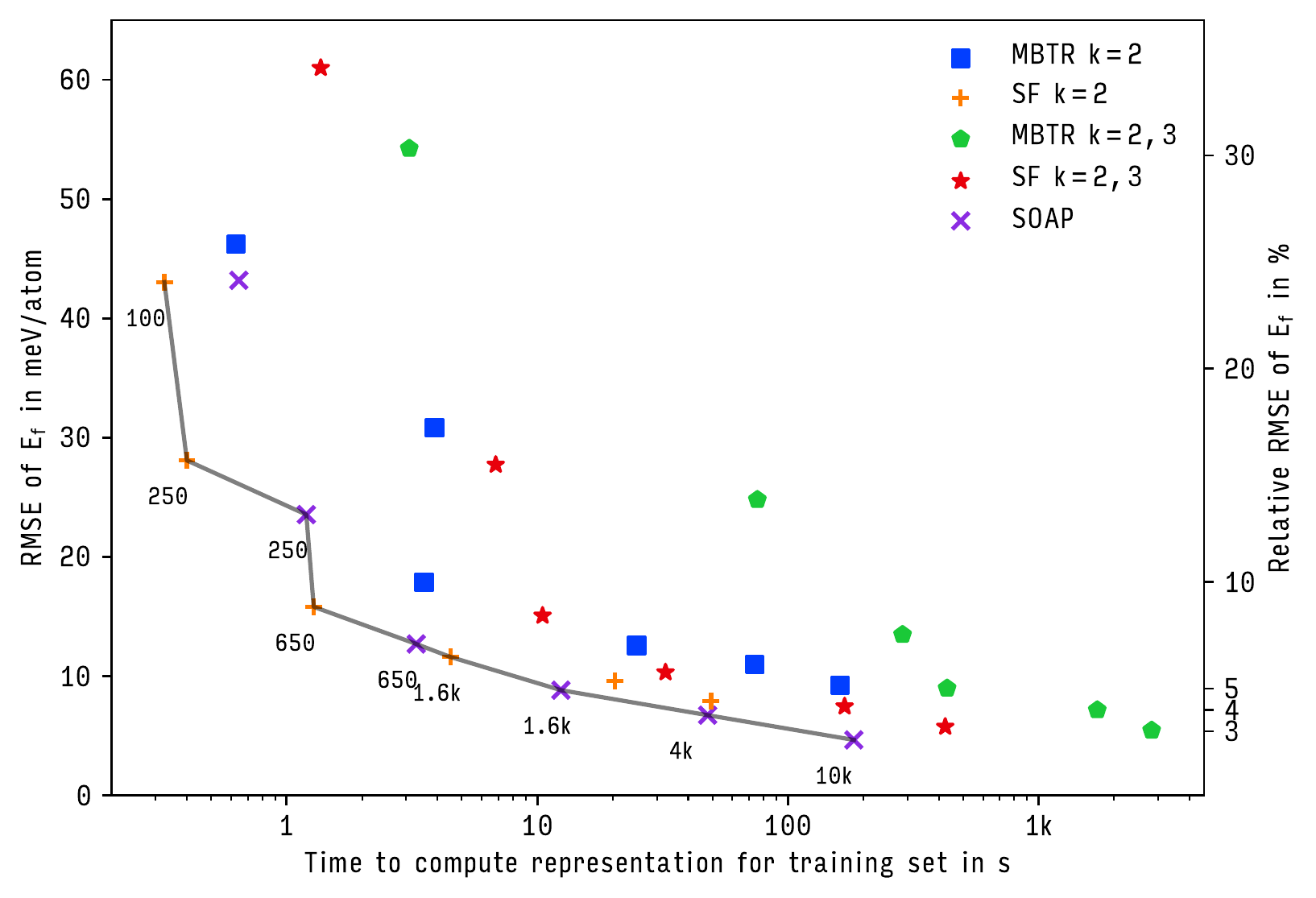}
		
		Dataset~\dsba.
	\end{minipage}
	
	\medskip
	
	\begin{minipage}[t]{0.5\linewidth-1ex}\centering
		\includegraphics[width=\linewidth]{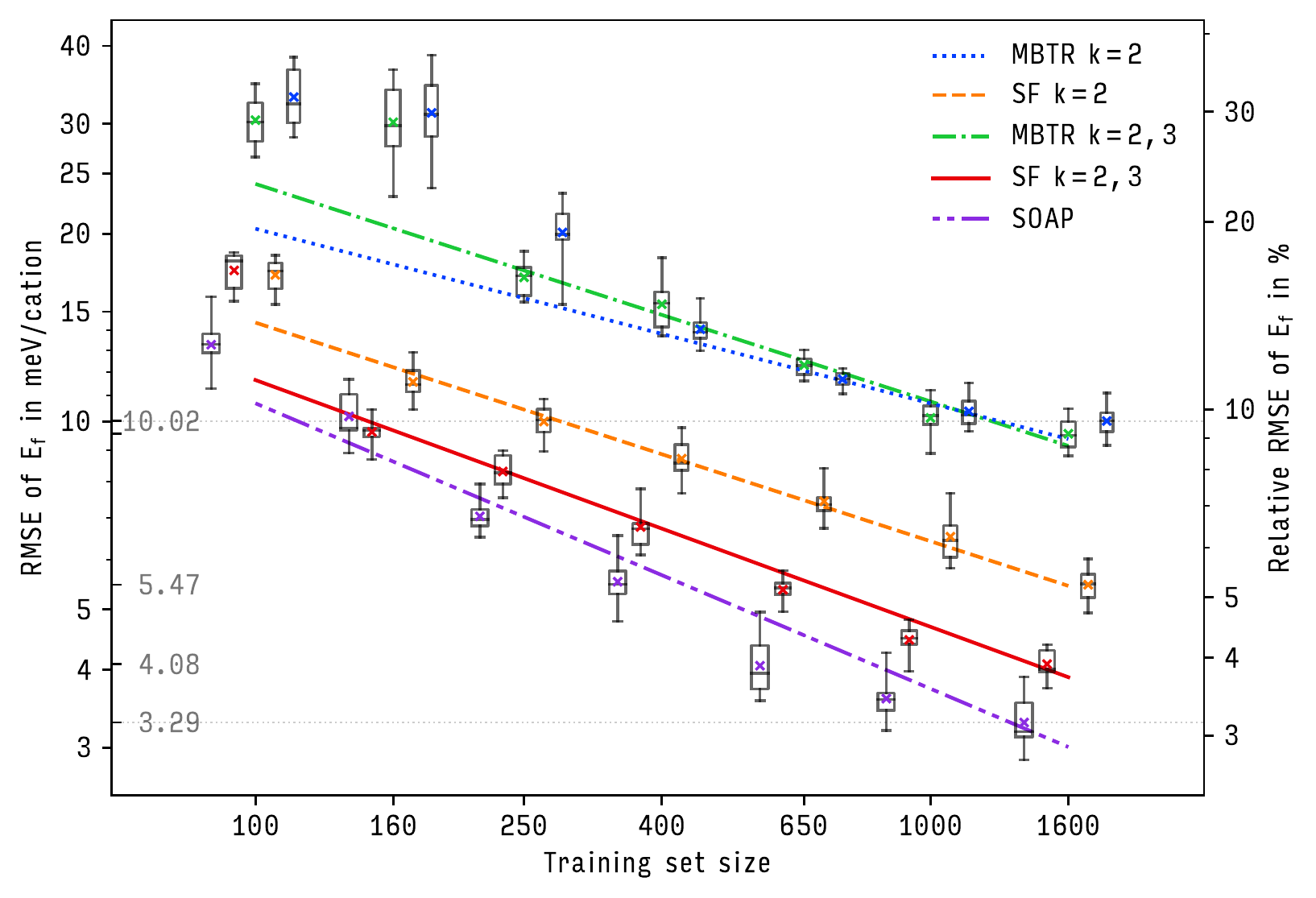}
		
		Dataset \dstcor.
	\end{minipage}%
	\hfill%
	\begin{minipage}[t]{0.5\linewidth-1ex}\centering
		\includegraphics[width=\linewidth]{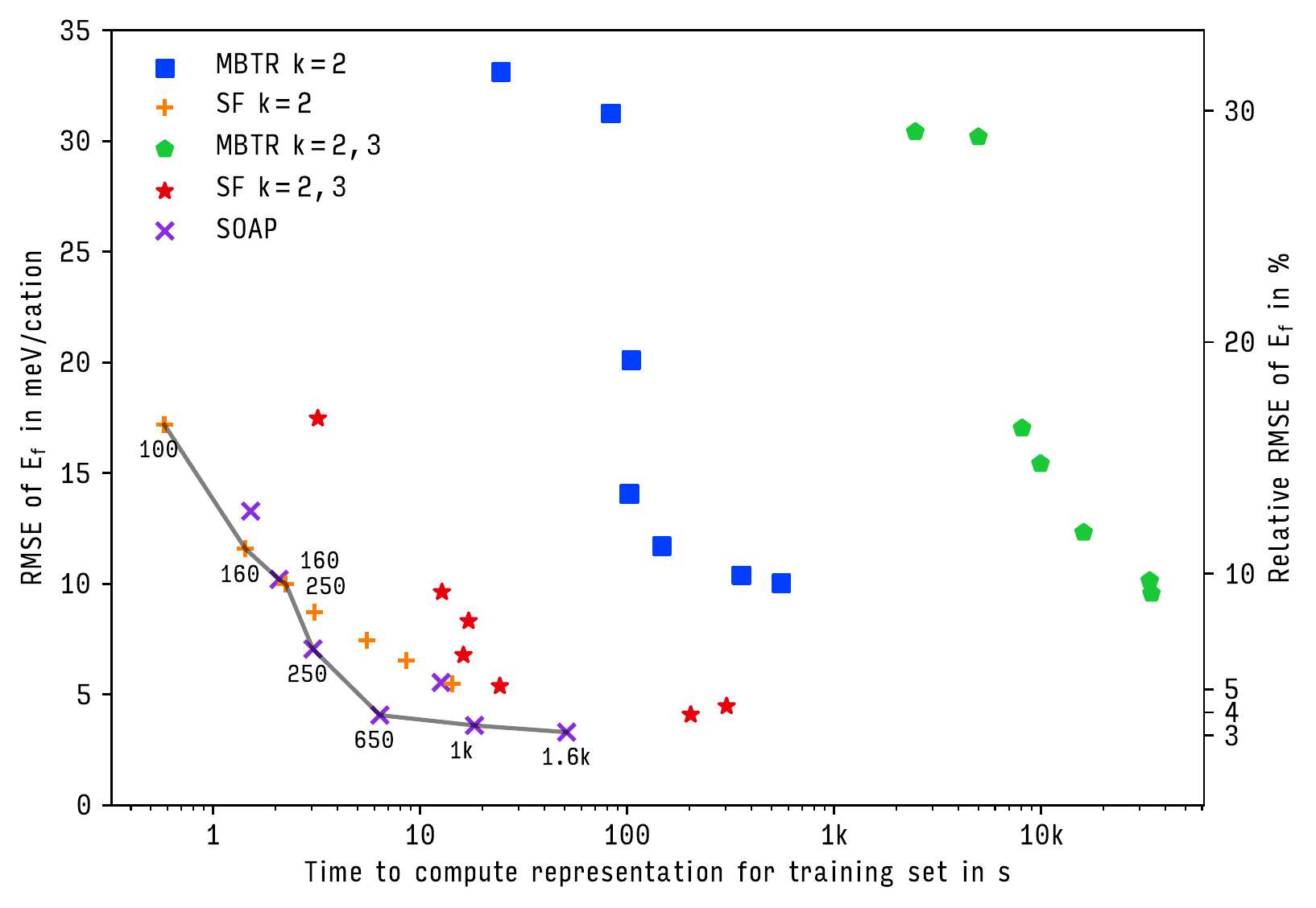}
		
		Dataset~\dstcor.
	\end{minipage}\\

	\vspace*{\captionskip}
	\begin{minipage}[t]{0.5\linewidth-1ex}
		\caption{\emph{Learning curves} for representations in \cref{sec:representations} on datasets \dsgdb{} (top), \dsba{} (middle), and \dstcor{} (bottom).
		Shown is root mean squared error (RMSE) of energy predictions on out-of-sample-data as a function of training set size.
		Boxes, whiskers, bars, crosses show interquartile range, total range, median, mean, respectively.
		Lines are fits to theoretical asymptotic RMSE. (\cref{si:learncurves})
		See Glossary for abbreviations.	
		\label{fig:learncurvesrmse}}
	\end{minipage}%
	\hfill%
	\begin{minipage}[t]{0.5\linewidth-1ex}
		\caption{\emph{Compute times} of representations in \cref{sec:representations} for datasets \dsgdb{}~(top), \dsba{}~(middle), and \dstcor{}~(bottom).
		Shown is root mean squared error (RMSE) of energy predictions on out-of-sample-data as a function of the time needed to compute all representations in a training set.
		Lines indicate Pareto frontiers; inset numbers show training set sizes.
		See Glossary for abbreviations.	
		\label{fig:timingsrmse}}
	\end{minipage}
\end{figure*}

\Cref{fig:learncurvesrmse} presents learning curves for SF, MBTR, SOAP on datasets \dsgdb{}, \dsba{}, \dstcor{} (see \cref{si:predacc} for tabulated values).
For each dataset, representation, and training set size, we trained a KRR model and evaluated its predictions on a separate hold-out validation set of size 10\,k (\dsgdb), 1\,k (\dsba), and 0.6\,k (\dstcor).
This procedure was repeated 10 times to estimate the variance of these experiments.

Boxes, whiskers, horizontal bars, and crosses show interquartile ranges, minimum\,/\,maximum value, median, and mean, respectively, of the root mean squared error (RMSE) of hold-out-set predictions over repetitions.
We show RMSE as it is the loss minimized by least-squares regression such as KRR, and thus a natural choice.
For other loss functions, see \cref{si:errormetrics}.
From statistical learning theory, RMSE decays as a negative power of training set size (a reason why learning curves are preferably shown on log-log plots). \cite{cjsvd1994,mfmsa1996,hsl2018q}
Lines show corresponding fits of mean RMSE, weighted by the standard deviation for each training set size.

\Cref{fig:timingsrmse} reveals dependencies between the time to compute representations for a training set (horizontal axis) and RMSE (vertical axis).
When comparing observations in two dimensions, here time~$t$ and error~$e$, there is no unique ordering~$<$,
and we resort to the usual notion of dominance:
Let $\vec{x}, \vec{x'} \in \mathbb{R}^d$;
then $\vec{x}$ dominates $\vec{x'}$ if $x_i \leq x'_i$ for all dimensions~$i$ and $x_i < x'_i$ for some~$i$.
The set of all non-dominated points is called the Pareto frontier,
shown by a line, with numbers indicating training set sizes.
\Cref{tab:compcosts} presents compute times for representations (\cref{si:computetimes} for kernel matrices).

\enlargethispage{-\baselineskip}

\subsection{Discussion}

Asymptotically, observed prediction errors for all representations on all datasets relate as
\begin{align*}
	\text{SF-2,3} & \prec   \text{SF-2} ,     & \text{MBTR-2,3} & \preceq \text{MBTR-2} ,   \\
	\text{SOAP}   & \prec   \text{SF-2,3} , & \text{SOAP}     & \prec   \text{MBTR-2,3} , \\
	\text{SF-2,3} & \preceq \text{MBTR-2,3} , & \text{SF-2}     & \prec   \text{MBTR-2} ,
\end{align*}
where $A \prec B$ ($A \preceq B$) indicates that $A$ has lower (or equal) estimated error than~$B$ asymptotically.
Except for $\text{MBTR-2,3} \not\preceq \text{SF-2}$ on dataset \dstcor{},
\begin{equation*}
	\text{SOAP} \prec \text{SF-2,3} \preceq \text{MBTR-2,3} \prec \text{SF-2} \prec \text{MBTR-2} .
\end{equation*}
\emph{We conclude that, for energy predictions, accuracy improves with modeled interaction order and for local representations over global ones.}
The magnitude of, and between, these effects varies across datasets.

Dependence of predictive accuracy on interaction order has been observed by others \cite{ptm2018q,hjmfrgrf2020q,fchl2018q,wt2018q,s2018cq}
and might be partially due to a higher resolution of structural features. \cite{pwbocc2020q}
The latter would only show for sufficient training data, such as for dataset \dsba{} in \cref{fig:learncurvesrmse}.
We do not observe this for dataset \dsgdb{}, possibly because angular terms might be immediately relevant for characterizing organic molecules' carbon scaffolds. \cite{ptm2018q}

Better performance of local representations might be due to higher resolution and better generalization 
(both from representing only a small part of the whole structure), and has also been observed by others \cite{jmfhf2018q,hxh2020q}.
The impact of assuming additivity is unclear but likely depends on the structure of the modeled property. (\cref{si:extensivity})
Our comparison includes only a single global representation (MBTR), warranting further study of the locality aspect.
For additional analysis details, see~\cref{si:analysisdetails,si:dstcounrel}.

Computational costs tend to increase with predictive accuracy.
Representations should therefore be selected based on a target accuracy, constrained by available computing resources.

Converged prediction errors are in reasonable agreement with the literature (\cref{si:complit})
considering the lack of standardized conditions such as sampling, regression method, HP optimization, and reported performance statistics.
In absolute terms, prediction errors of models trained on 10\,k samples are closer to the differences between DFT codes than the (systematic) differences between the underlying DFT reference and experimental measurements. (\cref{si:compdft})

\begin{table}
	\caption{\emph{Computational cost of calculating representations} in milliseconds of processor time.
		Shown are mean $\pm$ standard deviation over all training set sizes of a dataset for the time to compute the representation of a single molecule or unit cell.
		See \cref{si:computetimes} for details.
		\label{tab:compcosts}}

	\vspace*{\captionskip}	
	\begin{tabular}[b]{@{}llcr@{}}
		\toprule
		Time in ms& \multicolumn{3}{c}{Dataset} \\ \cmidrule(lr){2-4}
		Representation & \multicolumn{1}{c}{\dsgdb} & \multicolumn{1}{c}{\dsba} & \multicolumn{1}{c}{\dstco} \\
		\midrule
		MBTR $k=2$   & \num{0.76\pm 0.32 } &     {13 $\pm$ 5.1} & \num{340 \pm 99 } \\
SF $k=2$     & \num{1.4 \pm 0.18 } & \num{3.3 \pm 1.4 } & \num{8.2 \pm 1.1 } \\
MBTR $k=2,3$ & \num{12  \pm 6.9  } & \num{290 \pm 140 } & 28\,k $\pm$ 4.4\,k \\
SF $k=2,3$   & \num{2.8 \pm 0.85 } & \num{27  \pm 12  } & \num{98 \pm 89 } \\
SOAP         & \num{1.9 \pm 0.54 } & \num{9.1 \pm 4.8 } & \num{19 \pm 8.6 } \\

		\bottomrule
	\end{tabular}
\end{table}


\section{Conclusions and outlook}\label{sec:outlook}

We review representations of atomistic systems, such as molecules and crystalline materials, for machine-learning of ab initio quantum-mechanical simulations.
For this, we distinguish between local and global representations and between using invariant $k$-body functions and explicit symmetrization to deal with invariances.
Despite their apparent diversity, many representations can be formulated in a single mathematical framework based on $k$-atom terms, symmetrization, and tensor products.
Empirically, we observe that when controlling for other factors, including distribution of training and validation data, regression method, and HP optimization, both prediction errors and compute time of SFs, MBTR and SOAP improve with interaction order, and for local representations over global ones.

\clearpage

\noindent
Our findings suggest the following guidance:
\begin{itemize}
	\item If their prediction errors are sufficient for an application, we recommend two-body versions of simple representations such as SF and MBTR as they are fastest to compute. 
	\item For large systems, local representations should be used.
	\item For strong noise or bias on input structures, as in dataset~\dstcou{}, performance differences between representations vanish, (\cref{si:dstcounrel})
		and computationally cheaper features that do not satisfy the requirements in \cref{sec:requirements} (descriptors) suffice.
\end{itemize}
We conclude by providing related current research directions, grouped by topic.

\medskip

\noindent
Directly related to representations:

\begin{itemize}
	\item \emph{Systematic development of representations} via extending the mathematical framework (\cref{sec:framework}) to include more state-of-the-art representations.
		  This would enable deriving ``missing'' variants of representations (see \cref{tab:classification}), such as a global SOAP \cite{dbcc2016q} and local MBTR \cite{localmbtr2019}, on a principled basis,
		  as well as understanding and reformulating existing representations in a joint framework, perhaps to the extent of an efficient general implementation. \cite{mvgfwsjc2021}

	\item \emph{Representing more systems.}
		  Develop or extend representations for atomistic systems currently not representable, or only to a limited extent, such as
		  charged atoms and systems, \cite{ghsg2015q,nlsslirbt2018q,rag2018q,ns2018q,um2019bq,d2020q,kfgb2020,pfg2020} excited states~\cite{wm2020q,wm2020bq,wgm2020q,bdlrs2005q,wfclm2020q}, spin systems, isotopes, and systems in an applied external field~\cite{gsm2020,cfl2019q}.

	\item \emph{Alchemical learning.}
		  Further understand and develop alchemical representations \cite{fchl2018q,wmc2018q,hkyp2019q} that incorporate similarity between chemical species to improve sample efficiency. 
		  What are the salient features of chemical elements that need to be considered, also with respect to charges, excitations, spins, and isotopes?

	\item \emph{Analysis of representations} to better understand structure and data distribution in feature spaces and how they relate to physics and chemistry concepts.
		  Possible approaches include
		  quantitative measures of structure and distribution of datasets in these spaces,
		  dimensionality reduction methods, 
		  analysis of data-driven representations from deep neural networks,
		  and construction, or proof of non-existence, of non-distinguishable environments for representations employing terms of order higher than four.

	\item \emph{Explicit complexity control.}
		  Different applications require different trade-offs between computational cost and predictive accuracy.
		  This requires determination, and automatic adaptation as an HP, of the capacity (complexity, dimensionality) and computational cost of a representation to a dataset, for example, through selection, combination, \cite{gmm2019q} or systematic construction of features \cite{gzfd2020q,pwbocc2020q}.
\end{itemize}

\noindent
Related to benchmarking of representations:

\begin{itemize}
	\item \emph{Extended scope.}
		  We empirically compare one global and two local representations on three datasets to predict energies using KRR with a Gaussian kernel.
		  For a more systematic coverage, other representations (\cref{sec:other}) and datasets, training with forces, \cite{bc2015q,csmt2018q} and more properties should be included
		  while maintaining control over regression method, data distribution, and HP optimization.
		  Deep neural networks \cite{ssktm2018q,nlsslirbt2018q,sgtm2019q,sgtmm2019q} could be included via representation learning.
		  Comparison with simple baseline models such as $k$-nearest neighbors \cite{rc2018q} would be desirable.
		  
	\item \emph{Improved optimization of HPs:}
		  The stochastic optimizer used in this work required multiple restarts in practice to avoid sub-optimal results, and reached its limits for large HP search spaces.
		  It would be desirable to reduce the influence and computational cost of HP optimization.
		  Possible means include reducing the number of HPs in representations, employing more systematic and thus more robust optimization methods, and providing reliable heuristics for HP default values.
		  
	\item \emph{Multi-objective optimization.}
		  We optimize HPs for predictive accuracy on a single property.
		  In practice, though, parametrizations of similar accuracy but lower computational cost would be preferable, and more than one property can be of interest.
		  HPs should, therefore, be optimized for multiple properties and criteria, including computational cost and predictive uncertainties (see below).
		  How to balance these is part of the problem. \cite{rrkal2020q}
		  
	\item \emph{Predictive uncertainties.}
		  While prediction errors are frequently analyzed, and reasonable guidelines exist, this is not the case for predictive uncertainties.
	      These are becoming increasingly important as applications of ML mature, for example, for human assessment and decisions, learning on the fly, \cite{capdv2004q} and active learning.
	      Beyond global analysis of uncertainty estimates, local characterization (in input or feature space) of prediction errors is relevant. \cite{rrkal2020q,sbgrvs2020q}
\end{itemize}

\noindent
Related through context:

\begin{itemize}
	\item \emph{Long-range interactions.}
		  ML models appear to be well-suited for short- and medium-ranged interactions, but problematic for long-ranged interactions due to the increasing degrees of freedom of larger systems and larger necessary cut-off radii of atomic environments.
		  Two approaches are to integrate ML models with physical models for long-range interactions, \cite{amb2011q,ghsg2015q,kfgb2020} and to adapt ML models to learn long-range interactions directly. \cite{gc2019q}

	\item \emph{Relationships between QM and ML.}
		  A deeper understanding of the relationships between QM and kernel-based ML could lead to insights and technical progress in both fields.
		  As both share concepts from linear algebra, such relationships could be formal mathematical ones.
		  For example, QM concepts such as matrix product states can parameterize non-linear kernel models. \cite{ss2016}
\end{itemize}


\subsection{Acknowledgments}

This work received funding from the European Union's Horizon 2020 Research and Innovation Programme, Grant Agreements No.~676580, the NOMAD Laboratory CoE, and No.~740233, ERC:~TEC1P.
Part of the research was performed while the authors visited the Institute for Pure and Applied Mathematics (IPAM), which is supported by the National Science Foundation (Grant No. DMS-1440415).

The authors thank Profs.~Matthias Scheffler, 
Klaus-Robert M{\"u}ller,
J{\"o}rg Behler,
G{\'a}bor Cs{\'a}nyi,
O. Anatole von Lilienfeld,
Carsten Baldauf,
Matthew Hirn,
as well as
Emre Ahmetcik,
Lauri Himanen,
Yair Litman,
Dmitrii Maksimov,
Felix Mocanu,
Wiktor Pronobis,
and Christopher Sutton
for constructive discussions.

\subsection{Author contributions}

{M.F.L.} and {M.R.} designed numerical experiments and analyzed results.
{M.F.L.} developed software and conducted numerical experiments.
All authors contributed to writing, with emphasis by {A.G.} on the mathematical framework and {M.F.L.} on representations and benchmarking.
{M.R.} supervised the study.

\subsection{Competing interests}

The authors declare no competing financial or non-financial interests.

\subsection{Data and code availability}

The data that support the findings of this study are publicly available,
datasets at \url{https://qmml.org}, 
hyperparameter search spaces, machine-learning models, program code, and results at \url{https://marcel.science/repbench}.
A tutorial introduction to the \texttt{cmlkit} Python framework developed for this work is part of the NOMAD Analytics Toolkit. \cite{nomadanalyticstoolkit}


\setbiblabelwidth{999} 
\phantomsection\addcontentsline{toc}{section}{References}
\begingroup
\raggedright
\bibliography{abbreviations,reprrev}

\begin{thebibliography}{100}

\bibitem{br2009}
Lorenz~C. Blum, Jean-Louis Reymond:
  \href{https://doi.org/10.1021/ja902302h}{\textit{970 million druglike small
  molecules for virtual screening in the chemical universe database {GDB-13}}}.
  Journal of the American Chemical Society
  \href{https://doi.org/10.1021/ja902302h}{131(25):
  8732}\href{https://doi.org/10.1021/ja902302h}{, 2009}.

\bibitem{g2015}
Zoubin Ghahramani:
  \href{https://doi.org/10.1038/nature14541}{\textit{Probabilistic machine
  learning and artificial intelligence}}. Nature
  \href{https://doi.org/10.1038/nature14541}{521(7553):
  452}\href{https://doi.org/10.1038/nature14541}{, 2015}.

\bibitem{jm2015}
Michael~I. Jordan, Tom~M. Mitchell:
  \href{https://doi.org/10.1126/science.aaa8415}{\textit{Machine learning:
  trends, perspectives, and prospects}}. Science
  \href{https://doi.org/10.1126/science.aaa8415}{349(6245):
  255}\href{https://doi.org/10.1126/science.aaa8415}{, 2015}.

\bibitem{rtml2012q}
Matthias Rupp, Alexandre Tkatchenko, Klaus-Robert M{\"u}ller, O.~Anatole von
  Lilienfeld:
  \href{https://doi.org/10.1103/PhysRevLett.108.058301}{\textit{Fast and
  accurate modeling of molecular atomization energies with machine learning}}.
  Physical Review Letters
  \href{https://doi.org/10.1103/PhysRevLett.108.058301}{108(5):
  058301}\href{https://doi.org/10.1103/PhysRevLett.108.058301}{, 2012}.

\bibitem{b2017q}
J{\"o}rg Behler: \href{https://doi.org/10.1002/anie.201703114}{\textit{First
  principles neural network potentials for reactive simulations of large
  molecular and condensed systems}}. Angewandte Chemie International Edition
  \href{https://doi.org/10.1002/anie.201703114}{56(42):
  12828}\href{https://doi.org/10.1002/anie.201703114}{, 2017}.

\bibitem{cwc2018q}
Michele Ceriotti, Michael~J. Willatt, G{\'a}bor Cs{\'a}nyi:
  \href{https://doi.org/10.1007/978-3-319-42913-7_68-1}{\textit{Machine
  learning of atomic-scale properties based on physical principles}}. In Wanda
  Andreoni, Sidney Yip (editors), \textit{Handbook of Materials Modeling.
  Methods: Theory and Modeling}, Springer, 2018.

\bibitem{hl2020b}
Bing Huang, O.~Anatole von Lilienfeld:
  \textit{\href{https://arxiv.org/abs/arXiv:2012.07502}{Ab initio machine
  learning in chemical compound space}}. arXiv 2012.07502, 2020.

\bibitem{bp2007q}
J{\"o}rg Behler, Michele Parrinello:
  \href{https://doi.org/10.1103/PhysRevLett.98.146401}{\textit{Generalized
  neural-network representation of high-dimensional potential-energy
  surfaces}}. Physical Review Letters
  \href{https://doi.org/10.1103/PhysRevLett.98.146401}{98(14):
  146401}\href{https://doi.org/10.1103/PhysRevLett.98.146401}{, 2007}.

\bibitem{bpkc2010q}
Albert~P. Bart{\'o}k, Mike~C. Payne, Risi Kondor, G{\'a}bor Cs{\'a}nyi:
  \href{https://doi.org/10.1103/PhysRevLett.104.136403}{\textit{{G}aussian
  approximation potentials: the accuracy of quantum mechanics, without the
  electrons}}. Physical Review Letters
  \href{https://doi.org/10.1103/PhysRevLett.104.136403}{104(13):
  136403}\href{https://doi.org/10.1103/PhysRevLett.104.136403}{, 2010}.

\bibitem{cdklc2018q}
Miguel~A. Caro, Volker~L. Deringer, Jari Koskinen, Tomi Laurila, G{\'a}bor
  Cs{\'a}nyi:
  \href{https://doi.org/10.1103/PhysRevLett.120.166101}{\textit{Growth
  mechanism and origin of high $sp^3$ content in tetrahedral amorphous
  carbon}}. Physical Review Letters
  \href{https://doi.org/10.1103/PhysRevLett.120.166101}{120(16):
  166101}\href{https://doi.org/10.1103/PhysRevLett.120.166101}{, 2018}.

\bibitem{jlkkb2019q}
Ryosuke Jinnouchi, Jonathan Lahnsteiner, Ferenc Karsai, Georg Kresse, Menno
  Bokdam: \href{https://doi.org/10.1103/PhysRevLett.122.225701}{\textit{Phase
  transitions of hybrid perovskites simulated by machine-learning force fields
  trained on the fly with {B}ayesian inference}}. Physical Review Letters
  \href{https://doi.org/10.1103/PhysRevLett.122.225701}{122(2):
  225701}\href{https://doi.org/10.1103/PhysRevLett.122.225701}{, 2019}.

\bibitem{kotm2016q}
Shin Kiyohara, Hiromi Oda, Koji Tsuda, Teruyasu Mizoguchi:
  \href{https://doi.org/10.7567/JJAP.55.045502}{\textit{Acceleration of stable
  interface structure searching using a {K}riging approach}}. Japanese Journal
  of Applied Physics \href{https://doi.org/10.7567/JJAP.55.045502}{55(4):
  045502}\href{https://doi.org/10.7567/JJAP.55.045502}{, 2016}.

\bibitem{bkbc2018q}
Albert~P. Bart{\'o}k, James Kermode, Noam Bernstein, G{\'a}bor Cs{\'a}nyi:
  \href{https://doi.org/10.1103/PhysRevX.8.041048}{\textit{Machine learning a
  general-purpose interatomic potential for silicon}}. Physical Review~X
  \href{https://doi.org/10.1103/PhysRevX.8.041048}{8(4):
  041048}\href{https://doi.org/10.1103/PhysRevX.8.041048}{, 2018}.

\bibitem{scaccr2019q}
Austin~D. Sendek, Ekin~D. Cubuk, Evan~R. Antoniuk, Gowoon Cheon, Yi~Cui,
  Evan~J. Reed:
  \href{https://doi.org/10.1021/acs.chemmater.8b03272}{\textit{Machine
  learning-assisted discovery of solid {Li}-ion conducting materials}}.
  Chemistry of Materials
  \href{https://doi.org/10.1021/acs.chemmater.8b03272}{31(2):
  342}\href{https://doi.org/10.1021/acs.chemmater.8b03272}{, 2018}.

\bibitem{jkk2019q}
Ryosuke Jinnouchi, Ferenc Karsai, Georg Kresse:
  \href{https://doi.org/10.1103/physrevb.100.014105}{\textit{On-the-fly machine
  learning force field generation: application to melting points}}. Physical
  Review~B \href{https://doi.org/10.1103/physrevb.100.014105}{100(1):
  014105}\href{https://doi.org/10.1103/physrevb.100.014105}{, 2019}.

\bibitem{ss2002}
Bernhard Sch{\"o}lkopf, Alexander Smola:
  \textit{\href{https://mitpress.mit.edu/books/learning-kernels}{Learning with
  kernels}}. MIT Press, Cambridge, 2002.

\bibitem{hss2008}
Thomas Hofmann, Bernhard Sch{\"o}lkopf, Alexander Smola:
  \href{https://doi.org/10.1214/009053607000000677}{\textit{Kernel methods in
  machine learning}}. Annals of Statistics
  \href{https://doi.org/10.1214/009053607000000677}{36(3):
  1171}\href{https://doi.org/10.1214/009053607000000677}{, 2008}.

\bibitem{tc2009}
Roberto Todeschini, Viviana Consonni:
  \href{https://doi.org/10.1002/9783527613106}{\textit{Handbook of molecular
  descriptors}}. Wiley, Weinheim, Germany, 2nd edition, 2009.

\bibitem{gsrvd2017q}
Justin Gilmer, Samuel~S. Schoenholz, Patrick~F. Riley, Oriol Vinyals, George~E.
  Dahl: \textit{\href{http://proceedings.mlr.press/v70/gilmer17a.html}{Neural
  message passing for quantum chemistry}}. In \textit{Proceedings of the 34th
  International Conference on Machine Learning (ICML)}, 1263, 2017.

\bibitem{sacmt2017q}
Kristof~T. Sch{\"u}tt, Farhad Arbabzadah, Stefan Chmiela, Klaus-Robert
  M{\"u}ller, Alexandre Tkatchenko:
  \href{https://doi.org/10.1038/ncomms13890}{\textit{Quantum-chemical insights
  from deep tensor neural networks}}. Nature Communications
  \href{https://doi.org/10.1038/ncomms13890}{8:
  13890}\href{https://doi.org/10.1038/ncomms13890}{, 2017}.

\bibitem{sksctm2017q}
Kristof~T. Sch{\"u}tt, Pieter-Jan Kindermans, Huziel~E. Sauceda, Stefan
  Chmiela, Alexandre Tkatchenko, Klaus-Robert M{\"u}ller:
  \textit{\href{https://papers.nips.cc/paper/6700-schnet-a-continuous-filter-convolutional-neural-network-for-modeling-quantum-interactions}{{SchNet}:
  a continuous-filter convolutional neural network for modeling quantum
  interactions}}. In \textit{Advances in Neural Information Processing
  Systems~30 (NIPS)}, 2017.

\bibitem{k2017bq}
Risi Kondor: \textit{\href{https://arxiv.org/abs/1803.01588}{$n$-body networks:
  a covariant hierarchical neural network architecture for learning atomic
  potentials}}. In \textit{NeurIPS Workshop on Machine Learning for Molecules
  and Materials}, 2017.

\bibitem{ssktm2018q}
Kristof~T. Sch{\"u}tt, Huziel~E. Sauceda, Pieter-Jan Kindermans, Alexandre
  Tkatchenko, Klaus-Robert M{\"u}ller:
  \href{https://doi.org/10.1063/1.5019779}{\textit{{SchNet}---a deep learning
  architecture for molecules and materials}}. Journal of Chemical Physics
  \href{https://doi.org/10.1063/1.5019779}{148(24):
  241722}\href{https://doi.org/10.1063/1.5019779}{, 2018}.

\bibitem{zhwsce2018q}
Linfeng Zhang, Jiequn Han, Han Wang, Wissam~A. Saidi, Roberto Car, Weinan E:
  \textit{\href{https://papers.nips.cc/paper/7696-end-to-end-symmetry-preserving-inter-atomic-potential-energy-model-for-finite-and-extended-systems}{End-to-end
  symmetry preserving inter-atomic potential energy model for finite and
  extended systems}}. In \textit{Advances in Neural Information Processing
  Systems~31 (NeurIPS)}, 4436, 2018.

\bibitem{tskylkr2018q}
Nathaniel Thomas, Tess Smidt, Steven Kearnes, Lusann Yang, Li~Li, Kai Kohlhoff,
  Patrick Riley: \textit{\href{https://arxiv.org/abs/1802.08219}{Tensor field
  networks: rotation- and translation-equivariant neural networks for 3D point
  clouds}}. In \textit{NeurIPS Workshop on Machine Learning for Molecules and
  Materials}, 2018.

\bibitem{klt2018q}
Risi Kondor, Zhen Li, Shubhendu Trivedi:
  \textit{\href{https://papers.nips.cc/paper/8215-clebschgordan-nets-a-fully-fourier-space-spherical-convolutional-neural-network}{{C}lebsch-{G}ordan
  nets: a fully {F}ourier space spherical convolutional neural network}}. In
  \textit{Advances in Neural Information Processing Systems~31 (NeurIPS)},
  10117, 2018.

\bibitem{wgwbc2018q}
Maurice Weiler, Mario Geiger, Max Welling, Wouter Boomsma, Taco~S. Cohen:
  \textit{\href{https://papers.nips.cc/paper/8239-3d-steerable-cnns-learning-rotationally-equivariant-features-in-volumetric-data}{{3D}
  steerable {CNN}s: learning rotationally equivariant features in volumetric
  data}}. In \textit{Advances in Neural Information Processing Systems~31
  (NeurIPS)}, 10381, 2018.

\bibitem{um2019bq}
Oliver~T. Unke, Markus Meuwly:
  \href{https://doi.org/10.1021/acs.jctc.9b00181}{\textit{{PhysNet}: a neural
  network for predicting energies, forces, dipole moments, and partial
  charges}}. Journal of Chemical Theory and Computation
  \href{https://doi.org/10.1021/acs.jctc.9b00181}{15(6):
  3678}\href{https://doi.org/10.1021/acs.jctc.9b00181}{, 2019}.

\bibitem{ahk2019q}
Brandon Anderson, Truong-Son Hy, Risi Kondor:
  \textit{\href{https://papers.nips.cc/paper/9596-cormorant-covariant-molecular-neural-networks}{{C}ormorant:
  covariant molecular neural networks}}. In \textit{Advances in Neural
  Information Processing Systems~32 (NeurIPS)}, 14537, 2019.

\bibitem{zhj2019q}
Yaolong Zhang, Ce~Hu, Bin Jiang:
  \href{https://doi.org/10.1021/acs.jpclett.9b02037}{\textit{Embedded atom
  neural network potentials: efficient and accurate machine learning with a
  physically inspired representation}}. Journal of Physical Chemistry Letters
  \href{https://doi.org/10.1021/acs.jpclett.9b02037}{10(17):
  4962}\href{https://doi.org/10.1021/acs.jpclett.9b02037}{, 2019}.

\bibitem{kgg2020q}
Johannes Klicpera, Janek Gro{\ss}, Stephan G{\"u}nnemann:
  \textit{\href{https://iclr.cc/virtual_2020/poster_B1eWbxStPH.html}{Directional
  message passing for molecular graphs}}. In \textit{Proceedings of the {8th}
  International Conference on Learning Representations (ICLR)}, 2020.

\bibitem{mgsn2020q}
Benjamin~Kurt Miller, Mario Geiger, Tess~E. Smidt, Frank No{\'e}:
  \textit{Relevance of rotationally equivariant convolutions for predicting
  molecular properties}. In \textit{NeurIPS Workshop on Machine Learning for
  Molecules}, 2020.

\bibitem{fwfw2020}
Fabian~B. Fuchs, Daniel~E. Worrall, Volker Fischer, Max Welling:
  \textit{\href{https://arxiv.org/abs/arXiv:2006.10503}{{SE(3)}-transformers:
  {3D} roto-translation equivariant attention networks}}. arXiv 2006.10503,
  2020.

\bibitem{qwamm2020q}
Zhuoran Qiao, Matthew Welborn, Animashree Anandkumar, Frederick~R. Manby,
  Thomas~F. Miller, III:
  \href{https://doi.org/10.1063/5.0021955}{\textit{{OrbNet}: deep learning for
  quantum chemistry using symmetry-adapted atomic-orbital features}}. Journal
  of Chemical Physics \href{https://doi.org/10.1063/5.0021955}{153(12):
  124111}\href{https://doi.org/10.1063/5.0021955}{, 2020}.

\bibitem{czwe2020q}
Yixiao Chen, Linfeng Zhang, Han Wang, Weinan E:
  \href{https://doi.org/10.1021/acs.jpca.0c03886}{\textit{Ground state energy
  functional with {H}artree-{F}ock efficiency and chemical accuracy}}. Journal
  of Physical Chemistry~A
  \href{https://doi.org/10.1021/acs.jpca.0c03886}{124(35):
  7155}\href{https://doi.org/10.1021/acs.jpca.0c03886}{, 2020}.

\bibitem{hsn2020q}
Jan Hermann, Zeno Sch{\"a}tzle, Frank No{\'{e}}:
  \href{https://doi.org/10.1038/s41557-020-0544-y}{\textit{Deep-neural-network
  solution of the electronic {S}chr{\"o}dinger equation}}. Nature Chemistry
  \href{https://doi.org/10.1038/s41557-020-0544-y}{12(10):
  891}\href{https://doi.org/10.1038/s41557-020-0544-y}{, 2020}.

\bibitem{hr1996q}
Tak-San Ho, Herschel Rabitz: \href{https://doi.org/10.1063/1.470984}{\textit{A
  general method for constructing multidimensional molecular potential energy
  surfaces from \textit{ab initio} calculations}}. Journal of Chemical Physics
  \href{https://doi.org/10.1063/1.470984}{104(7):
  2584}\href{https://doi.org/10.1063/1.470984}{, 1996}.

\bibitem{um2017q}
Oliver~T. Unke, Markus Meuwly:
  \href{https://doi.org/10.1021/acs.jcim.7b00090}{\textit{A toolkit for the
  construction of reproducing kernel-based representations of data: application
  to multi-dimensional potential energy surfaces}}. Journal of Chemical
  Information and Modeling
  \href{https://doi.org/10.1021/acs.jcim.7b00090}{57(8):
  1923}\href{https://doi.org/10.1021/acs.jcim.7b00090}{, 2017}.

\bibitem{gzv2018q}
Aldo Glielmo, Claudio Zeni, Alessandro~De Vita:
  \href{https://doi.org/10.1103/PhysRevB.97.184307}{\textit{Efficient
  non-parametric $n$-body force fields from machine learning}}. Physical
  Review~B \href{https://doi.org/10.1103/PhysRevB.97.184307}{97(18):
  184307}\href{https://doi.org/10.1103/PhysRevB.97.184307}{, 2018}.

\bibitem{km2020q}
Debasish Koner, Markus Meuwly:
  \href{https://doi.org/10.1021/acs.jctc.0c00535}{\textit{Permutationally
  invariant, reproducing kernel-based potential energy surfaces for polyatomic
  molecules: from formaldehyde to acetone}}. Journal of Chemical Theory and
  Computation \href{https://doi.org/10.1021/acs.jctc.0c00535}{16(9):
  5474}\href{https://doi.org/10.1021/acs.jctc.0c00535}{, 2020}.

\bibitem{gzfd2020q}
Aldo Glielmo, Claudio Zeni, {\'{A}}d{\'{a}}m Fekete, Alessandro De~Vita:
  \href{https://doi.org/10.1007/978-3-030-40245-7_5}{\textit{Building
  nonparametric $n$-body force fields using {G}aussian process regression}}. In
  Kristof~T. Sch{\"{u}}tt, Stefan Chmiela, O.~Anatole von Lilienfeld, Alexandre
  Tkatchenko, Koji Tsuda, Klaus-Robert M{\"{u}}ller (editors), \textit{Machine
  Learning Meets Quantum Physics}, 67, Springer, Heidelberg, Germany, 2020.

\bibitem{fhhgsdvkrv2017q}
Felix~A. Faber, Luke Hutchison, Bing Huang, Justin Gilmer, Samuel~S.
  Schoenholz, George~E. Dahl, Oriol Vinyals, Steven Kearnes, Patrick~F. Riley,
  O.~Anatole von Lilienfeld:
  \href{https://doi.org/10.1021/acs.jctc.7b00577}{\textit{Prediction errors of
  molecular machine learning models lower than hybrid {DFT} error}}. Journal of
  Chemical Theory and Computation
  \href{https://doi.org/10.1021/acs.jctc.7b00577}{13(11):
  5255}\href{https://doi.org/10.1021/acs.jctc.7b00577}{, 2017}.

\bibitem{hjmfrgrf2020q}
Lauri Himanen, Marc~O.J. J{\"{a}}ger, Eiaki~V. Morooka, Filippo~Federici
  Canova, Yashasvi~S. Ranawat, David~Z. Gao, Patrick Rinke, Adam~S. Foster:
  \href{https://doi.org/10.1016/j.cpc.2019.106949}{\textit{{DScribe}: library
  of descriptors for machine learning in materials science}}. Computer Physics
  Communications \href{https://doi.org/10.1016/j.cpc.2019.106949}{247:
  106949}\href{https://doi.org/10.1016/j.cpc.2019.106949}{, 2020}.

\bibitem{zcldcbcstwo2020q}
Yunxing Zuo, Chi Chen, Xiangguo Li, Zhi Deng, Yiming Chen, J{\"o}rg Behler,
  G{\'a}bor Cs{\'a}nyi, Alexander~V. Shapeev, Aidan~P. Thompson, Mitchell~A.
  Wood, Shyue~Ping Ong:
  \href{https://doi.org/10.1021/acs.jpca.9b08723}{\textit{Performance and cost
  assessment of machine learning interatomic potentials}}. Journal of Physical
  Chemistry~A \href{https://doi.org/10.1021/acs.jpca.9b08723}{124(4):
  731}\href{https://doi.org/10.1021/acs.jpca.9b08723}{, 2020}.

\bibitem{sgc2019q}
Gunnar Schmitz, Ian~Heide Godtliebsen, Ove Christiansen:
  \href{https://doi.org/10.1063/1.5100141}{\textit{Machine learning for
  potential energy surfaces: an extensive database and assessment of methods}}.
  Journal of Chemical Physics \href{https://doi.org/10.1063/1.5100141}{150(24):
  244113}\href{https://doi.org/10.1063/1.5100141}{, 2019}.

\bibitem{nrbsmrcwh2019q}
Chandramouli Nyshadham, Matthias Rupp, Brayden Bekker, Alexander~V. Shapeev,
  Tim Mueller, Conrad~W. Rosenbrock, G{\'a}bor Cs{\'a}nyi, David~W. Wingate,
  Gus~L.W. Hart:
  \href{https://doi.org/10.1038/s41524-019-0189-9}{\textit{Machine-learned
  multi-system surrogate models for materials prediction}}. npj Computational
  Materials \href{https://doi.org/10.1038/s41524-019-0189-9}{5:
  51}\href{https://doi.org/10.1038/s41524-019-0189-9}{, 2019}.

\bibitem{strkghr2019q}
Annika Stuke, Milica Todorovi{\'c}, Matthias Rupp, Christian Kunkel, Kunal
  Ghosh, Lauri Himanen, Patrick Rinke:
  \href{https://doi.org/10.1063/1.5086105}{\textit{Chemical diversity in
  molecular orbital energy predictions with kernel ridge regression}}. Journal
  of Chemical Physics \href{https://doi.org/10.1063/1.5086105}{150(20):
  204121}\href{https://doi.org/10.1063/1.5086105}{, 2019}.

\bibitem{ook2020q}
Berk Onat, Christoph Ortner, James~R. Kermode:
  \href{https://doi.org/10.1063/5.0016005}{\textit{Sensitivity and
  dimensionality of atomic environment representations used for machine
  learning interatomic potentials}}. Journal of Chemical Physics
  \href{https://doi.org/10.1063/5.0016005}{153(14):
  144106}\href{https://doi.org/10.1063/5.0016005}{, 2020}.

\bibitem{kkclm2020q}
Silvan K{\"a}ser, Debasish Koner, Anders~S. Christensen, O.~Anatole von
  Lilienfeld, Markus Meuwly:
  \href{https://doi.org/10.1021/acs.jpca.0c05979}{\textit{{ML} models of
  vibrating {H$_2$CO}: Comparing reproducing kernels, {FCHL} and {PhysNet}}}.
  Journal of Physical Chemistry~A
  \href{https://doi.org/10.1021/acs.jpca.0c05979}{124(42):
  8853}\href{https://doi.org/10.1021/acs.jpca.0c05979}{, 2020}.

\bibitem{pdcfkdblg2020}
Behnam Parsaeifard, Deb~Sankar De, Anders~S. Christensen, Felix~A. Faber, Emir
  Kocer, Sandip De, J{\"o}rg Behler, Anatole von Lilienfeld, Stefan Goedecker:
  \href{https://doi.org/10.1088/2632-2153/abb212}{\textit{An assessment of the
  structural resolution of various fingerprints commonly used in machine
  learning}}. Machine Learning: Science and Technology
  \href{https://doi.org/10.1088/2632-2153/abb212}{in
  press}\href{https://doi.org/10.1088/2632-2153/abb212}{, 2020}.

\bibitem{jmfhf2018q}
Marc O.~J. J{\"a}ger, Eiaki~V. Morooka, Filippo Federici-Canova, Lauri Himanen,
  Adam~S. Foster:
  \href{https://doi.org/10.1038/s41524-018-0096-5}{\textit{Machine learning
  hydrogen adsorption on nanoclusters through structural descriptors}}. npj
  Computational Materials \href{https://doi.org/10.1038/s41524-018-0096-5}{4:
  37}\href{https://doi.org/10.1038/s41524-018-0096-5}{, 2018}.

\bibitem{gfc2020}
Alexander Goscinski, Guillaume Fraux, Michele Ceriotti:
  \textit{\href{https://arxiv.org/abs/arXiv:2009.02741}{The role of feature
  space in atomistic learning}}. arXiv 2009.02741, 2020.

\bibitem{bbm2008}
Mikio~L. Braun, Joachim~M. Buhmann, Klaus-Robert M{\"u}ller:
  \textit{\href{http://jmlr.org/papers/v9/braun08a.html}{On relevant dimensions
  in kernel feature spaces}}. Journal of Machine Learning Research 9(Aug):
  1875, 2008.

\bibitem{bkc2013q}
Albert~P. Bart{\'o}k, Risi Kondor, G{\'a}bor Cs{\'a}nyi:
  \href{https://doi.org/10.1103/PhysRevB.87.184115}{\textit{On representing
  chemical environments}}. Physical Review~B
  \href{https://doi.org/10.1103/PhysRevB.87.184115}{87(18):
  184115}\href{https://doi.org/10.1103/PhysRevB.87.184115}{, 2013}.

\bibitem{rrl2015q}
Matthias Rupp, Raghunathan Ramakrishnan, O.~Anatole von Lilienfeld:
  \href{https://doi.org/10.1021/acs.jpclett.5b01456}{\textit{Machine learning
  for quantum mechanical properties of atoms in molecules}}. Journal of
  Physical Chemistry Letters
  \href{https://doi.org/10.1021/acs.jpclett.5b01456}{6(16):
  3309}\href{https://doi.org/10.1021/acs.jpclett.5b01456}{, 2015}.

\bibitem{bbh1986q}
Joel~M. Bowman, Joseph~S. Bittman, Lawrence~B. Harding:
  \href{https://doi.org/10.1063/1.451246}{\textit{\textit{Ab initio}
  calculations of electronic and vibrational energies of {HCO} and {HOC}}}.
  Journal of Chemical Physics \href{https://doi.org/10.1063/1.451246}{85(2):
  911}\href{https://doi.org/10.1063/1.451246}{, 1986}.

\bibitem{dnu1991q}
Jerry~A. Darsey, Donald~W. Noid, Belle~R. Upadhyaya:
  \href{https://doi.org/10.1016/0009-2614(91)90066-I}{\textit{Application of
  neural network computing to the solution for the ground-state eigenenergy of
  two-dimensional harmonic oscillators}}. Chemical Physics Letters
  \href{https://doi.org/10.1016/0009-2614(91)90066-I}{177(2):
  189}\href{https://doi.org/10.1016/0009-2614(91)90066-I}{, 1991}.

\bibitem{hhlr1992q}
Hoon Heo, Tak-San Ho, Kevin~K. Lehmann, Herschel Rabitz:
  \href{https://doi.org/10.1063/1.463188}{\textit{Regularized inversion of
  diatomic vibration-rotation spectral data: a functional sensitivity analysis
  approach}}. Journal of Chemical Physics
  \href{https://doi.org/10.1063/1.463188}{97(2):
  852}\href{https://doi.org/10.1063/1.463188}{, 1992}.

\bibitem{hhr1999q}
Timothy Hollebeek, Tak-San Ho, Herschel Rabitz:
  \href{https://doi.org/10.1146/annurev.physchem.50.1.537}{\textit{Constructing
  multidimensional molecular potential energy surfaces from ab initio data}}.
  Annual Review of Physical Chemistry
  \href{https://doi.org/10.1146/annurev.physchem.50.1.537}{50:
  537}\href{https://doi.org/10.1146/annurev.physchem.50.1.537}{, 1999}.

\bibitem{lhwgsr2006q}
Genyuan Li, Jishan Hu, Sheng-Wei Wang, Panos~G. Georgopoulos, Jacqueline
  Schoendorf, Herschel Rabitz:
  \href{https://doi.org/10.1021/jp054148m}{\textit{Random sampling-high
  dimensional model representation ({RS-HDMR}) and orthogonality of its
  different order component functions}}. Journal of Physical Chemistry~A
  \href{https://doi.org/10.1021/jp054148m}{110(7):
  2474}\href{https://doi.org/10.1021/jp054148m}{, 2006}.

\bibitem{b2011eq}
J{\"o}rg Behler: \href{https://doi.org/10.1063/1.3553717}{\textit{Atom-centered
  symmetry functions for constructing high-dimensional neural network
  potentials}}. Journal of Chemical Physics
  \href{https://doi.org/10.1063/1.3553717}{134(7):
  074106}\href{https://doi.org/10.1063/1.3553717}{, 2011}.

\bibitem{sir2017q}
Justin~S. Smith, Olexandr Isayev, Adrian~E. Roitberg:
  \href{https://doi.org/10.1039/C6SC05720A}{\textit{{ANI-1}: an extensible
  neural network potential with {DFT} accuracy at force field computational
  cost}}. Chemical Science \href{https://doi.org/10.1039/C6SC05720A}{8(4):
  3192}\href{https://doi.org/10.1039/C6SC05720A}{, 2017}.

\bibitem{gsbbm2018q}
Michael Gastegger, Ludwig Schwiedrzik, Marius Bittermann, Florian Berzsenyi,
  Philipp Marquetand:
  \href{https://doi.org/10.1063/1.5019667}{\textit{w{ACSF}---weighted
  atom-centered symmetry functions as descriptors in machine learning
  potentials}}. Journal of Chemical Physics
  \href{https://doi.org/10.1063/1.5019667}{148(24):
  241709}\href{https://doi.org/10.1063/1.5019667}{, 2018}.

\bibitem{rag2018q}
Samare Rostami, Maximilian Amsler, S.~Alireza Ghasemi:
  \href{https://doi.org/10.1063/1.5040005}{\textit{Optimized symmetry functions
  for machine-learning interatomic potentials of multicomponent systems}}.
  Journal of Chemical Physics \href{https://doi.org/10.1063/1.5040005}{149(12):
  124106}\href{https://doi.org/10.1063/1.5040005}{, 2018}.

\bibitem{auc2018q}
Nongnuch Artrith, Alexander Urban, Gerbrand Ceder:
  \href{https://doi.org/10.1063/1.5017661}{\textit{Constructing
  first-principles phase diagrams of amorphous {Li$_x$Si} using
  machine-learning-assisted sampling with an evolutionary algorithm}}. Journal
  of Chemical Physics \href{https://doi.org/10.1063/1.5017661}{148(24):
  241711}\href{https://doi.org/10.1063/1.5017661}{, 2018}.

\bibitem{availsf}
Available as part of the software \texttt{RuNNer} at
  \url{http://www.uni-goettingen.de/de/560580.html}, GPL license, per email
  request).

\bibitem{wt2018q}
Mitchell~A. Wood, Aidan~P. Thompson:
  \href{https://doi.org/10.1063/1.5017641}{\textit{Extending the accuracy of
  the {SNAP} interatomic potential form}}. Journal of Chemical Physics
  \href{https://doi.org/10.1063/1.5017641}{148(24):
  241721}\href{https://doi.org/10.1063/1.5017641}{, 2018}.

\bibitem{stt2019q}
Atsuto Seko, Atsushi Togo, Isao Tanaka:
  \href{https://doi.org/10.1103/PhysRevB.99.214108}{\textit{Group-theoretical
  high-order rotational invariants for structural representations: application
  to linearized machine learning interatomic potential}}. Physical Review~B
  \href{https://doi.org/10.1103/PhysRevB.99.214108}{99(21):
  214108}\href{https://doi.org/10.1103/PhysRevB.99.214108}{, 2019}.

\bibitem{s2020bq}
Atsuto Seko: \href{https://doi.org/10.1103/PhysRevB.102.174104}{\textit{Machine
  learning potentials for multicomponent systems: The {Ti}-{Al} binary
  system}}. Physical Review~B
  \href{https://doi.org/10.1103/PhysRevB.102.174104}{102(17):
  174104}\href{https://doi.org/10.1103/PhysRevB.102.174104}{, 2020}.

\bibitem{availbs}
Available as part of the software \texttt{LAMMPS} (large-scale atomic/molecular
  massively parallel simulator, \url{http://lammps.sandia.gov}, GPL license,
  publicly accessible).

\bibitem{m2012eq}
Jonathan~E. Moussa:
  \href{https://doi.org/10.1103/PhysRevLett.109.059801}{\textit{Comment on
  ``{F}ast and accurate modeling of molecular atomization energies with machine
  learning''}}. Physical Review Letters
  \href{https://doi.org/10.1103/PhysRevLett.109.059801}{109(5):
  059801}\href{https://doi.org/10.1103/PhysRevLett.109.059801}{, 2012}.

\bibitem{rtml2012bq}
Matthias Rupp, Alexandre Tkatchenko, Klaus-Robert M{\"u}ller, O.~Anatole von
  Lilienfeld:
  \href{https://doi.org/10.1103/PhysRevLett.109.059802}{\textit{Reply to the
  comment by {J.E.~Moussa}}}. Physical Review Letters
  \href{https://doi.org/10.1103/PhysRevLett.109.059802}{109(5):
  059802}\href{https://doi.org/10.1103/PhysRevLett.109.059802}{, 2012}.

\bibitem{rdrl2015q}
Raghunathan Ramakrishnan, Pavlo~O. Dral, Matthias Rupp, O.~Anatole von
  Lilienfeld: \href{https://doi.org/10.1021/acs.jctc.5b00099}{\textit{Big data
  meets quantum chemistry approximations: the {$\Delta$}-machine learning
  approach}}. Journal of Chemical Theory and Computation
  \href{https://doi.org/10.1021/acs.jctc.5b00099}{11(5):
  2087}\href{https://doi.org/10.1021/acs.jctc.5b00099}{, 2015}.

\bibitem{bbhm2017q}
James Barker, Johannes Bulin, Jan Hamaekers, Sonja Mathias:
  \href{https://doi.org/10.1007/978-3-319-62458-7_2}{\textit{{LC-GAP}:
  localized {C}oulomb descriptors for the {G}aussian approximation potential}}.
  In Michael Griebel, Anton Sch{\"u}ller, Marc~Alexander Schweitzer (editors),
  \textit{\href{https://doi.org/10.1007/978-3-319-62458-7}{Scientific computing
  and algorithms in industrial simulations}}, 25, Springer, 2017.

\bibitem{availmbtr}
Available as part of the software \texttt{qmmlpack} (quantum mechanics machine
  learning package) at \url{https://gitlab.com/qmml/qmmlpack}, Apache 2.0
  license, publicly accessible).

\bibitem{dbcc2016q}
Sandip De, Albert~P. Bart{\'o}k, G{\'a}bor Cs{\'a}nyi, Michele Ceriotti:
  \href{https://doi.org/10.1039/C6CP00415F}{\textit{Comparing molecules and
  solids across structural and alchemical space}}. Physical Chemistry Chemical
  Physics \href{https://doi.org/10.1039/C6CP00415F}{18(20):
  13754}\href{https://doi.org/10.1039/C6CP00415F}{, 2016}.

\bibitem{bdpbkcc2017q}
Albert~P. Bart{\'o}k, Sandip De, Carl Poelking, Noam Bernstein, James~R.
  Kermode, G{\'a}bor Cs{\'a}nyi, Michele Ceriotti:
  \href{https://doi.org/10.1126/sciadv.1701816}{\textit{Machine learning
  unifies the modelling of materials and molecules}}. Science Advances
  \href{https://doi.org/10.1126/sciadv.1701816}{3(12):
  e1701816}\href{https://doi.org/10.1126/sciadv.1701816}{, 2017}.

\bibitem{c2019dq}
Miguel~A. Caro:
  \href{https://doi.org/10.1103/physrevb.100.024112}{\textit{Optimizing
  many-body atomic descriptors for enhanced computational performance of
  machine learning based interatomic potentials}}. Physical Review~B
  \href{https://doi.org/10.1103/physrevb.100.024112}{100(2):
  024112}\href{https://doi.org/10.1103/physrevb.100.024112}{, 2019}.

\bibitem{kme2019q}
Emir Kocer, Jeremy~K. Mason, Hakan Erturk:
  \href{https://doi.org/10.1063/1.5086167}{\textit{A novel approach to describe
  chemical environments in high-dimensional neural network potentials}}.
  Journal of Chemical Physics \href{https://doi.org/10.1063/1.5086167}{150(15):
  154102}\href{https://doi.org/10.1063/1.5086167}{, 2019}.

\bibitem{kme2020q}
Emir Kocer, Jeremy~K. Mason, Hakan Erturk:
  \href{https://doi.org/10.1063/1.5111045}{\textit{Continuous and optimally
  complete description of chemical environments using spherical {B}essel
  descriptors}}. AIP Advances \href{https://doi.org/10.1063/1.5111045}{10(1):
  015021}\href{https://doi.org/10.1063/1.5111045}{, 2020}.

\bibitem{availsoap}
Available as part of the software \texttt{libAtoms}
  (\url{http://www.libatoms.org}, custom license, per webform request).

\bibitem{sgsmlg2013q}
Ali Sadeghi, S.~Alireza Ghasemi, Bastian Schaefer, Stephan Mohr, Markus~A.
  Lill, Stefan Goedecker:
  \href{https://doi.org/10.1063/1.4828704}{\textit{Metrics for measuring
  distances in configuration spaces}}. Journal of Chemical Physics
  \href{https://doi.org/10.1063/1.4828704}{139(18):
  184118}\href{https://doi.org/10.1063/1.4828704}{, 2013}.

\bibitem{zafsfrgsgwg2016q}
Li~Zhu, Maximilian Amsler, Tobias Fuhrer, Bastian Schaefer, Somayeh Faraji,
  Samare Rostami, S.~Alireza Ghasemi, Ali Sadeghi, Migle Grauzinyte, Chris
  Wolverton, Stefan Goedecker:
  \href{https://doi.org/10.1063/1.4940026}{\textit{A fingerprint based metric
  for measuring similarities of crystalline structures}}. Journal of Chemical
  Physics \href{https://doi.org/10.1063/1.4940026}{144(3):
  034203}\href{https://doi.org/10.1063/1.4940026}{, 2016}.

\bibitem{hbrplmt2015q}
Katja Hansen, Franziska Biegler, Raghunathan Ramakrishnan, Wiktor Pronobis,
  O.~Anatole von Lilienfeld, Klaus-Robert M{\"u}ller, Alexandre Tkatchenko:
  \href{https://doi.org/10.1021/acs.jpclett.5b00831}{\textit{Machine learning
  predictions of molecular properties: accurate many-body potentials and
  nonlocality in chemical space}}. Journal of Physical Chemistry Letters
  \href{https://doi.org/10.1021/acs.jpclett.5b00831}{6(12):
  2326}\href{https://doi.org/10.1021/acs.jpclett.5b00831}{, 2015}.

\bibitem{hl2016q}
Bing Huang, O.~Anatole von Lilienfeld:
  \href{https://doi.org/10.1063/1.4964627}{\textit{Communication: understanding
  molecular representations in machine learning: the role of uniqueness and
  target similarity}}. Journal of Chemical Physics
  \href{https://doi.org/10.1063/1.4964627}{145(16):
  161102}\href{https://doi.org/10.1063/1.4964627}{, 2016}.

\bibitem{availqml}
Available as part of the software \texttt{QML} (quantum machine learning,
  \url{https://www.qmlcode.org/}, MIT license, publicly accessible).

\bibitem{hpm2015}
Matthew Hirn, Nicolas Poilvert, St{\'e}phane Mallat:
  \textit{\href{https://arxiv.org/abs/arXiv:1502.02077}{Quantum energy
  regression using scattering transforms}}. arXiv 1502.02077, 2015.

\bibitem{hmp2017q}
Matthew Hirn, St{\'e}phane Mallat, Nicolas Poilvert:
  \href{https://doi.org/10.1137/16M1075454}{\textit{Wavelet scattering
  regression of quantum chemical energies}}. Multiscale Modeling and Simulation
  \href{https://doi.org/10.1137/16M1075454}{15(2):
  827}\href{https://doi.org/10.1137/16M1075454}{, 2017}.

\bibitem{eehm2017q}
Michael Eickenberg, Georgios Exarchakis, Matthew Hirn, St{\'e}phane Mallat:
  \textit{\href{http://papers.nips.cc/paper/7232-solid-harmonic-wavelet-scattering-predicting-quantum-molecular-energy-from-invariant-descriptors-of-3d-electronic-densities}{Solid
  harmonic wavelet scattering: predicting quantum molecular energy from
  invariant descriptors of {3D} electronic densities}}. In \textit{Advances in
  Neural Information Processing Systems~30 (NeurIPS)}, 6522, 2017.

\bibitem{bskqh2018q}
Xavier Brumwell, Paul Sinz, Kwang~Jin Kim, Yue Qi, Matthew Hirn:
  \textit{\href{https://arxiv.org/abs/1812.02320}{Steerable wavelet scattering
  for {3D} atomic systems with application to {Li}-{Si} energy prediction}}. In
  \textit{NeurIPS Workshop on Machine Learning for Molecules and Materials},
  2018.

\bibitem{eehmt2018q}
Michael Eickenberg, Georgios Exarchakis, Matthew Hirn, St{\'e}phane Mallat,
  Louis Thiry: \href{https://doi.org/10.1063/1.5023798}{\textit{Solid harmonic
  wavelet scattering for predictions of molecule properties}}. Journal of
  Chemical Physics \href{https://doi.org/10.1063/1.5023798}{148(24):
  241732}\href{https://doi.org/10.1063/1.5023798}{, 2018}.

\bibitem{hhrnh2019q}
Eric~R. Homer, Derek~M. Hensley, Conrad~W. Rosenbrock, Andrew~H. Nguyen, Gus
  L.~W. Hart:
  \href{https://doi.org/10.3389/fmats.2019.00168}{\textit{Machine-learning
  informed representations for grain boundary structures}}. Frontiers in
  Materials \href{https://doi.org/10.3389/fmats.2019.00168}{6:
  168}\href{https://doi.org/10.3389/fmats.2019.00168}{, 2019}.

\bibitem{ssblkqh2020q}
Paul Sinz, Michael~W. Swift, Xavier Brumwell, Jialin Liu, Kwang~Jin Kim, Yue
  Qi, Matthew Hirn: \href{https://doi.org/10.1063/5.0016020}{\textit{Wavelet
  scattering networks for atomistic systems with extrapolation of material
  properties}}. Journal of Chemical Physics
  \href{https://doi.org/10.1063/5.0016020}{153(8):
  084109}\href{https://doi.org/10.1063/5.0016020}{, 2020}.

\bibitem{aaelrtzetal2020q}
Mathieu Andreux, Tom{\'a}s Angles, Georgios Exarchakis, Roberto Leonarduzzi,
  Gaspar Rochette, Louis Thiry, John Zarka, St{\'e}phane Mallat, Joakim
  And{\'e}n, Eugene Belilovsky, Joan Bruna, Vincent Lostanlen, Muawiz
  Chaudhary, Matthew~J. Hirn, Edouard Oyallon, Sixin Zhang, Carmine Cella,
  Michael Eickenberg:
  \textit{\href{http://jmlr.org/papers/v21/19-047.html}{{K}ymatio: scattering
  transforms in {P}ython}}. Journal of Machine Learning Research 21(60): 1,
  2020.

\bibitem{s2016q}
Alexander~V. Shapeev: \href{https://doi.org/10.1137/15M1054183}{\textit{Moment
  tensor potentials: a class of systematically improvable interatomic
  potentials}}. Multiscale Modeling and Simulation
  \href{https://doi.org/10.1137/15M1054183}{14(3):
  1153}\href{https://doi.org/10.1137/15M1054183}{, 2016}.

\bibitem{ps2017q}
Evgeny~V. Podryabinkin, Alexander~V. Shapeev:
  \href{https://doi.org/10.1016/j.commatsci.2017.08.031}{\textit{Active
  learning of linearly parametrized interatomic potentials}}. Computational
  Materials Science \href{https://doi.org/10.1016/j.commatsci.2017.08.031}{140:
  171}\href{https://doi.org/10.1016/j.commatsci.2017.08.031}{, 2017}.

\bibitem{gps2018q}
Konstantin Gubaev, Evgeny~V. Podryabinkin, Alexander~V. Shapeev:
  \href{https://doi.org/10.1063/1.5005095}{\textit{Machine learning of
  molecular properties: locality and active learning}}. Journal of Chemical
  Physics \href{https://doi.org/10.1063/1.5005095}{148(24):
  241727}\href{https://doi.org/10.1063/1.5005095}{, 2018}.

\bibitem{ns2018q}
Ivan~S. Novikov, Alexander~V. Shapeev:
  \href{https://doi.org/10.1016/j.mtcomm.2018.11.008}{\textit{Improving
  accuracy of interatomic potentials: more physics or more data? {A} case study
  of silica}}. Materials Today Communications
  \href{https://doi.org/10.1016/j.mtcomm.2018.11.008}{18:
  74}\href{https://doi.org/10.1016/j.mtcomm.2018.11.008}{, 2018}.

\bibitem{s2019q}
Alexander~V. Shapeev:
  \href{https://doi.org/10.1039/9781788010122}{\textit{Applications of machine
  learning for representing interatomic interactions}}. In Artem~R. Oganov,
  Gabriele Saleh, Alexander~G. Kvashnin (editors), \textit{Computational
  Materials Discovery}, chapter~3, 66, Royal Society of Chemistry, 2019.

\bibitem{ngps2020}
Ivan~S. Novikov, Konstantin Gubaev, Evgeny~V. Podryabinkin, Alexander~V.
  Shapeev: \href{https://doi.org/10.1088/2632-2153/abc9fe}{\textit{The {MLIP}
  package: Moment tensor potentials with {MPI} and active learning}}. Machine
  Learning: Science and Technology
  \href{https://doi.org/10.1088/2632-2153/abc9fe}{in
  press}\href{https://doi.org/10.1088/2632-2153/abc9fe}{, 2020}.

\bibitem{hr2017}
Haoyan Huo, Matthias Rupp:
  \textit{\href{https://arxiv.org/abs/1704.06439}{Unified representation of
  molecules and crystals for machine learning}}. arXiv 1704.06439, 2017.

\bibitem{tzk2018q}
Yu-Hang Tang, Dongkun Zhang, George~Em Karniadakis:
  \href{https://doi.org/10.1063/1.5008630}{\textit{An atomistic fingerprint
  algorithm for learning \textit{ab initio} molecular force fields}}. Journal
  of Chemical Physics \href{https://doi.org/10.1063/1.5008630}{148(3):
  034101}\href{https://doi.org/10.1063/1.5008630}{, 2018}.

\bibitem{availdecaf}
A reference implementation in Python can be found at
  \url{https://doi.org/10.5281/ZENODO.1054550}, publicly accessible.

\bibitem{fchl2018q}
Felix~A. Faber, Anders~S. Christensen, Bing Huang, O.~Anatole von Lilienfeld:
  \href{https://doi.org/10.1063/1.5020710}{\textit{Alchemical and structural
  distribution based representation for universal quantum machine learning}}.
  Journal of Chemical Physics \href{https://doi.org/10.1063/1.5020710}{148(24):
  241717}\href{https://doi.org/10.1063/1.5020710}{, 2018}.

\bibitem{cbfgl2020q}
Anders~S. Christensen, Lars~A. Bratholm, Felix~A. Faber, O.~Anatole von
  Lilienfeld: \href{https://doi.org/10.1063/1.5126701}{\textit{{FCHL}
  revisited: faster and more accurate quantum machine learning}}. Journal of
  Chemical Physics \href{https://doi.org/10.1063/1.5126701}{152(4):
  044107}\href{https://doi.org/10.1063/1.5126701}{, 2020}.

\bibitem{ptm2018q}
Wiktor Pronobis, Alexandre Tkatchenko, Klaus-Robert M{\"u}ller:
  \href{https://doi.org/10.1021/acs.jctc.8b00110}{\textit{Many-body descriptors
  for predicting molecular properties with machine learning: analysis of
  pairwise and three-body interactions in molecules}}. Journal of Chemical
  Theory and Computation \href{https://doi.org/10.1021/acs.jctc.8b00110}{14(6):
  2991}\href{https://doi.org/10.1021/acs.jctc.8b00110}{, 2018}.

\bibitem{availidmbr}
Pseudo-code is available as part of the supporting information at
  \url{http://pubs.acs.org/doi/abs/10.1021/acs.jctc.8b00110}.

\bibitem{wcm2018q}
Matthew Welborn, Lixue Cheng, Thomas~F. Miller, III:
  \href{https://doi.org/10.1021/acs.jctc.8b00636}{\textit{Transferability in
  machine learning for electronic structure via the molecular orbital basis}}.
  Journal of Chemical Theory and Computation
  \href{https://doi.org/10.1021/acs.jctc.8b00636}{14(9):
  4772}\href{https://doi.org/10.1021/acs.jctc.8b00636}{, 2018}.

\bibitem{cwcm2019q}
Lixue Cheng, Matthew Welborn, Anders~S. Christensen, Thomas~F. Miller, III:
  \href{https://doi.org/10.1063/1.5088393}{\textit{A universal density matrix
  functional from molecular orbital-based machine learning: transferability
  across organic molecules}}. Journal of Chemical Physics
  \href{https://doi.org/10.1063/1.5088393}{150(13):
  131103}\href{https://doi.org/10.1063/1.5088393}{, 2019}.

\bibitem{hsclm2020}
Tamara Husch, Jiace Sun, Lixue Cheng, Sebastian J.~R. Lee, Thomas~F. Miller:
  \textit{\href{https://arxiv.org/abs/arXiv:2010.03626}{Improved accuracy and
  transferability of molecular-orbital-based machine learning: organics,
  transition-metal complexes, non-covalent interactions, and transition
  states}}. arXiv 2010.03626, 2020.

\bibitem{lhdm2020}
Sebastian J.~R. Lee, Tamara Husch, Feizhi Ding, Thomas~F. Miller:
  \textit{\href{https://arxiv.org/abs/2012.08899}{Analytical Gradients for
  Molecular-Orbital-Based Machine Learning}}. arXiv 2012.08899, 2020.

\bibitem{d2019bq}
Ralf Drautz: \href{https://doi.org/10.1103/physrevb.99.014104}{\textit{Atomic
  cluster expansion for accurate and transferable interatomic potentials}}.
  Physical Review~B \href{https://doi.org/10.1103/physrevb.99.014104}{99(1):
  249901}\href{https://doi.org/10.1103/physrevb.99.014104}{, 2019}.

\bibitem{dbcdeoo2020}
Markus Bachmayr, G{\'a}bor Cs{\'a}nyi, Ralf Drautz, Genevieve Dusson, Simon
  Etter, Cas van~der Oord, Christoph Ortner:
  \textit{\href{https://arxiv.org/abs/1911.03550}{Atomic cluster expansion:
  completeness, efficiency and stability}}. arXiv 1911.03550, 2020.

\bibitem{d2020q}
Ralf Drautz: \href{https://doi.org/10.1103/physrevb.102.024104}{\textit{Atomic
  cluster expansion of scalar, vectorial, and tensorial properties including
  magnetism and charge transfer}}. Physical Review~B
  \href{https://doi.org/10.1103/physrevb.102.024104}{102(2):
  024104}\href{https://doi.org/10.1103/physrevb.102.024104}{, 2020}.

\bibitem{availace}
An implementation in Julia can be found at
  \url{https://github.com/ACEsuit/ACE.jl}, publicly accessible.

\bibitem{npc2020q}
Jigyasa Nigam, Sergey Pozdnyakov, Michele Ceriotti:
  \href{https://doi.org/10.1063/5.0021116}{\textit{Recursive evaluation and
  iterative contraction of $n$-body equivariant features}}. Journal of Chemical
  Physics \href{https://doi.org/10.1063/5.0021116}{153(12):
  121101}\href{https://doi.org/10.1063/5.0021116}{, 2020}.

\bibitem{availnice}
An implementation in Python can be found at
  \url{https://github.com/cosmo-epfl/nice}.

\bibitem{zk2020q}
Viktor Zaverkin, Johannes K{\"a}stner:
  \href{https://doi.org/10.1021/acs.jctc.0c00347}{\textit{{G}aussian moments as
  physically inspired molecular descriptors for accurate and scalable machine
  learning potentials}}. Journal of Chemical Theory and Computation
  \href{https://doi.org/10.1021/acs.jctc.0c00347}{16(8):
  5410}\href{https://doi.org/10.1021/acs.jctc.0c00347}{, 2020}.

\bibitem{gsd2017q}
Aldo Glielmo, Peter Sollich, Alessandro De~Vita:
  \href{https://doi.org/10.1103/PhysRevB.95.214302}{\textit{Accurate
  interatomic force fields via machine learning with covariant kernels}}.
  Physical Review~B \href{https://doi.org/10.1103/PhysRevB.95.214302}{95(21):
  214302}\href{https://doi.org/10.1103/PhysRevB.95.214302}{, 2017}.

\bibitem{gwcc2018q}
Andrea Grisafi, David~M. Wilkins, G{\'a}bor Cs{\'a}nyi, Michele Ceriotti:
  \href{https://doi.org/10.1103/PhysRevLett.120.036002}{\textit{Symmetry-adapted
  machine-learning for tensorial properties of atomistic systems}}. Physical
  Review Letters \href{https://doi.org/10.1103/PhysRevLett.120.036002}{120(3):
  036002}\href{https://doi.org/10.1103/PhysRevLett.120.036002}{, 2018}.

\bibitem{htpak2019q}
Truong~Son Hy, Shubhendu Trivedi, Horace Pan, Brandon~M. Anderson, Risi Kondor:
  \textit{\href{http://www.mlgworkshop.org/2019/papers/MLG2019_paper_36.pdf}{Covariant
  compositional networks for learning graphs}}. In \textit{Proceedings of the
  International Workshop on Mining and Learning with Graphs (MLG)}, 2019.

\bibitem{lrrk2015q}
O.~Anatole von Lilienfeld, Raghunathan Ramakrishnan, Matthias Rupp, Aaron
  Knoll: \href{https://doi.org/10.1002/qua.24912}{\textit{{F}ourier series of
  atomic radial distribution functions: a molecular fingerprint for machine
  learning models of quantum chemical properties}}. International Journal of
  Quantum Chemistry \href{https://doi.org/10.1002/qua.24912}{115(16):
  1084}\href{https://doi.org/10.1002/qua.24912}{, 2015}.

\bibitem{pwbocc2020q}
Sergey~N. Pozdnyakov, Michael~J. Willatt, Albert~P. Bart{\'{o}}k, Christoph
  Ortner, G{\'{a}}bor Cs{\'{a}}nyi, Michele Ceriotti:
  \href{https://doi.org/10.1103/physrevlett.125.166001}{\textit{Incompleteness
  of atomic structure representations}}. Physical Review Letters
  \href{https://doi.org/10.1103/physrevlett.125.166001}{125(16):
  166001}\href{https://doi.org/10.1103/physrevlett.125.166001}{, 2020}.

\bibitem{lhr2009q}
Hung~M. Le, Sau Huynh, Lionel~M. Raff:
  \href{https://doi.org/10.1063/1.3159748}{\textit{Molecular dissociation of
  hydrogen peroxide ({HOOH}) on a neural network \textit{ab initio} potential
  surface with a new configuration sampling method involving gradient
  fitting}}. Journal of Chemical Physics
  \href{https://doi.org/10.1063/1.3159748}{131(1):
  014107}\href{https://doi.org/10.1063/1.3159748}{, 2009}.

\bibitem{bc2015q}
Albert~P. Bart{\'o}k, G{\'a}bor Cs{\'a}nyi:
  \href{https://doi.org/10.1002/qua.24927}{\textit{{G}aussian approximation
  potentials: a brief tutorial introduction}}. International Journal of Quantum
  Chemistry \href{https://doi.org/10.1002/qua.24927}{116(13):
  1051}\href{https://doi.org/10.1002/qua.24927}{, 2015}.

\bibitem{csmt2018q}
Stefan Chmiela, Huziel~E. Sauceda, Klaus-Robert M{\"u}ller, Alexandre
  Tkatchenko: \href{https://doi.org/10.1038/s41467-018-06169-2}{\textit{Towards
  exact molecular dynamics simulations with machine-learned force fields}}.
  Nature Communications \href{https://doi.org/10.1038/s41467-018-06169-2}{9:
  3887}\href{https://doi.org/10.1038/s41467-018-06169-2}{, 2018}.

\bibitem{cgly2018q}
Christopher~R. Collins, Geoffrey~J. Gordon, O.~Anatole von Lilienfeld, David~J.
  Yaron: \href{https://doi.org/10.1063/1.5020441}{\textit{Constant size
  descriptors for accurate machine learning models of molecular properties}}.
  Journal of Chemical Physics \href{https://doi.org/10.1063/1.5020441}{148(24):
  241718}\href{https://doi.org/10.1063/1.5020441}{, 2018}.

\bibitem{jkvak2020q}
Ryosuke Jinnouchi, Ferenc Karsai, Carla Verdi, Ryoji Asahi, Georg Kresse:
  \href{https://doi.org/10.1063/5.0009491}{\textit{Descriptors representing
  two- and three-body atomic distributions and their effects on the accuracy of
  machine-learned inter-atomic potentials}}. Journal of Chemical Physics
  \href{https://doi.org/10.1063/5.0009491}{152(23)}\href{https://doi.org/10.1063/5.0009491}{,
  2020}.

\bibitem{wmc2019q}
Michael~J. Willatt, F{\'e}lix Musil, Michele Ceriotti:
  \href{https://doi.org/10.1063/1.5090481}{\textit{Atom-density representations
  for machine learning}}. Journal of Chemical Physics
  \href{https://doi.org/10.1063/1.5090481}{150(15):
  154110}\href{https://doi.org/10.1063/1.5090481}{, 2019}.

\bibitem{sgbsmg2014q}
Kristof~T. Sch{\"u}tt, Henning Glawe, Felix Brockherde, Antonio Sanna,
  Klaus-Robert M{\"u}ller, Eberhard~K.U. Gross:
  \href{https://doi.org/10.1103/PhysRevB.89.205118}{\textit{How to represent
  crystal structures for machine learning: towards fast prediction of
  electronic properties}}. Physical Review~B
  \href{https://doi.org/10.1103/PhysRevB.89.205118}{89(20):
  205118}\href{https://doi.org/10.1103/PhysRevB.89.205118}{, 2014}.

\bibitem{localmbtr2019}
The \texttt{DScribe} code contains a local MBTR example of this. See
  \url{https://github.com/SINGROUP/dscribe}.

\bibitem{lkv2015q}
Zhenwei Li, James~R. Kermode, Alessandro De~Vita:
  \href{https://doi.org/10.1103/PhysRevLett.114.096405}{\textit{Molecular
  dynamics with on-the-fly machine learning of quantum mechanical forces}}.
  Physical Review Letters
  \href{https://doi.org/10.1103/PhysRevLett.114.096405}{114(9):
  096405}\href{https://doi.org/10.1103/PhysRevLett.114.096405}{, 2015}.

\bibitem{rdrl2014q}
Raghunathan Ramakrishnan, Pavlo~O. Dral, Matthias Rupp, O.~Anatole von
  Lilienfeld: \href{https://doi.org/10.1038/sdata.2014.22}{\textit{Quantum
  chemistry structures and properties of 134 kilo molecules}}. Scientific Data
  \href{https://doi.org/10.1038/sdata.2014.22}{1:
  140022}\href{https://doi.org/10.1038/sdata.2014.22}{, 2014}.

\bibitem{availqm9dft10b}
Available at the \texttt{QM/ML} website (quantum mechanics/machine learning,
  \url{https://qmml.org}, publicly accessible).

\bibitem{kaggle}
Nomad2018 Predicting Transparent Conductors. Predict the key properties of
  novel transparent semiconductors. Available at
  \url{https://www.kaggle.com/c/nomad2018-predict-transparent-conductors}.

\bibitem{sgylbhglzs2019}
Christopher Sutton, Luca~M. Ghiringhelli, Takenori Yamamoto, Yury Lysogorskiy,
  Lars Blumenthal, Thomas Hammerschmidt, Jacek~R. Golebiowski, Xiangyue Liu,
  Angelo Ziletti, Matthias Scheffler:
  \href{https://doi.org/10.1038/s41524-019-0239-3}{\textit{Crowd-sourcing
  materials-science challenges with the {NOMAD} 2018 {K}aggle competition}}.
  npj Computational Materials
  \href{https://doi.org/10.1038/s41524-019-0239-3}{5:
  111}\href{https://doi.org/10.1038/s41524-019-0239-3}{, 2019}.

\bibitem{r2015fq}
Matthias Rupp: \href{https://doi.org/10.1002/qua.24954}{\textit{Machine
  learning for quantum mechanics in a nutshell}}. International Journal of
  Quantum Chemistry \href{https://doi.org/10.1002/qua.24954}{115(16):
  1058}\href{https://doi.org/10.1002/qua.24954}{, 2015}.

\bibitem{rw2006}
Carl Rasmussen, Christopher Williams:
  \textit{\href{http://www.gaussianprocess.org/gpml}{{G}aussian processes for
  machine learning}}. MIT Press, Cambridge, 2006.

\bibitem{bbbk2011}
James~S. Bergstra, R{\'e}mi Bardenet, Yoshua Bengio, Bal{\'a}zs K{\'e}gl:
  \textit{\href{https://papers.nips.cc/paper/4443-algorithms-for-hyper-parameter-optimization}{Algorithms
  for hyper-parameter optimization}}. In \textit{Advances in Neural Information
  Processing Systems~24 (NIPS)}, 2546, 2011.

\bibitem{byc2013}
James~S. Bergstra, Daniel Yamins, David~D. Cox:
  \textit{\href{http://proceedings.mlr.press/v28/bergstra13.pdf}{Making a
  science of model search: hyperparameter optimization in hundreds of
  dimensions for vision architectures}}. In \textit{Proceedings of the 30th
  International Conference on Machine Learning (ICML)}, 115, 2013.

\bibitem{cjsvd1994}
Corinna Cortes, Lawrence~D. Jackel, Sara~A. Solla, Vladimir Vapnik, John~S.
  Denker:
  \textit{\href{https://papers.nips.cc/paper/803-learning-curves-asymptotic-values-and-rate-of-convergence}{Learning
  curves: asymptotic values and rate of convergence}}. In \textit{Advances in
  Neural Information Processing Systems~6 (NIPS)}, 1993.

\bibitem{mfmsa1996}
Klaus-Robert M{\"u}ller, Michael Finke, Noboru Murata, Klaus Schulten,
  Shun{-}ichi Amari:
  \href{https://doi.org/10.1162/neco.1996.8.5.1085}{\textit{A numerical study
  on learning curves in stochastic multilayer feedforward networks}}. Neural
  Computation \href{https://doi.org/10.1162/neco.1996.8.5.1085}{8(5):
  1085}\href{https://doi.org/10.1162/neco.1996.8.5.1085}{, 1996}.

\bibitem{hsl2018q}
Bing Huang, Nadine~O. Symonds, O.~Anatole von Lilienfeld:
  \href{https://doi.org/10.1007/978-3-319-42913-7_67-1}{\textit{Quantum machine
  learning in chemistry and materials}}. In Wanda Andreoni, Sidney Yip
  (editors), \textit{\href{https://doi.org/10.1007/978-3-319-42913-7}{Handbook
  of materials modeling. Methods: Theory and modeling}}, Springer, 2018.

\bibitem{s2018cq}
Amit Samanta: \href{https://doi.org/10.1063/1.5055772}{\textit{Representing
  local atomic environment using descriptors based on local correlations}}.
  Journal of Chemical Physics \href{https://doi.org/10.1063/1.5055772}{149(24):
  244102}\href{https://doi.org/10.1063/1.5055772}{, 2018}.

\bibitem{hxh2020q}
Shreyas~J. Honrao, Stephen~R. Xie, Richard~G. Hennig:
  \href{https://doi.org/10.1063/5.0012407}{\textit{Augmenting machine learning
  of energy landscapes with local structural information}}. Journal of Applied
  Physics \href{https://doi.org/10.1063/5.0012407}{128(8):
  085101}\href{https://doi.org/10.1063/5.0012407}{, 2020}.

\bibitem{mvgfwsjc2021}
F{\'e}lix Musil, Max Veit, Alexander Goscinski, Guillaume Fraux, Michael~J.
  Willatt, Markus Stricker, Till Junge, Michele Ceriotti:
  \textit{\href{https://arxiv.org/abs/arXiv:2101.08814}{Efficient
  implementation of atom-density representations}}. arXiv 2101.08814, 2021.

\bibitem{ghsg2015q}
S.~Alireza Ghasemi, Albert Hofstetter, Santanu Saha, Stefan Goedecker:
  \href{https://doi.org/http://dx.doi.org/10.1103/PhysRevB.92.045131}{\textit{Interatomic
  potentials for ionic systems with density functional accuracy based on charge
  densities obtained by a neural network}}. Physical Review~B
  \href{https://doi.org/http://dx.doi.org/10.1103/PhysRevB.92.045131}{92(4):
  045131}\href{https://doi.org/http://dx.doi.org/10.1103/PhysRevB.92.045131}{,
  2015}.

\bibitem{nlsslirbt2018q}
Benjamin Nebgen, Nicholas Lubbers, Justin~S. Smith, Andrew~E. Sifain, Andrey
  Lokhov, Olexandr Isayev, Adrian~E. Roitberg, Kipton Barros, Sergei Tretiak:
  \href{https://doi.org/10.1021/acs.jctc.8b00524}{\textit{Transferable dynamic
  molecular charge assignment using deep neural networks}}. Journal of Chemical
  Theory and Computation \href{https://doi.org/10.1021/acs.jctc.8b00524}{14(9):
  4687}\href{https://doi.org/10.1021/acs.jctc.8b00524}{, 2018}.

\bibitem{kfgb2020}
Tsz~Wai Ko, Jonas~A. Finkler, Stefan Goedecker, J{\"o}rg Behler:
  \textit{\href{https://arxiv.org/abs/arXiv:2009.06484}{A fourth-generation
  high-dimensional neural network potential with accurate electrostatics
  including non-local charge transfer}}. arXiv 2009.06484, 2020.

\bibitem{pfg2020}
Behnam Parsaeifard, Jonas~A. Finkler, Stefan Goedecker:
  \textit{\href{https://arxiv.org/abs/arXiv:2008.11277}{Detecting non-local
  effects in the electronic structure of a simple covalent system with machine
  learning methods}}. arXiv 2008.11277, 2020.

\bibitem{wm2020q}
Julia Westermayr, Philipp Marquetand:
  \href{https://doi.org/10.1088/2632-2153/ab9c3e}{\textit{Machine learning and
  excited-state molecular dynamics}}. Machine Learning: Science and Technology
  \href{https://doi.org/10.1088/2632-2153/ab9c3e}{1(4):
  043001}\href{https://doi.org/10.1088/2632-2153/ab9c3e}{, 2020}.

\bibitem{wm2020bq}
Julia Westermayr, Philipp Marquetand:
  \href{https://doi.org/10.1063/5.0021915}{\textit{Deep learning for {UV}
  absorption spectra with {SchNarc}: first steps toward transferability in
  chemical compound space}}. Journal of Chemical Physics
  \href{https://doi.org/10.1063/5.0021915}{153(15):
  154112}\href{https://doi.org/10.1063/5.0021915}{, 2020}.

\bibitem{wgm2020q}
Julia Westermayr, Michael Gastegger, Philipp Marquetand:
  \href{https://doi.org/10.1021/acs.jpclett.0c00527}{\textit{Combining {SchNet}
  and {SHARC}: the {SchNarc} machine learning approach for excited-state
  dynamics}}. Journal of Physical Chemistry Letters
  \href{https://doi.org/10.1021/acs.jpclett.0c00527}{11(10):
  3828}\href{https://doi.org/10.1021/acs.jpclett.0c00527}{, 2020}.

\bibitem{bdlrs2005q}
J{\"o}rg Behler, Bernard Delley, S{\"o}nke Lorenz, Karsten Reuter, Matthias
  Scheffler:
  \href{https://doi.org/10.1103/physrevlett.94.036104}{\textit{Dissociation of
  {O$_2$} at {Al}(111): the role of spin selection rules}}. Physical Review
  Letters \href{https://doi.org/10.1103/physrevlett.94.036104}{94(3):
  036104}\href{https://doi.org/10.1103/physrevlett.94.036104}{, 2005}.

\bibitem{wfclm2020q}
Julia Westermayr, Felix~A. Faber, Anders~S. Christensen, O.~Anatole von
  Lilienfeld, Philipp Marquetand:
  \href{https://doi.org/10.1088/2632-2153/ab88d0}{\textit{Neural networks and
  kernel ridge regression for excited states dynamics of {CH$_2$NH$_2^+$}: from
  single-state to multi-state representations and multi-property machine
  learning models}}. Machine Learning: Science and Technology
  \href{https://doi.org/10.1088/2632-2153/ab88d0}{1(2):
  025009}\href{https://doi.org/10.1088/2632-2153/ab88d0}{, 2020}.

\bibitem{gsm2020}
Michael Gastegger, Kristof~T. Sch{\"u}tt, Klaus-Robert M{\"u}ller:
  \textit{\href{https://arxiv.org/abs/2010.14942}{Machine learning of solvent
  effects on molecular spectra and reactions}}. arXiv 2010.14942, 2020.

\bibitem{cfl2019q}
Anders~S. Christensen, Felix~A. Faber, O.~Anatole von Lilienfeld:
  \href{https://doi.org/10.1063/1.5053562}{\textit{Operators in quantum machine
  learning: response properties in chemical space}}. Journal of Chemical
  Physics \href{https://doi.org/10.1063/1.5053562}{150(6):
  064105}\href{https://doi.org/10.1063/1.5053562}{, 2019}.

\bibitem{wmc2018q}
Michael~J. Willatt, F{\'e}lix Musil, Michele Ceriotti:
  \href{https://doi.org/10.1039/C8CP05921G}{\textit{Feature optimization for
  atomistic machine learning yields a data-driven construction of the periodic
  table of the elements}}. Physical Chemistry Chemical Physics
  \href{https://doi.org/10.1039/C8CP05921G}{20(47):
  29661}\href{https://doi.org/10.1039/C8CP05921G}{, 2018}.

\bibitem{hkyp2019q}
John~E. Herr, Kevin Koh, Kun Yao, John Parkhill:
  \href{https://doi.org/10.1063/1.5108803}{\textit{Compressing physics with an
  autoencoder: creating an atomic species representation to improve machine
  learning models in the chemical sciences}}. Journal of Chemical Physics
  \href{https://doi.org/10.1063/1.5108803}{151(6--7):
  455}\href{https://doi.org/10.1063/1.5108803}{, 2019}.

\bibitem{gmm2019q}
Alexandra~M. Goryaeva, Jean-Bernard Maillet, Mihai-Cosmin Marinica:
  \href{https://doi.org/10.1016/j.commatsci.2019.04.043}{\textit{Towards better
  efficiency of interatomic linear machine learning potentials}}. Computational
  Materials Science \href{https://doi.org/10.1016/j.commatsci.2019.04.043}{166:
  200}\href{https://doi.org/10.1016/j.commatsci.2019.04.043}{, 2019}.

\bibitem{sgtm2019q}
Kristof~T. Sch{\"u}tt, Michael Gastegger, Alexandre Tkatchenko, Klaus-Robert
  M{\"u}ller:
  \href{https://doi.org/10.1007/978-3-030-28954-6_17}{\textit{Quantum-chemical
  insights from interpretable atomistic neural networks}}. In Wojciech Samek,
  Gr{\'{e}}goire Montavon, Andrea Vedaldi, Lars~Kai Hansen, Klaus-Robert
  M{\"u}ller (editors),
  \textit{\href{https://doi.org/10.1007/978-3-030-28954-6}{Explainable {AI}:
  interpreting, explaining and visualizing deep learning}}, 311--330, Springer,
  2019.

\bibitem{sgtmm2019q}
Kristof~T. Sch{\"u}tt, Michael Gastegger, Alexandre Tkatchenko, Klaus-Robert
  M{\"u}ller, Reinhard~J. Maurer:
  \href{https://doi.org/10.1038/s41467-019-12875-2}{\textit{Unifying machine
  learning and quantum chemistry with a deep neural network for molecular
  wavefunctions}}. Nature Communications
  \href{https://doi.org/10.1038/s41467-019-12875-2}{10:
  5024}\href{https://doi.org/10.1038/s41467-019-12875-2}{, 2019}.

\bibitem{rc2018q}
Mardochee Reveil, Paulette Clancy:
  \href{https://doi.org/10.1039/c8me00003d}{\textit{Classification of spatially
  resolved molecular fingerprints for machine learning applications and
  development of a codebase for their implementation}}. Molecular Systems
  Design \& Engineering \href{https://doi.org/10.1039/c8me00003d}{3(3):
  431}\href{https://doi.org/10.1039/c8me00003d}{, 2018}.

\bibitem{rrkal2020q}
Zachary del Rosario, Matthias Rupp, Yoolhee Kim, Erin Antono, Julia Ling:
  \href{https://doi.org/10.1063/5.0006124}{\textit{Assessing the frontier:
  active learning, model accuracy, and multi-objective candidate discovery and
  optimization}}. Journal of Chemical Physics
  \href{https://doi.org/10.1063/5.0006124}{153(2):
  024112}\href{https://doi.org/10.1063/5.0006124}{, 2020}.

\bibitem{capdv2004q}
G{\'a}bor Cs{\'a}nyi, Tristan Albaret, Mike~C. Payne, Alessandro De~Vita:
  \href{https://doi.org/10.1103/PhysRevLett.93.175503}{\textit{``{L}earn on the
  fly'': a hybrid classical and quantum-mechanical molecular dynamics
  simulation}}. Physical Review Letters
  \href{https://doi.org/10.1103/PhysRevLett.93.175503}{93(17):
  175503}\href{https://doi.org/10.1103/PhysRevLett.93.175503}{, 2004}.

\bibitem{sbgrvs2020q}
Christopher Sutton, Mario Boley, Luca~M. Ghiringhelli, Matthias Rupp, Jilles
  Vreeken, Matthias Scheffler:
  \href{https://doi.org/10.1038/s41467-020-17112-9}{\textit{Identifying domains
  of applicability of machine learning models for materials science}}. Nature
  Communications \href{https://doi.org/10.1038/s41467-020-17112-9}{11:
  4428}\href{https://doi.org/10.1038/s41467-020-17112-9}{, 2020}.

\bibitem{amb2011q}
Nongnuch Artrith, Tobias Morawietz, J{\"o}rg Behler:
  \href{https://doi.org/10.1103/PhysRevB.83.153101}{\textit{High-dimensional
  neural-network potentials for multicomponent systems: Applications to zinc
  oxide}}. Physical Review~B
  \href{https://doi.org/10.1103/PhysRevB.83.153101}{83(15):
  153101}\href{https://doi.org/10.1103/PhysRevB.83.153101}{, 2011}.

\bibitem{gc2019q}
Andrea Grisafi, Michele Ceriotti:
  \href{https://doi.org/10.1063/1.5128375}{\textit{Incorporating long-range
  physics in atomic-scale machine learning}}. Journal of Chemical Physics
  \href{https://doi.org/10.1063/1.5128375}{151(20):
  204105}\href{https://doi.org/10.1063/1.5128375}{, 2019}.

\bibitem{ss2016}
Edwin~Miles Stoudenmire, David~J. Schwab:
  \textit{\href{https://papers.nips.cc/paper/6211-supervised-learning-with-tensor-networks}{Supervised
  learning with tensor networks}}. In \textit{Advances in Neural Information
  Processing Systems~29 (NeurIPS)}, 4799, 2016.

\bibitem{nomadanalyticstoolkit}
Analytics Toolkit of the Novel Materials Discovery (NOMAD) Laboratory,
  \url{https://analytics-toolkit.nomad-coe.eu}.

\end{thebibliography}


\begin{thebibliography}{10}

\bibitem{pmv2013x}
Pavel~G. Polishchuk, Timur~I. Madzhidov, Alexandre Varnek:
  \href{https://doi.org/10.1007/s10822-013-9672-4}{\textit{Estimation of the
  size of drug-like chemical space based on {GDB-17} data}}. Journal of
  Computer-Aided Molecular Design
  \href{https://doi.org/10.1007/s10822-013-9672-4}{27(8):
  675}\href{https://doi.org/10.1007/s10822-013-9672-4}{, 2013}.

\bibitem{bmg1996x}
Regine~S. Bohacek, Colin McMartin, Wayne~C. Guida:
  \href{https://doi.org/10.1002/(SICI)1098-1128(199601)16:1\%3C3::AID-MED1\%3E3.0.CO;2-6}{\textit{The
  art and practice of structure-based drug design: a molecular modeling
  perspective}}. Medicinal Research Reviews
  \href{https://doi.org/10.1002/(SICI)1098-1128(199601)16:1\%3C3::AID-MED1\%3E3.0.CO;2-6}{16(1):
  3}\href{https://doi.org/10.1002/(SICI)1098-1128(199601)16:1\%3C3::AID-MED1\%3E3.0.CO;2-6}{,
  1996}.

\bibitem{fr2007x}
Tobias Fink, Jean-Louis Reymond:
  \href{https://doi.org/10.1021/ci600423u}{\textit{Virtual exploration of the
  chemical universe up to 11 atoms of {C}, {N}, {O}, {F}: assembly of 26.4
  million structures (110.9 million stereoisomers) and analysis for new ring
  systems, stereochemistry, physicochemical properties, compound classes, and
  drug discovery}}. Journal of Chemical Information and Modeling
  \href{https://doi.org/10.1021/ci600423u}{47(2):
  342}\href{https://doi.org/10.1021/ci600423u}{, 2007}.

\bibitem{br2009x}
Lorenz~C. Blum, Jean-Louis Reymond:
  \href{https://doi.org/10.1021/ja902302h}{\textit{970 million druglike small
  molecules for virtual screening in the chemical universe database {GDB-13}}}.
  Journal of the American Chemical Society
  \href{https://doi.org/10.1021/ja902302h}{131(25):
  8732}\href{https://doi.org/10.1021/ja902302h}{, 2009}.

\bibitem{rdbr2012x}
Lars Ruddigkeit, Ruud van Deursen, Lorenz~C. Blum, Jean-Louis Reymond:
  \href{https://doi.org/10.1021/ci300415d}{\textit{Enumeration of 166 billion
  organic small molecules in the chemical universe database {GDB-17}}}. Journal
  of Chemical Information and Modeling
  \href{https://doi.org/10.1021/ci300415d}{52(11):
  2864}\href{https://doi.org/10.1021/ci300415d}{, 2012}.

\bibitem{c2014cx}
Brian Cantor: \href{https://doi.org/10.3390/e16094749}{\textit{Multicomponent
  and high entropy alloys}}. Entropy
  \href{https://doi.org/10.3390/e16094749}{16(9):
  4749}\href{https://doi.org/10.3390/e16094749}{, 2014}.

\bibitem{fhhgsdvkrv2017x}
Felix~A. Faber, Luke Hutchison, Bing Huang, Justin Gilmer, Samuel~S.
  Schoenholz, George~E. Dahl, Oriol Vinyals, Steven Kearnes, Patrick~F. Riley,
  O.~Anatole von Lilienfeld:
  \href{https://doi.org/10.1021/acs.jctc.7b00577}{\textit{Prediction errors of
  molecular machine learning models lower than hybrid {DFT} error}}. Journal of
  Chemical Theory and Computation
  \href{https://doi.org/10.1021/acs.jctc.7b00577}{13(11):
  5255}\href{https://doi.org/10.1021/acs.jctc.7b00577}{, 2017}.

\bibitem{hjmfrgrf2020qx}
Lauri Himanen, Marc~O.J. J{\"{a}}ger, Eiaki~V. Morooka, Filippo~Federici
  Canova, Yashasvi~S. Ranawat, David~Z. Gao, Patrick Rinke, Adam~S. Foster:
  \href{https://doi.org/10.1016/j.cpc.2019.106949}{\textit{{DScribe}: library
  of descriptors for machine learning in materials science}}. Computer Physics
  Communications \href{https://doi.org/10.1016/j.cpc.2019.106949}{247:
  106949}\href{https://doi.org/10.1016/j.cpc.2019.106949}{, 2020}.

\bibitem{zcldcbcstwo2020x}
Yunxing Zuo, Chi Chen, Xiangguo Li, Zhi Deng, Yiming Chen, J{\"o}rg Behler,
  G{\'a}bor Cs{\'a}nyi, Alexander~V. Shapeev, Aidan~P. Thompson, Mitchell~A.
  Wood, Shyue~Ping Ong:
  \href{https://doi.org/10.1021/acs.jpca.9b08723}{\textit{Performance and cost
  assessment of machine learning interatomic potentials}}. Journal of Physical
  Chemistry~A \href{https://doi.org/10.1021/acs.jpca.9b08723}{124(4):
  731}\href{https://doi.org/10.1021/acs.jpca.9b08723}{, 2020}.

\bibitem{sgc2019x}
Gunnar Schmitz, Ian~Heide Godtliebsen, Ove Christiansen:
  \href{https://doi.org/10.1063/1.5100141}{\textit{Machine learning for
  potential energy surfaces: an extensive database and assessment of methods}}.
  Journal of Chemical Physics \href{https://doi.org/10.1063/1.5100141}{150(24):
  244113}\href{https://doi.org/10.1063/1.5100141}{, 2019}.

\bibitem{nrbsmrcwh2019x}
Chandramouli Nyshadham, Matthias Rupp, Brayden Bekker, Alexander~V. Shapeev,
  Tim Mueller, Conrad~W. Rosenbrock, G{\'a}bor Cs{\'a}nyi, David~W. Wingate,
  Gus~L.W. Hart:
  \href{https://doi.org/10.1038/s41524-019-0189-9}{\textit{Machine-learned
  multi-system surrogate models for materials prediction}}. npj Computational
  Materials \href{https://doi.org/10.1038/s41524-019-0189-9}{5:
  51}\href{https://doi.org/10.1038/s41524-019-0189-9}{, 2019}.

\bibitem{strkghr2019qx}
Annika Stuke, Milica Todorovi{\'c}, Matthias Rupp, Christian Kunkel, Kunal
  Ghosh, Lauri Himanen, Patrick Rinke:
  \href{https://doi.org/10.1063/1.5086105}{\textit{Chemical diversity in
  molecular orbital energy predictions with kernel ridge regression}}. Journal
  of Chemical Physics \href{https://doi.org/10.1063/1.5086105}{150(20):
  204121}\href{https://doi.org/10.1063/1.5086105}{, 2019}.

\bibitem{ook2020x}
Berk Onat, Christoph Ortner, James~R. Kermode:
  \href{https://doi.org/10.1063/5.0016005}{\textit{Sensitivity and
  dimensionality of atomic environment representations used for machine
  learning interatomic potentials}}. Journal of Chemical Physics
  \href{https://doi.org/10.1063/5.0016005}{153(14):
  144106}\href{https://doi.org/10.1063/5.0016005}{, 2020}.

\bibitem{pdcfkdblg2020x}
Behnam Parsaeifard, Deb~Sankar De, Anders~S. Christensen, Felix~A. Faber, Emir
  Kocer, Sandip De, J{\"o}rg Behler, Anatole von Lilienfeld, Stefan Goedecker:
  \href{https://doi.org/10.1088/2632-2153/abb212}{\textit{An assessment of the
  structural resolution of various fingerprints commonly used in machine
  learning}}. Machine Learning: Science and Technology
  \href{https://doi.org/10.1088/2632-2153/abb212}{in
  press}\href{https://doi.org/10.1088/2632-2153/abb212}{, 2020}.

\bibitem{kkclm2020qx}
Silvan K{\"a}ser, Debasish Koner, Anders~S. Christensen, O.~Anatole von
  Lilienfeld, Markus Meuwly:
  \href{https://doi.org/10.1021/acs.jpca.0c05979}{\textit{{ML} models of
  vibrating {H$_2$CO}: Comparing reproducing kernels, {FCHL} and {PhysNet}}}.
  Journal of Physical Chemistry~A
  \href{https://doi.org/10.1021/acs.jpca.0c05979}{124(42):
  8853}\href{https://doi.org/10.1021/acs.jpca.0c05979}{, 2020}.

\bibitem{jmfhf2018x}
Marc O.~J. J{\"a}ger, Eiaki~V. Morooka, Filippo Federici-Canova, Lauri Himanen,
  Adam~S. Foster:
  \href{https://doi.org/10.1038/s41524-018-0096-5}{\textit{Machine learning
  hydrogen adsorption on nanoclusters through structural descriptors}}. npj
  Computational Materials \href{https://doi.org/10.1038/s41524-018-0096-5}{4:
  37}\href{https://doi.org/10.1038/s41524-018-0096-5}{, 2018}.

\bibitem{gfc2020x}
Alexander Goscinski, Guillaume Fraux, Michele Ceriotti: \textit{The role of
  feature space in atomistic learning}. arXiv 2009.02741, 2020.

\bibitem{t1917x}
Richard~C. Tolman: \textit{The measurable quantities of physics}. Physical
  Review 9(3): 237, 1917.

\bibitem{hk1965x}
George~N. Hatsopoulos, Joseph~H. Keenan: \textit{Principles of General
  Thermodynamics}. Wiley, New York, 1965.

\bibitem{jskohrm2020qx}
Hyunwook Jung, Sina Stocker, Christian Kunkel, Harald Oberhofer, Byungchan Han,
  Karsten Reuter, Johannes~T. Margraf:
  \href{https://doi.org/10.1002/syst.201900052}{\textit{Size-extensive
  molecular machine learning with global representations}}. ChemSystemsChem
  \href{https://doi.org/10.1002/syst.201900052}{2(4):
  e1900052}\href{https://doi.org/10.1002/syst.201900052}{, 2020}.

\bibitem{bc2015x}
Albert~P. Bart{\'o}k, G{\'a}bor Cs{\'a}nyi:
  \href{https://doi.org/10.1002/qua.24927}{\textit{{G}aussian approximation
  potentials: a brief tutorial introduction}}. International Journal of Quantum
  Chemistry \href{https://doi.org/10.1002/qua.24927}{116(13):
  1051}\href{https://doi.org/10.1002/qua.24927}{, 2015}.

\bibitem{m2015x}
Sonja Mathias: \textit{A kernel-based learning method for an efficient
  approximation of the high-dimensional {B}orn-{O}ppenheimer potential energy
  hypersurface}. Master's thesis, Institute for Numerical Simulation,
  Mathematisch-Naturwissenschaftliche Fakult{\"a}t der Rheinischen
  Friedrich-Wilhelms-Universit{\"a}t Bonn, Germany, 2015.

\bibitem{d2019bqx}
Ralf Drautz: \href{https://doi.org/10.1103/physrevb.99.014104}{\textit{Atomic
  cluster expansion for accurate and transferable interatomic potentials}}.
  Physical Review~B \href{https://doi.org/10.1103/physrevb.99.014104}{99(1):
  249901}\href{https://doi.org/10.1103/physrevb.99.014104}{, 2019}.

\bibitem{c2019dqx}
Miguel~A. Caro:
  \href{https://doi.org/10.1103/physrevb.100.024112}{\textit{Optimizing
  many-body atomic descriptors for enhanced computational performance of
  machine learning based interatomic potentials}}. Physical Review~B
  \href{https://doi.org/10.1103/physrevb.100.024112}{100(2):
  024112}\href{https://doi.org/10.1103/physrevb.100.024112}{, 2019}.

\bibitem{npc2020qx}
Jigyasa Nigam, Sergey Pozdnyakov, Michele Ceriotti:
  \href{https://doi.org/10.1063/5.0021116}{\textit{Recursive evaluation and
  iterative contraction of $n$-body equivariant features}}. Journal of Chemical
  Physics \href{https://doi.org/10.1063/5.0021116}{153(12):
  121101}\href{https://doi.org/10.1063/5.0021116}{, 2020}.

\bibitem{dbcdeoo2020x}
Markus Bachmayr, G{\'a}bor Cs{\'a}nyi, Ralf Drautz, Genevieve Dusson, Simon
  Etter, Cas van~der Oord, Christoph Ortner:
  \textit{\href{https://arxiv.org/abs/1911.03550}{Atomic cluster expansion:
  completeness, efficiency and stability}}. arXiv 1911.03550, 2020.

\bibitem{fchl2018qx}
Felix~A. Faber, Anders~S. Christensen, Bing Huang, O.~Anatole von Lilienfeld:
  \href{https://doi.org/10.1063/1.5020710}{\textit{Alchemical and structural
  distribution based representation for universal quantum machine learning}}.
  Journal of Chemical Physics \href{https://doi.org/10.1063/1.5020710}{148(24):
  241717}\href{https://doi.org/10.1063/1.5020710}{, 2018}.

\bibitem{gzv2018x}
Aldo Glielmo, Claudio Zeni, Alessandro~De Vita:
  \href{https://doi.org/10.1103/PhysRevB.97.184307}{\textit{Efficient
  non-parametric $n$-body force fields from machine learning}}. Physical
  Review~B \href{https://doi.org/10.1103/PhysRevB.97.184307}{97(18):
  184307}\href{https://doi.org/10.1103/PhysRevB.97.184307}{, 2018}.

\bibitem{gzfd2020x}
Aldo Glielmo, Claudio Zeni, {\'{A}}d{\'{a}}m Fekete, Alessandro De~Vita:
  \href{https://doi.org/10.1007/978-3-030-40245-7_5}{\textit{Building
  nonparametric $n$-body force fields using {G}aussian process regression}}. In
  Kristof~T. Sch{\"{u}}tt, Stefan Chmiela, O.~Anatole von Lilienfeld, Alexandre
  Tkatchenko, Koji Tsuda, Klaus-Robert M{\"{u}}ller (editors), \textit{Machine
  Learning Meets Quantum Physics}, \textit{Lecture Notes in Physics}, volume
  968, 67--98, Springer, Heidelberg, Germany, 2020.

\bibitem{jkvak2020x}
Ryosuke Jinnouchi, Ferenc Karsai, Carla Verdi, Ryoji Asahi, Georg Kresse:
  \href{https://doi.org/10.1063/5.0009491}{\textit{Descriptors representing
  two- and three-body atomic distributions and their effects on the accuracy of
  machine-learned inter-atomic potentials}}. Journal of Chemical Physics
  \href{https://doi.org/10.1063/5.0009491}{152(23)}\href{https://doi.org/10.1063/5.0009491}{,
  2020}.

\bibitem{dbcc2016qx}
Sandip De, Albert~P. Bart{\'o}k, G{\'a}bor Cs{\'a}nyi, Michele Ceriotti:
  \href{https://doi.org/10.1039/C6CP00415F}{\textit{Comparing molecules and
  solids across structural and alchemical space}}. Physical Chemistry Chemical
  Physics \href{https://doi.org/10.1039/C6CP00415F}{18(20):
  13754}\href{https://doi.org/10.1039/C6CP00415F}{, 2016}.

\bibitem{bkc2013qx}
Albert~P. Bart{\'o}k, Risi Kondor, G{\'a}bor Cs{\'a}nyi:
  \href{https://doi.org/10.1103/PhysRevB.87.184115}{\textit{On representing
  chemical environments}}. Physical Review~B
  \href{https://doi.org/10.1103/PhysRevB.87.184115}{87(18):
  184115}\href{https://doi.org/10.1103/PhysRevB.87.184115}{, 2013}.

\bibitem{rdrl2014x}
Raghunathan Ramakrishnan, Pavlo~O. Dral, Matthias Rupp, O.~Anatole von
  Lilienfeld: \href{https://doi.org/10.1038/sdata.2014.22}{\textit{Quantum
  chemistry structures and properties of 134 kilo molecules}}. Scientific Data
  \href{https://doi.org/10.1038/sdata.2014.22}{1:
  140022}\href{https://doi.org/10.1038/sdata.2014.22}{, 2014}.

\bibitem{rdrl2015x}
Raghunathan Ramakrishnan, Pavlo~O. Dral, Matthias Rupp, O.~Anatole von
  Lilienfeld: \href{https://doi.org/10.1021/acs.jctc.5b00099}{\textit{Big data
  meets quantum chemistry approximations: the {$\Delta$}-machine learning
  approach}}. Journal of Chemical Theory and Computation
  \href{https://doi.org/10.1021/acs.jctc.5b00099}{11(5):
  2087}\href{https://doi.org/10.1021/acs.jctc.5b00099}{, 2015}.

\bibitem{sdcf1994x}
Philip~J. Stephens, Frank~J. Devlin, Cary~F. Chabalowski, Michael~J. Frisch:
  \href{https://doi.org/10.1021/j100096a001}{\textit{\textit{Ab initio}
  calculation of vibrational absorption and circular dichroism spectra using
  density functional force fields}}. Journal of Physical Chemistry
  \href{https://doi.org/10.1021/j100096a001}{98(45):
  11623}\href{https://doi.org/10.1021/j100096a001}{, 1994}.

\bibitem{v1921x}
Lars Vegard: \href{https://doi.org/10.1007/BF01349680}{\textit{Die
  {K}onstitution der {M}ischkristalle und die {R}aumf{\"u}llung der {A}tome}}.
  Zeitschrift f{\"u}r Physik \href{https://doi.org/10.1007/BF01349680}{5(1):
  17}\href{https://doi.org/10.1007/BF01349680}{, 1921}.

\bibitem{da1991x}
Alan~R. Denton, Neil~W. Ashcroft:
  \href{https://doi.org/10.1103/PhysRevA.43.3161}{\textit{{V}egard's law}}.
  Physical Review~A \href{https://doi.org/10.1103/PhysRevA.43.3161}{43(6):
  3161}\href{https://doi.org/10.1103/PhysRevA.43.3161}{, 1991}.

\bibitem{hf2008x}
Gus L.~W. Hart, Rodney~W. Forcade:
  \href{https://doi.org/10.1103/PhysRevB.77.224115}{\textit{Algorithm for
  generating derivative structures}}. Physical Review~B
  \href{https://doi.org/10.1103/PhysRevB.77.224115}{77(22):
  224115}\href{https://doi.org/10.1103/PhysRevB.77.224115}{, 2008}.

\bibitem{wmm2016x}
Pandu Wisesa, Kyle~A. McGill, Tim Mueller: \textit{Efficient generation of
  generalized {M}onkhorst-{P}ack grids through the use of informatics}.
  Physical Review~B 93(15): 155109, 2016.

\bibitem{mjhh2018x}
Wiley~S. Morgan, Jeremy~J. Jorgensen, Bret~C. Hess, Gus~L.W. Hart:
  \href{https://doi.org/10.1016/j.commatsci.2018.06.031}{\textit{Efficiency of
  generalized regular $k$-point grids}}. Computational Materials Science
  \href{https://doi.org/10.1016/j.commatsci.2018.06.031}{153:
  424}\href{https://doi.org/10.1016/j.commatsci.2018.06.031}{, 2018}.

\bibitem{sgylbhglzs2019x}
Christopher Sutton, Luca~M. Ghiringhelli, Takenori Yamamoto, Yury Lysogorskiy,
  Lars Blumenthal, Thomas Hammerschmidt, Jacek~R. Golebiowski, Xiangyue Liu,
  Angelo Ziletti, Matthias Scheffler:
  \href{https://doi.org/10.1038/s41524-019-0239-3}{\textit{Crowd-sourcing
  materials-science challenges with the {NOMAD} 2018 {K}aggle competition}}.
  npj Computational Materials
  \href{https://doi.org/10.1038/s41524-019-0239-3}{5:
  111}\href{https://doi.org/10.1038/s41524-019-0239-3}{, 2019}.

\bibitem{kagglex}
Nomad2018 Predicting Transparent Conductors. Predict the key properties of
  novel transparent semiconductors. Available at
  \url{https://www.kaggle.com/c/nomad2018-predict-transparent-conductors}.

\bibitem{bghhhrrs2009x}
Volker Blum, Ralf Gehrke, Felix Hanke, Paula Havu, Ville Havu, Xinguo Ren,
  Karsten Reuter, Matthias Scheffler:
  \href{https://doi.org/10.1016/j.cpc.2009.06.022}{\textit{\textit{Ab initio}
  molecular simulations with numeric atom-centered orbitals}}. Computer Physics
  Communications \href{https://doi.org/10.1016/j.cpc.2009.06.022}{180(11):
  2175}\href{https://doi.org/10.1016/j.cpc.2009.06.022}{, 2009}.

\bibitem{afs1992x}
Shun{-}ichi Amari, Naotake Fujita, Shigeru Shinomoto:
  \href{https://doi.org/10.1162/neco.1992.4.4.605}{\textit{Four types of
  learning curves}}. Neural Computation
  \href{https://doi.org/10.1162/neco.1992.4.4.605}{4(4):
  605}\href{https://doi.org/10.1162/neco.1992.4.4.605}{, 1992}.

\bibitem{hl2016x}
Bing Huang, O.~Anatole von Lilienfeld:
  \href{https://doi.org/10.1063/1.4964627}{\textit{Communication: Understanding
  molecular representations in machine learning: the role of uniqueness and
  target similarity}}. Journal of Chemical Physics
  \href{https://doi.org/10.1063/1.4964627}{145(16):
  161102}\href{https://doi.org/10.1063/1.4964627}{, 2016}.

\bibitem{htf2009x}
Trevor Hastie, Robert Tibshirani, Jerome Friedman:
  \href{https://doi.org/10.1007/978-0-387-84858-7}{\textit{The Elements of
  Statistical Learning. Data Mining, Inference, and Prediction}}. Springer, New
  York, 2nd edition, 2009.

\bibitem{rw2006x}
Carl Rasmussen, Christopher Williams: \textit{{G}aussian Processes for Machine
  Learning}. MIT Press, Cambridge, 2006.

\bibitem{r2015fx}
Matthias Rupp: \href{https://doi.org/10.1002/qua.24954}{\textit{Machine
  learning for quantum mechanics in a nutshell}}. International Journal of
  Quantum Chemistry \href{https://doi.org/10.1002/qua.24954}{115(16):
  1058}\href{https://doi.org/10.1002/qua.24954}{, 2015}.

\bibitem{qmmlpackx}
The \texttt{qmmlpack} (quantum mechanics machine learning package) library is
  publicly available at \url{https://gitlab.com/qmml/qmmlpack} under the
  Apache-2.0 license.

\bibitem{bbbk2011x}
James~S. Bergstra, R{\'e}mi Bardenet, Yoshua Bengio, Bal{\'a}zs K{\'e}gl:
  \textit{Algorithms for hyper-parameter optimization}. In John Shawe-Taylor,
  Richard~S. Zemel, Peter~L. Bartlett, Fernando~C.N. Pereira, Kilian~Q.
  Weinberger (editors), \textit{Advances in Neural Information Processing
  Systems~24 (NIPS~2011), Granada, Spain, December~12--15}, 2546--2554, 2011.

\bibitem{byc2013x}
James~S. Bergstra, Daniel Yamins, David~D. Cox: \textit{Making a science of
  model search: hyperparameter optimization in hundreds of dimensions for
  vision architectures}. In Sanjoy Dasgupta, David McAllester (editors),
  \textit{Proceedings of the 30th International Conference on Machine Learning
  (ICML~2013), Atlanta, Georgia, USA, June~16--21}, 115--123, Proceedings of
  Machine Learning Research 28, 2013.

\bibitem{cmlkitx}
The \texttt{cmlkit} Python package is publicly available at
  \url{https://marcel.science/cmlkit} under an MIT license and as part of the
  Nomad Analytics Toolkit (\url{https://analytics-toolkit.nomad-coe.eu/}).

\bibitem{hr2017x}
Haoyan Huo, Matthias Rupp: \textit{Unified representation for machine learning
  of molecules and materials}. arXiv 1704.06439, 2017.

\bibitem{pmrrv2001x}
Tomaso Poggio, Sayan Mukherjee, Ryan Rifkin, Alexander Rakhlin, Alessandro
  Verri: \textit{b}. Technical Report {AI} Memo 2001-011, {CBCL} Memo 198,
  Massachusetts Institute of Technology, 2001.

\bibitem{reproducibilityx}
HP search spaces and HP values for optimized models are available at
  \url{https://marcel.science/repbench} and \url{https://qmml.org}.

\bibitem{b2011ex}
J{\"o}rg Behler: \href{https://doi.org/10.1063/1.3553717}{\textit{Atom-centered
  symmetry functions for constructing high-dimensional neural network
  potentials}}. Journal of Chemical Physics
  \href{https://doi.org/10.1063/1.3553717}{134(7):
  074106}\href{https://doi.org/10.1063/1.3553717}{, 2011}.

\bibitem{arunnerx}
The \texttt{RuNNer} software
  (\url{https://www.uni-goettingen.de/de/560580.html}, GPL license) is
  available from its author J{\"o}rg Behler
  (\url{joerg.behler@uni-goettingen.de}) on request.

\bibitem{b2015qx}
J{\"o}rg Behler: \href{https://doi.org/10.1002/qua.24890}{\textit{Constructing
  high-dimensional neural network potentials: a tutorial review}}.
  International Journal of Quantum Chemistry
  \href{https://doi.org/10.1002/qua.24890}{115(16):
  1032}\href{https://doi.org/10.1002/qua.24890}{, 2015}.

\bibitem{b2017qx}
J{\"o}rg Behler: \href{https://doi.org/10.1002/anie.201703114}{\textit{First
  principles neural network potentials for reactive simulations of large
  molecular and condensed systems}}. Angewandte Chemie International Edition
  \href{https://doi.org/10.1002/anie.201703114}{56(42):
  12828}\href{https://doi.org/10.1002/anie.201703114}{, 2017}.

\bibitem{gsbbm2018x}
Michael Gastegger, Ludwig Schwiedrzik, Marius Bittermann, Florian Berzsenyi,
  Philipp Marquetand:
  \href{https://doi.org/10.1063/1.5019667}{\textit{{WACSF}---weighted
  atom-centered symmetry functions as descriptors in machine learning
  potentials}}. Journal of Chemical Physics
  \href{https://doi.org/10.1063/1.5019667}{148(24):
  241709}\href{https://doi.org/10.1063/1.5019667}{, 2018}.

\bibitem{adscribex}
The \texttt{DScribe} software library is publicly available at
  \url{https://github.com/SINGROUP/dscribe} under the Apache-2.0 license.

\bibitem{aquippyx}
The \texttt{quippy} software is publicly available at
  \url{http://libatoms.github.io/QUIP/} under the GNU General Public license~2.

\bibitem{ctg2017x}
Chao Chen, Jamie Twycross, Jonathan~M. Garibaldi:
  \href{https://doi.org/10.1371/journal.pone.0174202}{\textit{A new accuracy
  measure based on bounded relative error for time series forecasting}}. PLoS
  ONE \href{https://doi.org/10.1371/journal.pone.0174202}{12(3):
  e0174202}\href{https://doi.org/10.1371/journal.pone.0174202}{, 2017}.

\bibitem{b2019x}
Alexei Botchkarev: \href{https://doi.org/10.28945/4184}{\textit{A new typology
  design of performance metrics to measure errors in machine learning
  regression algorithms}}. Interdisciplinary Journal of Information, Knowledge,
  and Management \href{https://doi.org/10.28945/4184}{14:
  45}\href{https://doi.org/10.28945/4184}{, 2019}.

\bibitem{tj2008x}
Julian Tirado-Rives, William~L. Jorgensen:
  \href{https://doi.org/10.1021/ct700248k}{\textit{Performance of {B3LYP}
  density functional methods for a large set of organic molecules}}. Journal of
  Chemical Theory and Computation
  \href{https://doi.org/10.1021/ct700248k}{4(2):
  297}\href{https://doi.org/10.1021/ct700248k}{, 2008}.

\bibitem{l2008x}
Stephan Lany:
  \href{https://doi.org/10.1103/physrevb.78.245207}{\textit{Semiconductor
  thermochemistry in density functional calculations}}. Physical Review~B
  \href{https://doi.org/10.1103/physrevb.78.245207}{78(24):
  245207}\href{https://doi.org/10.1103/physrevb.78.245207}{, 2008}.

\bibitem{lvvc2014x}
Kurt Lejaeghere, Veronique~Van Speybroeck, Guido~Van Oost, Stefaan Cottenier:
  \href{https://doi.org/10.1080/10408436.2013.772503}{\textit{Error estimates
  for solid-state density-functional theory predictions: an overview by means
  of the ground-state elemental crystals}}. Critical Reviews in Solid State and
  Materials Sciences \href{https://doi.org/10.1080/10408436.2013.772503}{39(1):
  1}\href{https://doi.org/10.1080/10408436.2013.772503}{, 2014}.

\bibitem{zrts2018x}
Guo-Xu Zhang, Anthony~M. Reilly, Alexandre Tkatchenko, Matthias Scheffler:
  \href{https://doi.org/10.1088/1367-2630/aac7f0}{\textit{Performance of
  various density-functional approximations for cohesive properties of 64 bulk
  solids}}. New Journal of Physics
  \href{https://doi.org/10.1088/1367-2630/aac7f0}{20(6):
  063020}\href{https://doi.org/10.1088/1367-2630/aac7f0}{, 2018}.

\bibitem{lbbbbbcccdetal2016x}
Kurt Lejaeghere, Gustav Bihlmayer, Torbj{\"o}rn Bj{\"o}rkman, Peter Blaha,
  Stefan Bl{\"u}gel, Volker Blum, Damien Caliste, Ivano~E. Castelli, Stewart~J.
  Clark, Andrea Dal~Corso, Stefano de~Gironcoli, Thierry Deutsch, John~Kay
  Dewhurst, Igor Di~Marco, Claudia Draxl, Marcin Du{\l}ak, Olle Eriksson,
  Jos{\'e}~A. Flores-Livas, Kevin~F. Garrity, Luigi Genovese, Paolo Giannozzi,
  Matteo Giantomassi, Stefan Goedecker, Xavier Gonze, Oscar Gr{\r a}n{\"a}s,
  Eberhard K.~U. Gross, Andris Gulans, Fran{\c c}ois Gygi, Donald~R. Hamann,
  Phil~J. Hasnip, Natalie A.~W. Holzwarth, Diana Iu{\c s}an, Dominik~B. Jochym,
  Fran{\c c}ois Jollet, Daniel Jones, Georg Kresse, Klaus Koepernik, Emine
  K{\"u}{\c c}{\"u}kbenli, Yaroslav~O. Kvashnin, Inka L.~M. Locht, Sven Lubeck,
  Martijn Marsman, Nicola Marzari, Ulrike Nitzsche, Lars Nordstr{\"o}m, Taisuke
  Ozaki, Lorenzo Paulatto, Chris~J. Pickard, Ward Poelmans, Matt I.~J. Probert,
  Keith Refson, Manuel Richter, Gian-Marco Rignanese, Santanu Saha, Matthias
  Scheffler, Martin Schlipf, Karlheinz Schwarz, Sangeeta Sharma, Francesca
  Tavazza, Patrik Thunstr{\"o}m, Alexandre Tkatchenko, Marc Torrent, David
  Vanderbilt, Michiel~J. van Setten, Veronique Van~Speybroeck, John~M. Wills,
  Jonathan~R. Yates, Guo-Xu Zhang, Stefaan Cottenier:
  \href{https://doi.org/10.1126/science.aad3000}{\textit{Reproducibility in
  density functional theory calculations of solids}}. Science
  \href{https://doi.org/10.1126/science.aad3000}{351(6280):
  aad3000}\href{https://doi.org/10.1126/science.aad3000}{, 2016}.

\bibitem{ptm2018qx}
Wiktor Pronobis, Alexandre Tkatchenko, Klaus-Robert M{\"u}ller:
  \href{https://doi.org/10.1021/acs.jctc.8b00110}{\textit{Many-body descriptors
  for predicting molecular properties with machine learning: analysis of
  pairwise and three-body interactions in molecules}}. Journal of Chemical
  Theory and Computation \href{https://doi.org/10.1021/acs.jctc.8b00110}{14(6):
  2991}\href{https://doi.org/10.1021/acs.jctc.8b00110}{, 2018}.

\bibitem{cbfgl2020x}
Anders~S. Christensen, Lars~A. Bratholm, Felix~A. Faber, O.~Anatole von
  Lilienfeld: \href{https://doi.org/10.1063/1.5126701}{\textit{{FCHL}
  revisited: faster and more accurate quantum machine learning}}. Journal of
  Chemical Physics \href{https://doi.org/10.1063/1.5126701}{152(4):
  044107}\href{https://doi.org/10.1063/1.5126701}{, 2020}.

\bibitem{ssktm2018x}
Kristof~T. Sch{\"u}tt, Huziel~E. Sauceda, Pieter-Jan Kindermans, Alexandre
  Tkatchenko, Klaus-Robert M{\"u}ller:
  \href{https://doi.org/10.1063/1.5019779}{\textit{{SchNet}---a deep learning
  architecture for molecules and materials}}. Journal of Chemical Physics
  \href{https://doi.org/10.1063/1.5019779}{148(24):
  241722}\href{https://doi.org/10.1063/1.5019779}{, 2018}.

\bibitem{wmc2018x}
Michael~J. Willatt, F{\'e}lix Musil, Michele Ceriotti:
  \href{https://doi.org/10.1039/C8CP05921G}{\textit{Feature optimization for
  atomistic machine learning yields a data-driven construction of the periodic
  table of the elements}}. Physical Chemistry Chemical Physics
  \href{https://doi.org/10.1039/C8CP05921G}{20(47):
  29661}\href{https://doi.org/10.1039/C8CP05921G}{, 2018}.

\bibitem{sgtm2019x}
Kristof~T. Sch{\"u}tt, Michael Gastegger, Alexandre Tkatchenko, Klaus-Robert
  M{\"u}ller:
  \href{https://doi.org/10.1007/978-3-030-28954-6_17}{\textit{Quantum-chemical
  insights from interpretable atomistic neural networks}}. In Wojciech Samek,
  Gr{\'{e}}goire Montavon, Andrea Vedaldi, Lars~Kai Hansen, Klaus-Robert
  M{\"u}ller (editors), \textit{Explainable {AI}: interpreting, explaining and
  visualizing deep learning}, 311--330, Springer, 2019.

\bibitem{fhhgsdvkrv2017qx}
Felix~A. Faber, Luke Hutchison, Bing Huang, Justin Gilmer, Samuel~S.
  Schoenholz, George~E. Dahl, Oriol Vinyals, Steven Kearnes, Patrick~F. Riley,
  O.~Anatole von Lilienfeld:
  \href{https://doi.org/10.1021/acs.jctc.7b00577}{\textit{Prediction errors of
  molecular machine learning models lower than hybrid {DFT} error}}. Journal of
  Chemical Theory and Computation
  \href{https://doi.org/10.1021/acs.jctc.7b00577}{13(11):
  5255}\href{https://doi.org/10.1021/acs.jctc.7b00577}{, 2017}.

\end{thebibliography}
\hypersetup{allcolors=linkcolor} 
\endgroup

\clearpage

\stopcontents[main]


\clearpage

\title{Supplementary material for\\[0.5ex]\titletext}
\maketitle

\setcounter{secnumdepth}{2}
\counterwithout{subsection}{section}

\setcounter{section}{0}
\setcounter{subsection}{0}
\setcounter{page}{1}
\setcounter{figure}{0}
\setcounter{table}{0}

\renewcommand{\thepage}{S\arabic{page}}
\renewcommand{\thefigure}{S\arabic{figure}}
\renewcommand{\thetable}{S\arabic{table}}

\makeatletter
\renewcommand\subsection{\@startsection{subsection}{2}{\z@}%
	{3.25ex \@plus1ex \@minus7pt}%
	{-1em}%
	{\normalfont\normalsize\bfseries}}
\makeatother

\makeatletter
\def\@seccntformat#1{\@ifundefined{#1@cntformat}%
	{\csname the#1\endcsname\quad}
	{\csname #1@cntformat\endcsname}}
\newcommand\subsection@cntformat{\arabic{subsection}$\,|\,$}	
\makeatother

\renewcommand{\eref}[1]{Reference~\citenumsi{#1}} 
\renewcommand{\erefs}[1]{References~\citenumsi{#1}} 

\renewcommand{\topfraction}{1}%
\renewcommand{\bottomfraction}{1}%
\renewcommand{\textfraction}{0}
\renewcommand{\dbltopfraction}{1}
\renewcommand{\floatpagefraction}{1}


\startcontents[si]
\section*{Contents}
\titlecontents{section}[0em]
	{\vspace{0em}\relax}
	{\contentslabel[\relax]{0em}}{}{\hfill\contentspage}
\dottedcontents{subsection}[2em]{}{1.5em}{1pc}
\printcontents[si]{}{1}[2]{}


\section{Introduction}

\subsection{Cost of electronic structure calculations}\label{si:abinitiocosts}
Although the computational cost of ab initio methods scales only polynomially in system size~$N$ (measured, for example, in number of electrons or orbitals), it remains a strongly limiting factor. For example, the currently most-widely used approach, Kohn-Sham density functional theory, scales as $O(N^3)$ for (semi)local and $O(N^4)$ for hybrid functionals: 
Doubling~$N$ thus increases compute time by roughly an order of magnitude, and a few such doublings will exhaust any computational resource. 
Advances in large-scale computing facilities, such as current exascale computing inititatives, will move this ``computational wall'' to larger systems, but cannot remove it. 
In practice, the large prefactor hidden in the asymptotic runtimes is relevant as well.

\subsection{Size of molecular and materials spaces}\label{si:sizeccs}
Various estimates of the size of chemical compound spaces exist, popular ones \citesi{pmv2013x,bmg1996x} including $10^{33}$ and $10^{60}$ molecules. 
Reymond et~al.{} \citesi{fr2007x,br2009x,rdbr2012x} systematically enumerate all small organic molecules with up to 11 C, N, O, F atoms, 13 C, N, O, S, Cl atoms, and 17 C, N, O, S, F, Cl, Br, I atoms, yielding 26 million, 970 million and 166 billion molecules, respectively. 
Following Cantor, \citesi{c2014cx} we estimate the number of possible compositions (not considering unit cell size or symmetry) for an alloy system to be 
the multinomial coefficient $(n-1,k)! = \binom{n-1+k}{k}$, where $n$ is number of components and $k = 100/x$ is determined by the tolerance $x\,\%$ to which the amount of a species is specified.
For $n = 5$ and a very conservative choice of $x=5\,\%$, removing combinations that contain only 4 or fewer components and multiplying by all ways to choose 5 out of 30 elements yields 
$( \binom{5-1+20}{20} - \binom{4-1+20}{20} ) \binom{30}{5} \approx 1.5\cdot10^{9}$. 

\subsection{The role of data quality for QM/ML models}\label{si:dataquality}
Data are the basis for data-driven models, and errors in them can only be corrected to a limited extent (``garbage in, garbage out'').
Even dealing with simple errors like independent identically distributed noise requires additional data, and more severe errors lead to qualitative problems such as outliers.
Conversely, problems in fitting a ML model can be indicative of problems in the data.

\subsection{Explicit and implicit features}\label{si:scope}
Features used for regression can be defined explicitly via representations, or implicitly, for example, via kernels or deep neural networks.

In this work, we focus on explicit Hilbert-space representations in conjunction with kernel-based regression with a Gaussian kernel.
Technically, the features used for regression are the components of the kernel feature space, that is, the non-linear transformations of the representations' components via the Gaussian kernel.
While used implicitly in this sense, the representations are still defined explicitly. 

This is in contrast to implicitly defined representations, for example, feature spaces of kernels defined directly on ``raw inputs`` such as atomic coordinates and numbers, without an intermediate explicit Hilbert-space representation,
or, the layers of deep neural networks (end-to-end learning).
For the latter, the requirements in \cref{sec:requirements} can be imposed via the network architecture,
which can be seen as the conceptual analog to explicitly conformant representations or kernels.

\subsection{Related work}\label{si:relatedwork}

Faber et~al.{} \citesi{fhhgsdvkrv2017x} compare combinations of representations and regression methods for atomization energies of organic molecules (\dsgdb{} dataset, see \cref{sec:benchmark}).
Only some of the tested features are representations that satisfy the requirements in \cref{sec:requirements}; their HPs are not optimized.

Himanen et~al.{} \citesi{hjmfrgrf2020qx}
investigate the representations in \cref{sec:representations}, also using kernel regression, 
to predict ionic charges of molecules from the \dsgdb{} dataset, as well as formation energies in a custom dataset of inorganic crystals obtained from the Open Quantum Materials Database. 
They optimize numerical HPs of representations and regression method, but not structural ones.

Zuo et~al.{} \citesi{zcldcbcstwo2020x} focus on dynamics simulations, and therefore include forces and stresses in training and evaluation.
They also evaluate predictions of derived physical quantities, such as elastic constants or equation-of-state curves.
Different combinations of representation, regression method, and HP tuning 
are evaluated on a dataset of elemental solids. 
Timings are discussed.

Schmitz et~al.{} \citesi{sgc2019x} compare regression methods for potential energy surfaces of 15 small organic molecules, using non-redundant internal coordinates as features. 
HPs of the representation are not optimized.

Nyshadham et~al.{} \citesi{nrbsmrcwh2019x} compare selected combinations of representations and regression methods on binary alloys (\dsba{} dataset, see \cref{sec:benchmark}).
HPs of the representations are not optimized.

Stuke et~al.{} \citesi{strkghr2019qx} evaluate prediction of molecular orbital energies with kernel regression on three datasets: 
organic molecules (\dsgdb{} dataset, see \cref{sec:benchmark}), amino acids and dipeptides, as well as opto-electronically active molecules. 
Numerical HPs of representations and regression method are optimised via local grid search.

Onat et~al.{} \citesi{ook2020x} empirically investigate effective dimensionality and sensitivity (to small perturbations of the underlying system) of representations using both materials and molecular datasets.
HPs are not optimized.
They do not investigate prediction errors.

Parsaeifard et~al.{} \citesi{pdcfkdblg2020x} also study the sensitivity of representations to infinitesimal geometric perturbations (uniqueness), as well as correlations between the distances induced by different representations and with physical properties.

K{\"a}ser et~al.{} \citesi{kkclm2020qx} empirically evaluate representations with kernel-based regression and a deep neural network for prediction of energies, forces, vibrational modes and infrared spectra of formaldehyde at different levels of theory.

J{\"a}ger et~al.{} \citesi{jmfhf2018x} compare SOAP, MBTR, CM and SFs with KRR for prediction of the adsorption free energy of hydrogen on the surface of nanoclusters.
Numerical HPs of KRR, and some numerical HPs of representations are optimized.

Goscinski et~al.{} \citesi{gfc2020x} develop metrics to compare the feature spaces generated by different representations (SFs, SOAP, NICE), exploring the impact of HP choices in datasets of methane and solid carbon.


\section{Role and types of representations}

\subsection{Structure and distribution of data}\label{si:strucdata}
\begin{figure}[b]
	\includegraphics[width=\linewidth]{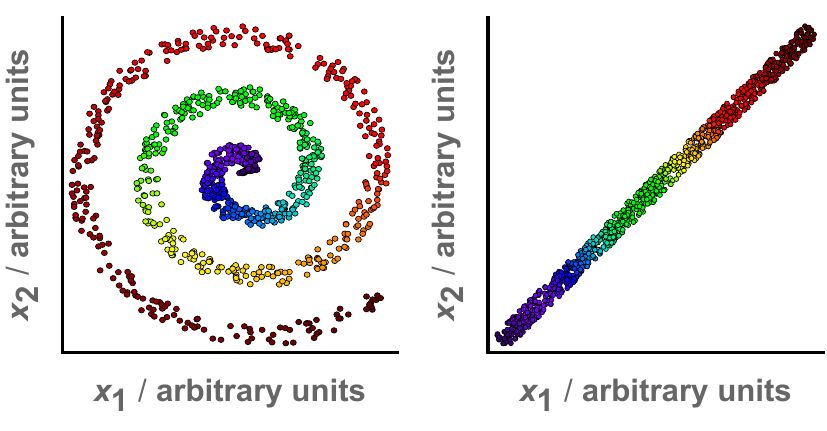}
	\caption{%
		\emph{Structure of representation determines suitability for regression.}
		Almost one-dimensional data is embedded into a two-dimensional space.
		The spiral embedding (left) is not suited for linear regression, but the ``unrolled'' embedding (right) is.
		\label{si:fig:spiral}}
\end{figure}

\Cref{si:fig:spiral} illustrates the importance of representation space structure for regression with a toy example.
Low-dimensional (here, essentially one-dimensional) data is embedded into a high-dimensional (here, two-dimensional) space.
The spiral embedding is not suited for linear regression, whereas the linear embedding is.

\subsection{Extensive and intensive properties}\label{si:extensivity}
A property whose magnitude is additive in the size (extent or mass) of an object is called \emph{extensive}; 
a property whose magnitude is independent of the size of an object is called \emph{intensive}.
For example, internal energy is an extensive property, band gap energy an intensive one.

Originating from thermodynamics, \citesi{t1917x,hk1965x} the application of these terms to microscopic quantities is limited by allowed changes in ``size`` of a system:
For finite systems such as molecules, a property $p$ is extensive if for any two \emph{non-interacting} systems $A$ and $B$, $p(A+B) = p(A) + p(B)$, \citesi{jskohrm2020qx}
and intensive if $p(A) = p(A+A)$.
For periodic systems such as bulk crystals, we take $A$ and $B$ to be supercells of the same unit cell.
In this minimal sense, total and atomization energy of atomistic systems are extensive.

However, energies are not additive for general changes in a system, such as changes in atomic position, and addition or removal of atoms.
With respect to the requirements in \cref{sec:requirements},
\emph{ML models for energies should be size-extensive in the (minimal) sense above.}
For global representations, this can be achieved via normalization in conjunction with the linear kernel, \citesi{jskohrm2020qx}
whereas local representations as described in \cref{si:atomiccontr} automatically satisfy this requirement.

\subsection{Learning with atomic contributions}\label{si:atomiccontr}
One ansatz to scale prediction of global properties to large atomistic systems is to predict atomic contributions. 
This assumes additivity, as the predicted property is a sum of predicted atomic contributions, and locality, as efficient scaling requires representations of atoms in their environment to have local support, often achieved through a finite-radius cut-off function.

Predicting atomic contributions requires a modification of the basic kernel regression scheme, 
which we derive here building on References~\citenumsi{bc2015x} and~\citenumsi{m2015x}:

Let $\mathcal{A}_1,\ldots,\mathcal{A}_{a}$ denote atoms of systems $\mathcal{M}_1,$ $\ldots,$ $\mathcal{M}_{m}$ and let $\mat{D} \in \{0,1\}^{m \times a}$ be their incidence matrix, that is $D_{i,j} = 1$ if $\mathcal{A}_j$ belongs to $\mathcal{M}_i$ and 0 otherwise.
Let $\tilde{k}$ denote a kernel function on atoms.
The prediction for the \mbox{$i$-th} system is the sum of its predicted atomic contributions,
\[ f(\mathcal{M}_i) = \sum_{j=1}^a f(\mathcal{A}_j) D_{i,j} = \sum_{j,\ell=1}^{a} \tilde{\alpha}_{\ell} \tilde{k}(\mathcal{A}_{\ell},\mathcal{A}_{j}) D_{i,j} . \]
Minimizing quadratic loss yields
\begin{gather*} 
	\argmin_{\vec{\tilde{\alpha}} \in \RR^{a}} \sum_{i=1}^{m} \Bigl( y_i - \sum_{j,\ell=1}^{a} \tilde{\alpha}_{\ell} \tilde{k}(\mathcal{A}_{\ell}, \mathcal{A}_{j}) D_{i,j} \Bigr)^2 + \lambda ||f||^2_{\mathcal{H}} \\
	= \argmin_{\vec{\tilde{\alpha}} \in \RR^{a}} \Braket{ \vec{y} - \mat{D}\mat{\tilde{K}}\vec{\tilde{\alpha}} | \vec{y} - \mat{D}\mat{\tilde{K}}\vec{\tilde{\alpha}} } + \lambda \, \vec{\tilde{\alpha}}^T \mat{\tilde{K}} \vec{\tilde{\alpha}} .
\end{gather*}
Since this is a quadratic form, it suffices to set its gradient to zero and solve for $\vec{\tilde{\alpha}}$:
\begin{gather}
	\nabla_{ \vec{\tilde{\alpha}} } = - 2 \vec{\tilde{\alpha}}^T \mat{\tilde{K}} \mat{D}^T \vec{y} + 2 \mat{\tilde{K}} \mat{D}^T \mat{D} \mat{\tilde{K}} \vec{\tilde{\alpha}} + 2 \lambda \mat{\tilde{K}} \vec{\tilde{\alpha}} = \vec{0} \nonumber \\
	\Leftrightarrow \; \vec{\tilde{\alpha}} = ( \mat{D}^T \mat{D} \mat{\tilde{K}} + \lambda \mat{I} )^{-1} \mat{D}^T \vec{y} \nonumber \\
	\Leftrightarrow \; \vec{\tilde{\alpha}} = \mat{D}^T ( \mat{D} \mat{\tilde{K}} \mat{D}^T + \lambda \mat{I} )^{-1} \vec{y} , \label{equ:atomicalpha}
\end{gather}
where the last expression is preferable for numerical evaluation.
Predictions for $m'$ new systems $\mathcal{M}'$ with $a'$ atoms $\mathcal{A}'$ can be expressed efficiently as
\begin{equation}
	{}\!\!\vec{y'} \!= \Bigl( \sum_{j=1}^{a'} D'_{i,j} \sum_{\ell=1}^a \tilde{\alpha}_{\ell} \tilde{k}(\mathcal{A}_{\ell}, \mathcal{A}'_j) \Bigr)_{i=1,\ldots,m'} 
	\!= \mat{D}' \mat{\tilde{L}}^T \vec{\tilde{\alpha}} , \label{equ:atomickrrpred}
\end{equation}
where $\mat{D}'$ is the incidence matrix for the predicted systems and $\mat{\tilde{L}}$ is the $a \times a'$ kernel matrix between atoms $\mathcal{A}$ and $\mathcal{A}'$.

\medskip

This approach is equivalent to kernel regression (\cref{si:krr}) on whole systems with a kernel~$k$ given by the sum of the atom kernel~$\tilde{k}$ over all pairs of atoms in two systems,
\begin{equation*}
	k(\mathcal{M}_i,\mathcal{M}_j) = \sum_{p,q=1}^a \mat{D}_{i,p} \mat{\tilde{K}}_{p,q} \mat{D}^T_{q,j} .
\end{equation*}
This follows from $\mat{K} = \mat{D} \mat{\tilde{K}} \mat{D}^T$, $\mat{L} = \mat{D} \mat{\tilde{L}} \mat{D'}^T$ and $\vec{\tilde{\alpha}} = \mat{D}^T \vec{\alpha}$ (\cref{equ:atomicalpha}):
Predictions for whole systems using~$k$ are identical to predictions using~$\tilde{k}$:
$\vec{y'} = \mat{L}^T \vec{\alpha} = \mat{D'} \mat{\tilde{L}}^T \mat{D}^T \vec{\alpha} = \mat{D'} \mat{\tilde{L}} \vec{\tilde{\alpha}}$.
In particular, atomic weights~$\vec{\tilde{\alpha}}$ are blocks of system weights~$\vec{\alpha}$.

Computing full atom kernel matrices~$\mat{\tilde{K}}$ and incidence matrices~$\mat{D}$ can require large amounts of memory.
In practice, we compute blocks of~$\mat{\tilde{K}}$ on the fly and directly sum over its entries.
Learning with atomic contributions is extensive (\cref{si:extensivity}).


\section{Requirements}

\subsection{Computational cost}\label{si:compcosts}
Let $f^{\text{QM}}$ and $f^{\text{ML}}$ denote the total computational cost when using only ab initio simulations and a ML-augmented model, respectively:
\begin{align*}
	f^{\text{QM}}(n,m) & = (n+m) \, \text{ref} \\
	f^{\text{ML}}(n,m) & = n \, \text{ref} + (n+m) \, \text{repr} + \text{train}(n) + m \, \text{pred},
\end{align*}
where $n$ and $m$ are number of training and predicted systems, $\text{ref}$, $\text{repr}$, $\text{pred}$ are the cost of one simulation, representation calculation and prediction,
and $\text{train}$ is the cost of training the ML model.
If $n \approx \epsilon \, m$ for small~$\epsilon$, and costs of training and prediction are negligible, total savings in compute time are
\begin{equation*}
	\label{si:equ:compcosts}
	f^{\text{QM}}(n,m) - f^{\text{ML}} (n,m) \approx m \, (\text{ref} - \text{repr}) .
\end{equation*}
Both $\text{ref}$ and $\text{repr}$ depend on system size, often polynomially, with differences in asymptotic runtime as well as constant factors relevant in practice.
Local representations require computing more kernel matrix entries (\cref{si:atomiccontr}) than global representations, which can noticeably influence compute time (\cref{si:tab:compcostkernelm}), but enable scaling with system size:

Let $c$ denote the (average) number of atoms per system, and $d$ the (average) number of atoms in a local environment.
In the following, we assume $d$ to be constant (bounded from above), and representations to have constant size.
Total computational effort to compute representations (first term) and kernel matrices (second term) is then given by
\begin{equation*}
	\mathcal{O} \bigl( (n+m) \, c^k + n m \bigr) 
	\quad \text{and} \quad
	\mathcal{O} \bigl( (n+m) \, c \, d^k + n m c^2 \bigr)
\end{equation*}
for global (left) and local (right) representations, where $d^k$ is constant.
For small systems $c \approx d$, and the additional overhead in computing kernel matrices will dominate runtime for small~$k$. 
In the limit $c \rightarrow \infty$ of large systems, the $c^k$ term will dominate for global representations, while local representations enjoy quadratic scaling.
This can be observed to some extent in \cref{si:tab:compcostrepr,si:tab:compcostkernelm,si:tab:compcostoverview}.

The above analysis applies to representations that explicitly compute $k$-body terms.
Representations that compute $k$-body terms as contractions of basis set coefficients enjoy better scaling in the number of atoms. \citesi{d2019bqx}
Recursion relations between $k$-body terms can improve scaling as well. \citesi{c2019dqx,npc2020qx,dbcdeoo2020x}

\clearpage

\subsection{Role of representations}\label{si:rolerepr}
The role of the representation is to map atomistic systems into a space amenable to regression (linear interpolation).
Strictly speaking, for kernel regression this is the kernel feature space, that is, representation space transformed by the kernel.
We limit our discussion to the representation itself---for the linear kernel this is exact as the transformation is the identity,
and many non-linear kernels like the Gaussian kernel act on the representation space, relying on its structure and implied metric.


\section{A unified framework}

\subsection{$k$-body functions}\label{si:kbodyf}
Informally, a $k$-body function maps information about $k$ distinct atoms $\ket{\alpha_1}, \ldots, \ket{\alpha_k}$, where order can matter, to an output space, here the real numbers, or a distribution on them.
Atom information $\ket{\alpha}$ typically includes coordinates and proton number, but is not limited to those; for example, it could include neutron number to model isotopes.

Typical $k$-body functions include 
atomic number counts ($k=1$),
distances, sometimes inverted or squared ($k=2$),
angles or their cosine ($k=3$),
dihedral or torsional angles, volume-related terms ($k=4$). 
Less common, (al)chemical relationships can be included, for example, based on atoms' period and group in the periodic table. \citesi{fchl2018qx}

In this work, we do not use $k=4$ or higher-order interactions due to the computational cost from combinatorial growth of number of terms when enumerated directly,
which becomes a limiting factor for larger systems, such as in the \dstco{} dataset.

More formally, Glielmo et~al.{} \citesi{gzv2018x,gzfd2020x} define the order of a kernel of two local atomic environments as the smallest integer~$k$ for which differentiating by $k$ different atomic coordinates always yields zero.
Conceptually, in our notation, the body-order of a global representation is
\[ 
	k = \argmin_{k} \frac{ \partial^{k+1} \ket{\mathcal{A}} }{ \partial \ket{\alpha_{i_1}} \ldots \partial \ket{\alpha_{i_{k+1}}} } = 0
\]
for all distinct $\ket{\alpha_{i_1}}, \ldots, \ket{\alpha_{i_{k+1}}}$.
For local representations, one of the atoms $\alpha_i$ is fixed, reducing the order of the derivative by one.

This definition is not concerned with the extent to which $k$-body terms utilise $k$-body information. For instance, in a local representation, products of 2-body terms, say, distances from the central atom to one other atom, depend on two distances (from the central atom to two other atoms), and are therefore formally $k=3$, but cannot resolve angular information because the distance between the two other atoms is not known. \citesi{gzv2018x,gzfd2020x}

Powers of $k$-body terms \citesi{gzv2018x,gzfd2020x} and products of $k$, $k'$-body terms \citesi{jkvak2020x} can improve performance, but are less expressive than full $k^\zeta$-body and $k k'$-body terms, respectively.


\section{Representations}

\subsection{Local atomic neighbourhoods}\label{si:centralatomneighborhood}
Local representations are computed for a local neighbourhood of a central atom,
usually defined as $\{ \, \ket{\alpha_i} | \, d_i \leq r_c \}$, where $\ket{\alpha_i}$ denotes atom~$i$, $d_i$ is distance of $\ket{\alpha_i}$ to the central atom, and $r_c \geq 0$ is a cut-off radius. 
Usually, a cut-off function that smoothly approaches zero for $d_i \rightarrow r_c$ is employed to prevent discontinuities at the threshold.

Both the \texttt{quippy} and \texttt{DScribe} implementations of SOAP include the central atom in the neighbourhood, and thus in the neighbourhood density, \citesi{dbcc2016qx}
in contrast to the original definition \citesi{bkc2013qx}.
SFs do not take the central atom into account explicitly. 

In periodic systems, the unit cell is replicated up to the cut-off radius to ensure that all interactions within the neighbourhood are included.
In practice, some implementations may internally use a modified \emph{effective} cut-off radius.
For instance, \texttt{DScribe} ensures that atoms up to the tail of the radial basis function are taken into account. 


\stepcounter{section}


\stepcounter{section}


\section{Empirical comparison}

\subsection{\dsgdb{} dataset}\label{si:dsgdb}
The \dsgdb{} dataset, \citesi{rdrl2014x,rdrl2015x} also known as \texttt{gdb9-14}, contains 133\,885 small organic molecules composed of H, C, N, O, F with up to 9 non-H atoms.
It is a subset of the ``generated database~17'' (\mbox{GDB-17}). \citesi{rdbr2012x}
Molecular ground state geometries and properties, including energetics, are computed at density functional level of theory using the Becke 3-parameter Lee-Yang-Parr (B3LYP) \citesi{sdcf1994x} hybrid functional with 6-31G(2df,p) basis set.

We use the version available at \url{qmml.org}, which offers a convenient format for parsing, and exclude all structures in the \texttt{uncharacterized.txt} file and those listed in the \texttt{readme.txt} file as ``difficult to converge'', as those are potentially problematic.
Total energies were converted to energies of atomization by subtracting the atomic contributions given in file \texttt{atomref.txt}.

\subsection{\dsba{} dataset}\label{si:dsba}
The \dsba{} dataset, \citesi{nrbsmrcwh2019x} also known as \texttt{dft-10b}, contains unrelaxed geometries and their enthalpies of formation for the 10 binary alloys AgCu, AlFe, AlMg, AlNi, AlTi, CoNi, CuFe, CuNi, FeV, and NbNi.
For each alloy system, unrelaxed geometries with lattice parameters from Vegard's rule \citesi{v1921x,da1991x} and energies are computed for all possible unit cells \citesi{hf2008x} with 1--8 atoms for FCC and BCC lattices, and 2--8 atoms for HCP lattices, 
using the generalized gradient approximation (GGA) of Perdew, Burke and Ernzerhof (PBE) with projector-augmented wave (PAW) potentials and generalized regular $k$-point grids \citesi{wmm2016x,mjhh2018x}.
The dataset contains 631 FCC, 631 BCC, and 333 HCP structures per alloy system, yielding 15\,950 structures in total. 
We use the version available at \url{qmml.org}.

\subsection{\dstco{} dataset}\label{si:dstco}
The \dstco{} dataset \citesi{sgylbhglzs2019x} is a Kaggle challenge \citesi{kagglex} dataset containing 3\,000 ternary (Al$_x$-Ga$_y$-In$_z$)$_2$O$_3$ oxides, $x+y+z=1$, of potential interest as transparent conducting oxides.
We predict formation and band-gap energies of relaxed structures, using either relaxed (\dstcor) or approximate (\dstcou) structures from Vegard's rule as input.
Geometries and energies are computed at the density functional level of theory using the PBE functional as implemented in the all-electron code \texttt{FHI-aims} \citesi{bghhhrrs2009x} with \texttt{tight} settings.

The challenge scenario is to predict formation and band-gap energies of relaxed structures from unrelaxed geometries obtained via Vegard's rule.
This is equivalent to strong noise or bias in the inputs.
Unlike pure benchmarking scenarios, where computationally expensive relaxed geometries are given, the challenge scenario is closer to a virtual screening application in that Vegard's rule geometries are computationally inexpensive to obtain.

The dataset contains all structures from the challenge training and leaderboard data.
Unless otherwise noted, we report RMSE, not the root mean square logarithmic error used in the challenge.

\enlargethispage*{1\baselineskip}

\subsection{Learning curves}\label{si:learncurves}
Plots of empirical prediction error~$\epsilon$ as a function of training set size~$n$ are called ``learning curves''. 
Asymptotically, we assume the error to decay as a negative power, \citesi{afs1992x} $\epsilon = a' n^{-b}$. 
On a log-log plot, $\epsilon$ is therefore linear, $\log \epsilon = a - b \log(n)$, and the offset~$a = \log a'$ and slope~$b$ can be used to characterize predictive performance of models. \citesi{hl2016x}
For QM/ML models the estimated quantities are noise-free (except for numerical noise, which is negligible for converged calculations)
and representations are unique.
We use base-10 logarithms for learning curves.
For asymptotic fits we weight training set sizes by the standard deviation over their respective splits to attenuate for small sample effects,
as the above equation is valid only in the limit $n \rightarrow \infty$.
All learning curves in \cref{fig:learncurvesrmse,si:fig:learncurvesmae} show linear behaviour after at most a few hundred samples.
See \cref{si:tab:offsetsslopes} for estimated offsets~$a$ and slopes~$b$.

\begin{table}
	\caption{\emph{Estimated offsets and slopes for learning curves} in \cref{fig:learncurvesrmse,si:fig:learncurvesnmd18u,si:fig:learncurvesmae}.
		Shown are estimated offsets~$a$ and slopes~$b$ for learning curves $\log_{10} \epsilon = a - b \log_{10} n$.
		See \cref{si:learncurves} for details of the fitting procedure.
		\label{si:tab:offsetsslopes}}

	\bigskip
	\centering

	(a) \dsgdb{} dataset.
	
	\medskip

	\begin{tabular}{@{}lllll@{}}
		\toprule
		               & \multicolumn{2}{c}{RMSE} & \multicolumn{2}{c}{MAE} \\ \cmidrule(lr){2-3} \cmidrule(lr){4-5}
		Representation & Offset & Slope & Offset & Slope \\
		\midrule
		MBTR $k=2$ & 1.82 & 0.35 & 1.67 & 0.36 \\
SF $k=2$ & 1.83 & 0.38 & 1.61 & 0.37 \\
MBTR $k=2,3$ & 1.97 & 0.46 & 1.64 & 0.43 \\
SF $k=2,3$ & 1.68 & 0.40 & 1.56 & 0.42 \\
SOAP & 1.63 & 0.42 & 1.58 & 0.47 \\
		\bottomrule
	\end{tabular}
	
	\bigskip

	(b) \dsba{} dataset.

	\medskip

	\begin{tabular}{@{}lllll@{}}
		\toprule
		               & \multicolumn{2}{c}{RMSE} & \multicolumn{2}{c}{MAE} \\ \cmidrule(lr){2-3} \cmidrule(lr){4-5}
		Representation & Offset & Slope & Offset & Slope \\
		\midrule
		MBTR $k=2$ & 1.74 & 0.20 & 1.66 & 0.24 \\
SF $k=2$ & 1.81 & 0.23 & 1.69 & 0.26 \\
MBTR $k=2,3$ & 1.94 & 0.30 & 1.77 & 0.31 \\
SF $k=2,3$ & 2.07 & 0.33 & 1.83 & 0.32 \\
SOAP & 2.14 & 0.37 & 2.00 & 0.39 \\
		\bottomrule
	\end{tabular}
	
	\bigskip

	(c) \dstcou{} dataset.

	\medskip

	\begin{tabular}{@{}lllll@{}}
		\toprule
		               & \multicolumn{2}{c}{RMSE} & \multicolumn{2}{c}{MAE} \\ \cmidrule(lr){2-3} \cmidrule(lr){4-5}
		Representation & Offset & Slope & Offset & Slope \\
		\midrule
		MBTR $k=2$ & 1.98 & 0.18 & 1.89 & 0.23 \\
SF $k=2$ & 2.07 & 0.22 & 1.89 & 0.23 \\
MBTR $k=2,3$ & 1.93 & 0.17 & 1.79 & 0.20 \\
SF $k=2,3$ & 2.01 & 0.20 & 1.86 & 0.22 \\
SOAP & 1.94 & 0.17 & 1.73 & 0.19 \\
		\bottomrule
	\end{tabular}

	\bigskip

	(d) \dstcor{} dataset.

	\medskip

	\begin{tabular}{@{}lllll@{}}
		\toprule
		               & \multicolumn{2}{c}{RMSE} & \multicolumn{2}{c}{MAE} \\ \cmidrule(lr){2-3} \cmidrule(lr){4-5}
		Representation & Offset & Slope & Offset & Slope \\
		\midrule
		MBTR $k=2$ & 1.87 & 0.28 & 1.65 & 0.29 \\
SF $k=2$ & 1.86 & 0.35 & 1.79 & 0.41 \\
MBTR $k=2,3$ & 2.08 & 0.35 & 1.86 & 0.35 \\
SF $k=2,3$ & 1.86 & 0.40 & 1.73 & 0.44 \\
SOAP & 1.94 & 0.46 & 1.80 & 0.50 \\
		\bottomrule
	\end{tabular}
\end{table}

\subsection{Subsets}\label{si:subsets}
For training and validation, data subsets were sampled as follows:
An \emph{outer validation set}%
\footnote{In the literature, the terms ``test set'' and ``validation set'' are sometimes used with different meaning.
To avoid confusion, we use ``outer'' for the subset employed to measure performance, and ``inner'' for the subset employed to optimize HPs.}
was randomly drawn (10\,k molecules for \dsgdb{}, 1\,k structures for \dsba{}, 600 structures for \dstco{}).
From the remaining entries, \emph{outer training sets} of sizes 100, 250, 650, 1\,600, 4\,000 and 10\,000 for datasets \dsgdb{}, \dsba{} and 100, 160, 250, 400, 650, 1\,000 and 1\,600 for dataset \dstco{} were randomly drawn.
These sizes were chosen to be equidistant in log-space.
Each outer training set was then split into an \emph{inner training set} and an \emph{inner validation set} by randomly drawing the latter.
We used an 80\,\,/\,20\,\% split, yielding inner validation sets of size 20, 50, 130, 320, 800, 2\,000 for datasets \dsgdb{}, \dsba{} and 20, 32, 50, 80, 130, 200, 320 for \dstco{}.
The whole procedure was repeated 10 times.
We excluded structures with few atoms (6 or fewer non-H atoms for \dsgdb, 5 or fewer atoms per unit cell for \dsba, 10 atoms per unit cell for \dstco{}) as there are not enough of these for statistical learning.

\subsection{Sampling}\label{si:sampling}
To reduce variance, remove bias and ensure that subsets faithfully represent the distribution of the whole dataset, subsets were drawn using Monte-Carlo sampling such that differences to the parent dataset in selected statistics were below pre-defined fractional thresholds.

For dataset \dsgdb, these were number of N, O and F atoms, number of molecules with 7, 8 and 9 non-H atoms, binned number of atoms (with H), and binned energy.
For dataset \dsba, these were number of all constituting elements, unit cells with 6, 7, 8, and 9 atoms, binned sizes and energies. 
For dataset \dstco, these were number of Al, Ga, In, O atoms, unit cells with 20, 30, 40, 60, 80 atoms, and binned energies.

\subsection{Kernel regression}\label{si:krr}
We use kernel ridge regression \citesi{htf2009x} or Gaussian process regression \citesi{rw2006x} (the two are equivalent in terms of predictions).
A detailed derivation can be found in \eref{r2015fx}.
In summary, predictions are basis set expansions of the form
$f(\vec{x}) = \sum_{i=1}^n \alpha_i k(\vec{x_i},\vec{x})$,
where $\vec{x}$ is the system to predict, $\vec{x_1},\ldots,\vec{x_n}$ are the training systems, and $k$ is a symmetric positive definite function (kernel).
The regression coefficients $\vec{\alpha}$ are obtained by minimizing the regularized quadratic loss
$\sum_{i=1}^n ( y_i - f(\vec{x_i}) )^2 + \lambda ||f||^2_{\mathcal{H}}$,
where $\vec{y}$ are property values of the training data and the regularization strength~$\lambda$ is a HP that controls the smoothness of the predictor.
In this work, we use the Gaussian kernel $k(\vec{x},\vec{z}) = \exp (-||\vec{x}-\vec{z}||_2^2 / 2\sigma^2)$, where the length scale $\sigma$ is a HP (\cref{si:hpregression}).
We used the \texttt{qmmlpack} \citesi{r2015fx,qmmlpackx} implementation.

\subsection{HP optimization}\label{si:hpopt}
For model selection we optimized the HPs of representations, kernel and regression method, 
including structural HPs (for example, which $k$-body functions to use) and numerical HPs (for example, the Gaussian kernel length scale).
Specifically, the RMSE of an inner validation set was minimized using tree-structured Parzen estimators \citesi{bbbk2011x,byc2013x} in combination with local grid search.
The same optimization scheme was used for all representations, using consistent grid spacings and parameter ranges to reduce human bias.
Our corresponding \texttt{cmlkit} package \citesi{cmlkitx} provides interfaces to the \texttt{hyperopt} optimization package \citesi{bbbk2011x} and to each representation's implementation(s);
it is freely available under an open source license. 

The space of possible models (``HP search space'') is a tree-structured set of choices, for instance, between different $k$-body functions, or different values of a numerical HP.
Tree-structured Parzen estimators treat this search space as a prior distribution over HPs, updated every time a loss is computed to increase prior weight around HP settings with better loss. 
We use uniform priors throughout, discretizing numerical HPs on logarithmic or linear grids as necessary.
Once a HP search space has been defined, model selection is fully automatic.

HPs were optimized for each training set size as follows:
For each trial, representation HPs and starting values for regression method HPs (Gaussian kernel length scale and regularization strength) were drawn from the prior. 
The latter were then refined through a randomized local grid search and the resulting HP values used to update the prior.
All optimizations were run for 2\,000 steps, and rerun three times, to minimize variance from stochastic optimization.
To reduce computational cost, HPs were optimized on only one outer split.

\subsection{Kernel regression HPs}\label{si:hpregression}
We used KRR with a single Gaussian kernel, a frequently used combination in the literature.
Note that due to Requirement~\ref{req:Continuity}, the Gaussian kernel is better suited than less smooth kernels such as the Laplacian kernel. \citesi{hr2017x}

No post-processing of the kernel was performed.
In particular, centering of kernel and labels, which together is equivalent to having an explicit bias term~$b$ in the regression, were not performed, as this is not necessary for the Gaussian kernel. \citesi{pmrrv2001x}
Depending on the representation used, labels were normalized for training as needed to either represent values per atom or per entire system. (\cref{si:extensivity})

This setup entails two HPs:
The width~$\sigma$ of the Gaussian kernel, and the regularization strength~$\lambda$.
Search spaces for these two HPs (\cref{si:tab:regrhps}) were held constant across all representations and learning curves.
See \eref{reproducibilityx} for HP search spaces and optimized model HPs. 

\begin{table}[hbtp]
	\caption{\emph{Kernel regression hyperparameter search space.}
		Both parameters optimized on a base-2 logarithmic grid.
		{\small TPE = tree-structured Parzen estimators;
		LGS = local grid search.}
		\label{si:tab:regrhps}}
	\mbox{}

	\begin{center}
	    \begin{tabular}{l rrr rrr}
    	    \toprule
        	Hyper-    & \multicolumn{3}{c}{TPE} & \multicolumn{3}{c}{LGS} \\ \cmidrule(lr){2-4} \cmidrule(lr){5-7}
	        parameter & min & max & step & min & max & step \\
    	    \midrule
        	
	        $\log_2{\lambda}$ & -18 & 0 & 1.0 & -20 & 2 & 0.5\\
    	    $\log_2{\sigma}$ & -13 & 13 & 1.0 & -15 & 15 & 0.5\\
        
        	\bottomrule
	    \end{tabular}
	\end{center}
\end{table}

\enlargethispage*{2\baselineskip}

\subsection{Symmetry function HPs}\label{si:hpsf}
The five SFs proposed in \eref{b2011ex} are
\begin{align}
	G^1_i =& \sum_{j} f_c(d_{ij})  \label{equ:sisfone}\\
	G^2_i =& \sum_{j} \exp \bigl(-\eta (d_{ij}-\mu)^2 \bigr) \, f_c(d_{ij}) \label{equ:sisftwo}\\
	G^3_i =& \sum_{j} \cos ( \kappa \, d_{ij} ) \, f_c(d_{ij})  \label{equ:sisfthree}\\
	G^4_i =& \,2^{1-\zeta} \sum_{j,k \neq i} (1 + \lambda \cos \theta_{ijk})^{\zeta}\notag\\
	       & \exp{\bigl(-\eta (d_{ij}^2 + d_{ik}^2 + d_{jk}^2)\bigr)} \label{equ:sisffour}\\
	       & f_c(d_{ij}) \, f_c(d_{ik}) \, f_c(d_{jk}) \notag\\
	G^5_i =& \,2^{1-\zeta} \sum_{j,k \neq i} (1 + \lambda \cos \theta_{ijk})^{\zeta} \notag\\
	       & \exp{\bigl( -\eta (d_{ij}^2 + d_{ik}^2) \bigr)} \, f_c(d_{ij}) \, f_c(d_{ik}) \,, \label{equ:sisffive}
\end{align}
with the cut-off function
\begin{equation}
	\label{equ:cutoff}
	f_c(d_{ij}) = \begin{cases}
		0.5 \cos \bigl(\pi d_{ij} / c \bigr) & \text{for}\quad d_{ij} \leq c\\
			0 & \text{for}\quad d_{ij} > c \, .\\
	\end{cases}
\end{equation}
In \cref{equ:sisfone,equ:sisftwo,equ:sisfthree,equ:sisffour,equ:sisffive,equ:cutoff},
index $i$ is the central atom, $j, k$ run over all atoms in the local environment around~$i$ with cut-off radius~$c$,
$d_{lm}$ indicates pairwise distance, $\theta_{lmn}$ the angle between three atoms;
$\eta$ and $\kappa$ are broadening parameters, $\mu$ a shift, $\zeta$ determines angular resolution. 
$\lambda = \pm 1$ determines whether the angular part of $G_i^4$ and $G_i^5$ peaks at $\SI{0}{\degree}$ or $\SI{180}{\degree}$. 

We utilize the \texttt{RuNNer} \citesi{arunnerx,b2015qx,b2017qx} software to compute SFs and restrict ourselves to the radial SFs~$G_i^2$ and angular SFs~$G_i^4$ (\texttt{RuNNer} functions~2 and~3). 
We use the same SFs for all element combinations to minimize size of HP search space.
This can result in constant or low-variance features,
which are unproblematic for kernel regression with a Gaussian kernel (\cref{si:krr}) as they enter only through the norm of feature vector differences
(for neural networks, these features could be more problematic).

Similarly, we use an empirical parametrization scheme \citesi{gsbbm2018x} to choose HPs $\mu$ and $\eta$ for~$G_i^2$ and HPs $\eta, \zeta$, and $\lambda$ for~$G_i^4$. 

For radial SFs we use two schemes, \emph{shifted} and \emph{centered}. 
For \emph{shifted}, $\mu$ is chosen on a linear grid while $\eta$ is held fixed. 
For \emph{centered}, $\mu=0$ and $\eta$ is chosen such that the standard deviation of each SF lies on the same grid points. 
For $i \in \{0, 1, \dots, n \}$ a point on a one-dimensional grid, $\Delta = \frac{c - 1.5}{n-1}$, and $r_i = 1 + \Delta i$, 
in the \emph{centered} scheme, $\mu_i = 0$ and $\eta_i = \frac{1}{2 r_i^2}$, 
and in the \emph{shifted} scheme, $\mu_i = r_i$ and $\eta_i = (2 \Delta^2)^{-1}$.
In this setting, the only HP is the number of grid points~$n+1$, which we allow to vary from 2 to 10 for each scheme.

For angular SFs, we choose $\lambda=\pm 1$ and $\zeta = 1, 2, 4$.
The only HP remaining is the broadening~$\eta$, optimized on a $\log_2$ grid between \num{-20} and \num{1} with spacing 0.5.
The radial SFs and two angular SFs with $\lambda=\pm 1$ and $\zeta = 1$ are always included, but the optimizer can enable or disable any of the remaining $k=3$ SFs with $\lambda=\pm 1$ and $\zeta = 2, 4$.
Cut-off radii are varied in integer steps, starting from the integer above the smallest distance found in the dataset, up until at most \SI{9}{\angstrom}.

The output of \texttt{RuNNer} is post-processed to be suitable for KRR, placing all SFs for a given type of central atom in separate blocks of an atomic feature vector with $\#\,{\text{elements}} \times \#\,{\text{SFs}}$ components. 
For the Gaussian kernel, this leads to negligible kernel values between representations belonging to different elements.
As SFs are local representations, labels were normalized to (extensive) per-system values.

See \eref{reproducibilityx} for HP search spaces and optimized model HPs. 

\subsection{Many-body tensor representation HPs}\label{si:hpmbtr}
We employed the MBTR implementation in \texttt{qmmlpack}, \citesi{qmmlpackx} 
adding optional normalization by $\ell_1$ or $\ell_2$ norm.
For $k=2,3$, representations for $k=2$ and $k=3$ were concatenated.
MBTR exhibits several categorical HPs, with subsequent numerical HPs conditional on prior choices.

We used the $k$-body functions \texttt{1/distance}, \texttt{1/dot} ($k$=2), and \texttt{angle}, \texttt{cos\_angle}, \texttt{dot/dotdot} ($k$=3).
No one-body terms were used as atomization and formation energies already contain linear contributions of element counts.
Histogram ranges were chosen based on the whole dataset, as inter-atomic distance ranges are similar for all subsets.
\num{100} discretization bins were used throughout. 
Broadening parameters were restricted to at least a single bin and at most a quarter of the range of the corresponding geometry function.

From the weighting functions, we used \texttt{identity\^{}2}, \texttt{exp\_-1/identity}, \texttt{exp\_-1/identity\^{}2} ($k$=2), and \texttt{1/dotdotdot}, \texttt{exp\_-1/normnormnorm}, \texttt{exp\_-1/norm+norm+norm} ($k$=3). 
The latter two in each set introduce conditional HPs. 
For periodic systems, in particular the \dstco{} dataset, the ranges of these parameters were manually restricted to avoid excessive computation times (above \SI{30}{\s} for one trial). 
The convergence threshold was set to \num{0.001}.

We used the \texttt{full} indexing scheme, which generates all permutations of elements (as opposed to \texttt{noreversals}, which does not double-count element combinations, for example, CH and HC).
This seems to lead to more consistent behaviour and higher predictive accuracy for supercells, or unit cells of different sizes, 
and similar accuracy for molecules, at the expense of higher computational cost.
We used per-system energies for the \dsgdb{} dataset and per-atom energies for datasets \dsba{} and \dstco.

\subsection{Smooth Overlap of Atomic Positions HPs}\label{si:hpsoap}
We used the \texttt{DScribe} implementation of SOAP with Gaussian-type orbitals, \citesi{adscribex,hjmfrgrf2020qx} 
which we found to provide more accurate predictions at lower computational cost than the \texttt{quippy} \citesi{aquippyx} implementation.
Results are already structured by element types; 
no post-processing was applied.
HPs $l_{\text{max}}$ and $n_{\text{max}}$ were chosen between 2 and 8. 
Cut-off radii were chosen as for SFs, and the broadening adapted to the resulting ranges (53 steps from -20 to 6 on a $\log_2$ grid).
We report results for the \texttt{gto} basis set, which resulted in lower prediction errors than the \texttt{polynomial} one, and was faster to compute. 
Labels were normalized to per-system values.
See \eref{reproducibilityx} for HP search spaces and optimized model HPs. 

\subsection{Prediction errors}\label{si:predacc}
\Cref{si:tab:perfrmse,si:tab:perfmae} present numerical values underlying the learning curves for RMSE (\cref{si:tab:perfrmse}) and MAE (\cref{si:tab:perfmae}).
For rRMSE (\cref{si:errormetrics}), standard deviations of 
\SI{239.31}{\kilo\cal\per\mol}, 
\SI{178.86}{\milli\eV}, 
\SI{104.57}{\milli\eV}, 
were used for datasets \dsgdb, \dsba, \dstco, computed over the whole dataset 
(differences to standard deviations over validation sets were around 1\,\% or less in all cases).

\begin{table*}[hbtp]
	\caption{\emph{Prediction errors (RMSE)} for representations of \cref{sec:representations}.
		Shown is mean $\pm$ standard deviation over ten outer splits for energy predictions, measured on an out-of-sample validation set.
		\label{si:tab:perfrmse}}
		
	\centering
	\small
		
	\bigskip
	\medskip
	
	(a) Dataset~\dsgdb{}.
	
	\medskip

	\begin{tabular}{@{} l rrrrrr @{}}
		\toprule
		& \multicolumn{6}{c}{Training set size} \\ 
		\cmidrule(lr){2-7}
		Representation & \multicolumn{1}{c}{\num{100}} & \multicolumn{1}{c}{\num{250}} & \multicolumn{1}{c}{\num{650}} & \multicolumn{1}{c}{\num{1600}} & \multicolumn{1}{c}{\num{4000}} & \multicolumn{1}{c}{\num{10000}} \\
		\midrule
		MBTR $k=2$ & \num{12.19 \pm 1.83 } & \num{9.50 \pm 0.92 } & \num{6.60 \pm 0.29 } & \num{5.68 \pm 1.33 } & \num{3.52 \pm 0.12 } & \num{2.53 \pm 0.06 } \\
SF $k=2$ & \num{9.63 \pm 0.94 } & \num{7.73 \pm 0.54 } & \num{5.74 \pm 0.17 } & \num{4.10 \pm 0.36 } & \num{2.80 \pm 0.08 } & \num{1.98 \pm 0.04 } \\
MBTR $k=2,3$ & \num{11.90 \pm 1.86 } & \num{6.48 \pm 0.46 } & \num{6.03 \pm 0.40 } & \num{3.55 \pm 0.92 } & \num{1.98 \pm 0.10 } & \num{1.34 \pm 0.05 } \\
SF $k=2,3$ & \num{12.45 \pm 5.98 } & \num{6.75 \pm 0.60 } & \num{3.76 \pm 0.16 } & \num{2.97 \pm 0.48 } & \num{1.75 \pm 0.09 } & \num{1.27 \pm 0.05 } \\
SOAP & \num{7.77 \pm 1.53 } & \num{4.75 \pm 0.75 } & \num{2.77 \pm 0.16 } & \num{2.25 \pm 0.31 } & \num{1.29 \pm 0.07 } & \num{0.90 \pm 0.05 } \\
		\bottomrule
    \end{tabular}
    
    \bigskip
	\medskip
    
    (b) Dataset~\dsba{}.
    
    \medskip
    
	\begin{tabular}{@{} l rrrrrr @{}}
		\toprule
		& \multicolumn{6}{c}{Training set size} \\ 
		\cmidrule(lr){2-7}
		Representation & \multicolumn{1}{c}{\num{100}} & \multicolumn{1}{c}{\num{250}} & \multicolumn{1}{c}{\num{650}} & \multicolumn{1}{c}{\num{1600}} & \multicolumn{1}{c}{\num{4000}} & \multicolumn{1}{c}{\num{10000}} \\
		\midrule
		MBTR $k=2$ & \num{46.22 \pm 6.76 } & \num{30.81 \pm 2.90 } & \num{17.87 \pm 1.69 } & \num{12.57 \pm 0.33 } & \num{10.95 \pm 0.45 } & \num{9.20 \pm 0.46 } \\
SF $k=2$ & \num{43.03 \pm 5.47 } & \num{28.08 \pm 3.09 } & \num{15.80 \pm 0.98 } & \num{11.60 \pm 0.34 } & \num{9.59 \pm 0.37 } & \num{7.91 \pm 0.27 } \\
MBTR $k=2,3$ & \num{54.24 \pm 5.10 } & \num{24.79 \pm 2.70 } & \num{13.47 \pm 1.22 } & \num{8.98 \pm 0.42 } & \num{7.15 \pm 0.24 } & \num{5.45 \pm 0.20 } \\
SF $k=2,3$ & \num{60.99 \pm 6.66 } & \num{27.72 \pm 2.34 } & \num{15.07 \pm 0.89 } & \num{10.32 \pm 0.48 } & \num{7.45 \pm 0.24 } & \num{5.75 \pm 0.17 } \\
SOAP & \num{43.18 \pm 5.95 } & \num{23.54 \pm 2.06 } & \num{12.69 \pm 0.64 } & \num{8.82 \pm 0.37 } & \num{6.72 \pm 0.42 } & \num{4.64 \pm 0.25 } \\
		\bottomrule
    \end{tabular}
    
    \bigskip
	\medskip
    
    (c) Dataset~\dstcor{}.
    
    \medskip
    
	\begin{tabular}{@{} l rrrrrrr @{}}
		\toprule
		& \multicolumn{7}{c}{Training set size} \\ 
		\cmidrule(lr){2-8}
		Representation & 
		\multicolumn{1}{c}{\num{100}}  & 
		\multicolumn{1}{c}{\num{160}}  & 
		\multicolumn{1}{c}{\num{250}}  & 
		\multicolumn{1}{c}{\num{400}}  & 
		\multicolumn{1}{c}{\num{650}}  & 
		\multicolumn{1}{c}{\num{1000}} &
		\multicolumn{1}{c}{\num{1600}} \\
		\midrule
		MBTR $k=2$ & \num{33.12 \pm 3.55 } & \num{31.24 \pm 4.51 } & \num{20.09 \pm 2.08 } & \num{14.06 \pm 0.82 } & \num{11.69 \pm 0.33 } & \num{10.38 \pm 0.58 } & \num{10.02 \pm 0.53 } \\
SF $k=2$ & \num{17.18 \pm 0.99 } & \num{11.57 \pm 0.73 } & \num{10.00 \pm 0.61 } & \num{8.72 \pm 0.66 } & \num{7.45 \pm 0.46 } & \num{6.54 \pm 0.57 } & \num{5.47 \pm 0.32 } \\
MBTR $k=2,3$ & \num{30.41 \pm 2.78 } & \num{30.18 \pm 4.45 } & \num{17.03 \pm 1.10 } & \num{15.42 \pm 1.34 } & \num{12.31 \pm 0.47 } & \num{10.13 \pm 0.73 } & \num{9.56 \pm 0.57 } \\
SF $k=2,3$ & \num{17.47 \pm 1.20 } & \num{9.62 \pm 0.44 } & \num{8.31 \pm 0.51 } & \num{6.78 \pm 0.55 } & \num{5.37 \pm 0.24 } & \num{4.47 \pm 0.24 } & \num{4.08 \pm 0.22 } \\
SOAP & \num{13.28 \pm 1.19 } & \num{10.19 \pm 0.91 } & \num{7.05 \pm 0.38 } & \num{5.54 \pm 0.47 } & \num{4.06 \pm 0.43 } & \num{3.60 \pm 0.30 } & \num{3.29 \pm 0.34 } \\
		\bottomrule
    \end{tabular}
    
    \bigskip
	\medskip

    (d) Dataset~\dstcou{}.
    
    \medskip
    
	\begin{tabular}{@{} l rrrrrrr @{}}
		\toprule
		& \multicolumn{7}{c}{Training set size} \\ 
		\cmidrule(lr){2-8}
		Representation & 
		\multicolumn{1}{c}{\num{100}}  & 
		\multicolumn{1}{c}{\num{160}}  & 
		\multicolumn{1}{c}{\num{250}}  & 
		\multicolumn{1}{c}{\num{400}}  & 
		\multicolumn{1}{c}{\num{650}}  & 
		\multicolumn{1}{c}{\num{1000}} &
		\multicolumn{1}{c}{\num{1600}} \\
		\midrule
		MBTR $k=2$ & \num{45.31 \pm 5.43 } & \num{40.87 \pm 3.93 } & \num{33.42 \pm 3.08 } & \num{31.52 \pm 2.14 } & \num{29.93 \pm 2.07 } & \num{27.77 \pm 1.87 } & \num{25.46 \pm 2.01 } \\
SF $k=2$ & \num{42.94 \pm 4.61 } & \num{40.53 \pm 3.83 } & \num{37.91 \pm 1.95 } & \num{30.63 \pm 1.43 } & \num{28.64 \pm 1.82 } & \num{25.98 \pm 1.95 } & \num{24.22 \pm 1.52 } \\
MBTR $k=2,3$ & \num{39.27 \pm 5.65 } & \num{38.38 \pm 2.16 } & \num{31.71 \pm 3.01 } & \num{31.25 \pm 2.07 } & \num{27.63 \pm 1.92 } & \num{26.16 \pm 1.43 } & \num{25.89 \pm 1.76 } \\
SF $k=2,3$ & \num{37.31 \pm 3.64 } & \num{42.52 \pm 4.21 } & \num{35.94 \pm 2.28 } & \num{31.39 \pm 2.53 } & \num{29.83 \pm 1.78 } & \num{25.95 \pm 1.65 } & \num{24.30 \pm 1.46 } \\
SOAP & \num{42.02 \pm 5.04 } & \num{39.42 \pm 3.64 } & \num{30.31 \pm 2.92 } & \num{30.21 \pm 3.09 } & \num{28.21 \pm 1.81 } & \num{26.29 \pm 2.11 } & \num{24.46 \pm 1.88 } \\
		\bottomrule
    \end{tabular}
    
\end{table*}

\begin{table*}[hbtp]
	\caption{\emph{Prediction errors (MAE)} for representations of \cref{sec:representations}.
		Shown is mean $\pm$ standard deviation over ten outer splits for energy predictions, measured on an out-of-sample validation set.
		\label{si:tab:perfmae}}
		
	\centering
	\small
		
	\bigskip
	\medskip
	
	(a) Dataset~\dsgdb{}.
	
	\medskip

	\begin{tabular}{@{} l rrrrrr @{}}
		\toprule
		& \multicolumn{6}{c}{Training set size} \\ 
		\cmidrule(lr){2-7}
		Representation & 
		\multicolumn{1}{c}{\num{100}} & 
		\multicolumn{1}{c}{\num{250}} & 
		\multicolumn{1}{c}{\num{650}} & 
		\multicolumn{1}{c}{\num{1600}} & 
		\multicolumn{1}{c}{\num{4000}} & 
		\multicolumn{1}{c}{\num{10000}} \\
		\midrule
		MBTR $k=2$ & \num{8.54 \pm 0.85 } & \num{5.93 \pm 0.26 } & \num{4.66 \pm 0.17 } & \num{3.28 \pm 0.14 } & \num{2.33 \pm 0.03 } & \num{1.67 \pm 0.03 } \\
SF $k=2$ & \num{6.72 \pm 0.78 } & \num{5.34 \pm 0.28 } & \num{3.86 \pm 0.12 } & \num{2.64 \pm 0.05 } & \num{1.87 \pm 0.03 } & \num{1.34 \pm 0.02 } \\
MBTR $k=2,3$ & \num{8.25 \pm 0.87 } & \num{4.28 \pm 0.18 } & \num{3.88 \pm 0.12 } & \num{1.91 \pm 0.09 } & \num{1.21 \pm 0.03 } & \num{0.87 \pm 0.02 } \\
SF $k=2,3$ & \num{7.34 \pm 1.35 } & \num{4.18 \pm 0.23 } & \num{2.49 \pm 0.06 } & \num{1.80 \pm 0.05 } & \num{1.09 \pm 0.02 } & \num{0.78 \pm 0.02 } \\
SOAP & \num{4.93 \pm 0.59 } & \num{2.79 \pm 0.20 } & \num{1.70 \pm 0.05 } & \num{1.26 \pm 0.04 } & \num{0.73 \pm 0.02 } & \num{0.49 \pm 0.01 } \\
		\bottomrule
    \end{tabular}
    
    \bigskip
	\medskip
    
    (b) Dataset~\dsba{}.
    
    \medskip
    
	\begin{tabular}{@{} l rrrrrr @{}}
		\toprule
		& \multicolumn{6}{c}{Training set size} \\ 
		\cmidrule(lr){2-7}
		Representation &
		\multicolumn{1}{c}{\num{100}} & 
		\multicolumn{1}{c}{\num{250}} & 
		\multicolumn{1}{c}{\num{650}} & 
		\multicolumn{1}{c}{\num{1600}} & 
		\multicolumn{1}{c}{\num{4000}} & 
		\multicolumn{1}{c}{\num{10000}} \\
		\midrule
		MBTR $k=2$ & \num{27.01 \pm 1.99 } & \num{18.22 \pm 1.21 } & \num{10.74 \pm 0.60 } & \num{7.49 \pm 0.19 } & \num{6.35 \pm 0.19 } & \num{5.16 \pm 0.19 } \\
SF $k=2$ & \num{28.02 \pm 2.21 } & \num{18.23 \pm 1.24 } & \num{9.76 \pm 0.56 } & \num{6.98 \pm 0.16 } & \num{5.52 \pm 0.18 } & \num{4.49 \pm 0.12 } \\
MBTR $k=2,3$ & \num{36.93 \pm 2.62 } & \num{15.49 \pm 1.32 } & \num{8.47 \pm 0.42 } & \num{5.56 \pm 0.16 } & \num{4.34 \pm 0.07 } & \num{3.26 \pm 0.07 } \\
SF $k=2,3$ & \num{40.67 \pm 2.22 } & \num{18.18 \pm 1.20 } & \num{9.43 \pm 0.48 } & \num{6.38 \pm 0.23 } & \num{4.43 \pm 0.07 } & \num{3.45 \pm 0.08 } \\
SOAP & \num{27.68 \pm 2.06 } & \num{14.82 \pm 0.78 } & \num{7.89 \pm 0.34 } & \num{5.43 \pm 0.19 } & \num{3.96 \pm 0.09 } & \num{2.78 \pm 0.11 } \\
		\bottomrule
    \end{tabular}
    
    \bigskip
	\medskip
    
    (c) Dataset~\dstcor{}.
    
    \medskip
    
	\begin{tabular}{@{} l rrrrrrr @{}}
		\toprule
		& \multicolumn{7}{c}{Training set size} \\ 
		\cmidrule(lr){2-8}
		Representation &
		\multicolumn{1}{c}{\num{100}}  & 
		\multicolumn{1}{c}{\num{160}}  & 
		\multicolumn{1}{c}{\num{250}}  & 
		\multicolumn{1}{c}{\num{400}}  & 
		\multicolumn{1}{c}{\num{650}}  & 
		\multicolumn{1}{c}{\num{1000}} &
		\multicolumn{1}{c}{\num{1600}} \\
		\midrule
		MBTR $k=2$ & \num{21.06 \pm 2.10 } & \num{18.94 \pm 2.05 } & \num{11.73 \pm 0.91 } & \num{8.24 \pm 0.35 } & \num{6.73 \pm 0.19 } & \num{5.69 \pm 0.27 } & \num{5.63 \pm 0.17 } \\
SF $k=2$ & \num{11.10 \pm 0.64 } & \num{7.43 \pm 0.42 } & \num{6.10 \pm 0.48 } & \num{4.99 \pm 0.35 } & \num{4.30 \pm 0.14 } & \num{3.52 \pm 0.19 } & \num{2.98 \pm 0.12 } \\
MBTR $k=2,3$ & \num{19.77 \pm 1.41 } & \num{18.30 \pm 1.99 } & \num{10.51 \pm 0.66 } & \num{9.49 \pm 0.60 } & \num{7.23 \pm 0.30 } & \num{5.60 \pm 0.25 } & \num{5.52 \pm 0.14 } \\
SF $k=2,3$ & \num{11.49 \pm 0.75 } & \num{6.10 \pm 0.22 } & \num{5.07 \pm 0.29 } & \num{3.93 \pm 0.27 } & \num{3.07 \pm 0.10 } & \num{2.50 \pm 0.13 } & \num{2.21 \pm 0.08 } \\
SOAP & \num{8.38 \pm 1.05 } & \num{6.18 \pm 0.43 } & \num{4.24 \pm 0.21 } & \num{3.19 \pm 0.30 } & \num{2.29 \pm 0.18 } & \num{1.89 \pm 0.14 } & \num{1.70 \pm 0.11 } \\
		\bottomrule
    \end{tabular}
    
    \bigskip
	\medskip

    (d) Dataset~\dstcou{}.
    
    \medskip
    
	\begin{tabular}{@{} l rrrrrrr @{}}
		\toprule
		& \multicolumn{7}{c}{Training set size} \\ 
		\cmidrule(lr){2-8}
		Representation &
		\multicolumn{1}{c}{\num{100}}  & 
		\multicolumn{1}{c}{\num{160}}  & 
		\multicolumn{1}{c}{\num{250}}  & 
		\multicolumn{1}{c}{\num{400}}  & 
		\multicolumn{1}{c}{\num{650}}  & 
		\multicolumn{1}{c}{\num{1000}} &
		\multicolumn{1}{c}{\num{1600}} \\
		\midrule
		MBTR $k=2$ & \num{29.00 \pm 2.65 } & \num{25.94 \pm 2.79 } & \num{20.64 \pm 2.24 } & \num{19.46 \pm 1.30 } & \num{18.38 \pm 1.16 } & \num{16.19 \pm 0.67 } & \num{14.63 \pm 0.62 } \\
SF $k=2$ & \num{27.06 \pm 2.49 } & \num{24.03 \pm 1.71 } & \num{22.23 \pm 1.32 } & \num{18.17 \pm 1.04 } & \num{16.83 \pm 1.16 } & \num{14.87 \pm 0.78 } & \num{13.90 \pm 0.48 } \\
MBTR $k=2,3$ & \num{24.11 \pm 3.47 } & \num{23.49 \pm 1.19 } & \num{19.41 \pm 2.12 } & \num{19.33 \pm 1.40 } & \num{15.76 \pm 0.93 } & \num{15.26 \pm 0.53 } & \num{14.46 \pm 0.64 } \\
SF $k=2,3$ & \num{23.98 \pm 2.80 } & \num{25.83 \pm 3.34 } & \num{21.09 \pm 1.29 } & \num{19.51 \pm 1.76 } & \num{17.30 \pm 0.97 } & \num{15.21 \pm 0.73 } & \num{13.97 \pm 0.49 } \\
SOAP & \num{25.91 \pm 3.32 } & \num{24.28 \pm 2.29 } & \num{17.47 \pm 1.77 } & \num{17.68 \pm 2.10 } & \num{16.02 \pm 1.08 } & \num{14.67 \pm 0.82 } & \num{14.02 \pm 0.68 } \\
		\bottomrule
    \end{tabular}
    
\end{table*}

\subsection{Error metrics}\label{si:errormetrics}
We measure predictive performance by two metrics, an absolute one and a relative one that facilitates comparison across datasets.
In addition, we also provide a metric for qualitative comparison with the literature.

\begin{figure*}[hp]
	\begin{minipage}[t]{0.5\linewidth-1ex}\centering
		\includegraphics[width=\linewidth]{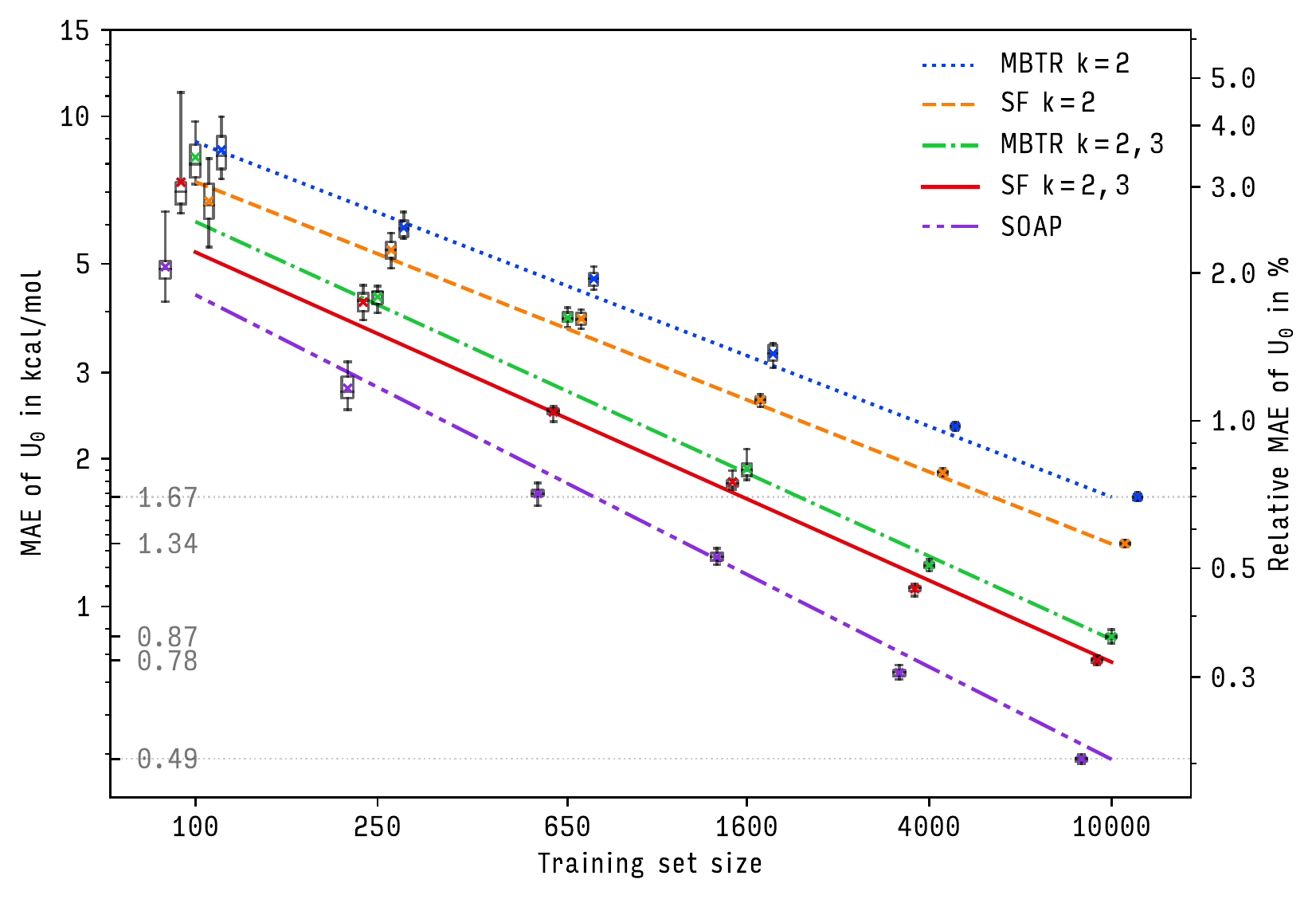}
		
		Dataset \dsgdb.
	\end{minipage}%
	\hfill%
	\begin{minipage}[t]{0.5\linewidth-1ex}\centering
		\includegraphics[width=\linewidth]{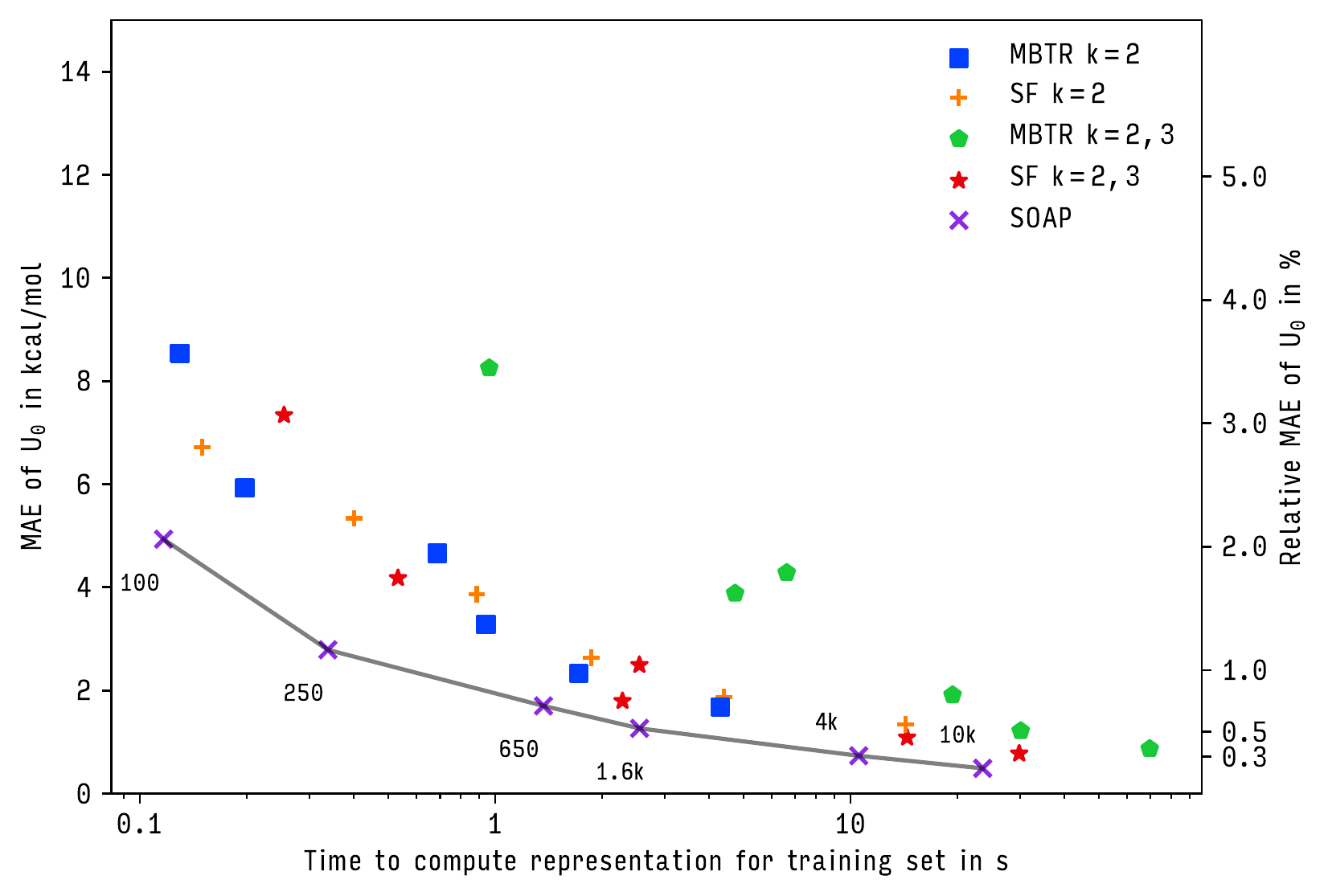}
		
		Dataset \dsgdb.
	\end{minipage}
	
	\medskip
	
	\begin{minipage}[t]{0.5\linewidth-1ex}\centering
		\includegraphics[width=\linewidth]{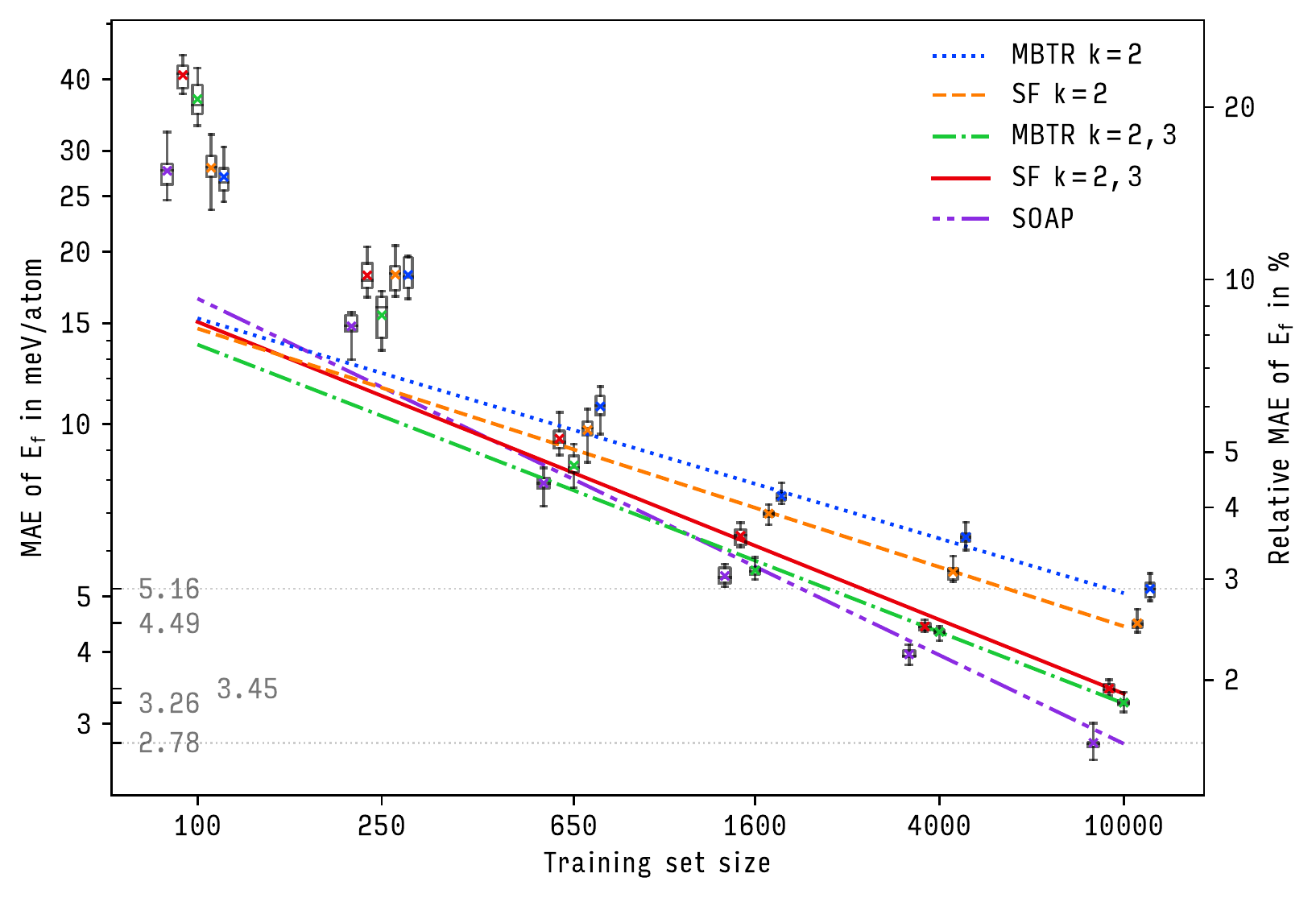}
		
		Dataset \dsba.
	\end{minipage}%
	\hfill%
	\begin{minipage}[t]{0.5\linewidth-1ex}\centering
		\includegraphics[width=\linewidth]{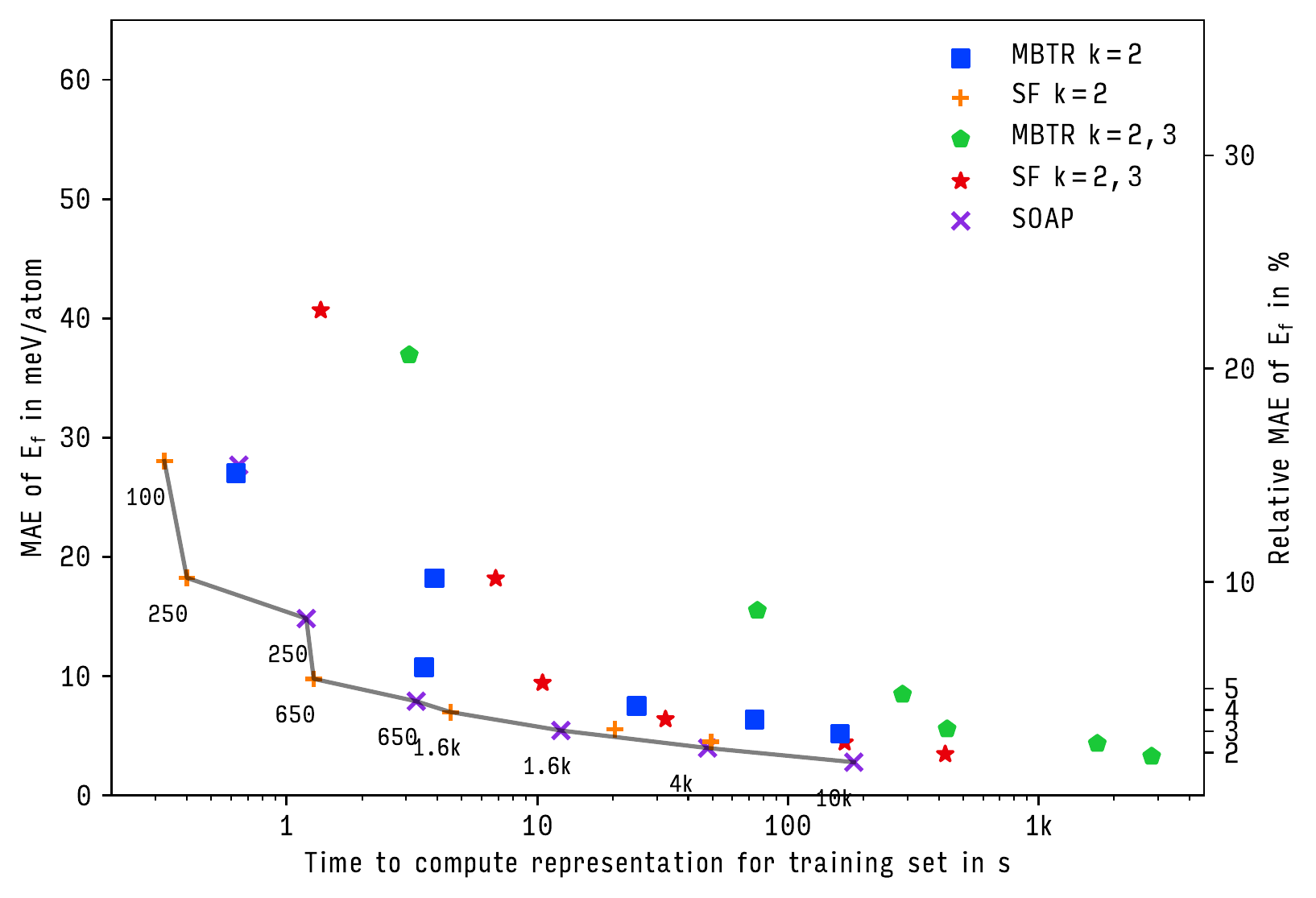}
		
		Dataset~\dsba.
	\end{minipage}
	
	\medskip
	
	\begin{minipage}[t]{0.5\linewidth-1ex}\centering
		\includegraphics[width=\linewidth]{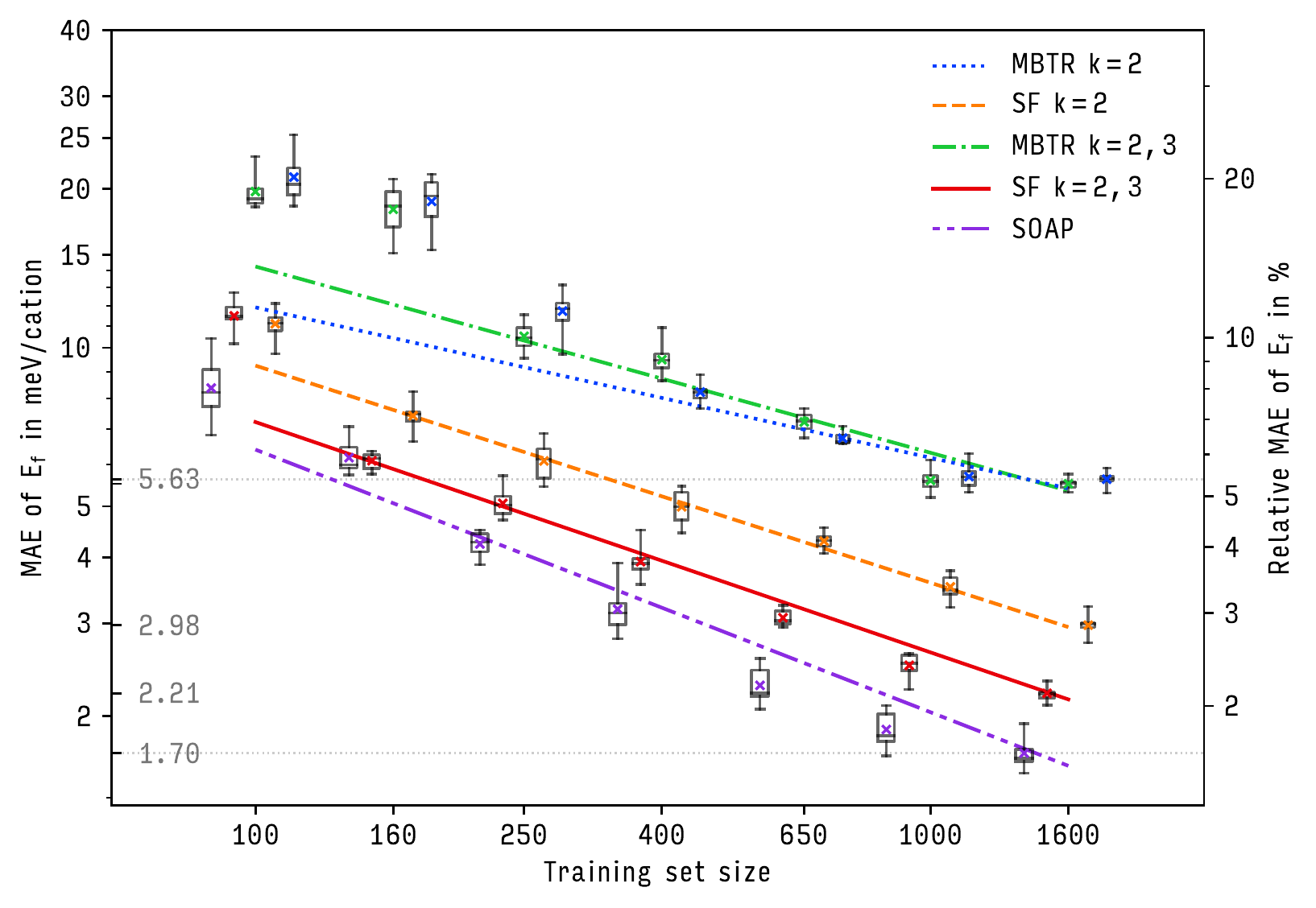}
		
		Dataset \dstcor.
	\end{minipage}%
	\hfill%
	\begin{minipage}[t]{0.5\linewidth-1ex}\centering
		\includegraphics[width=\linewidth]{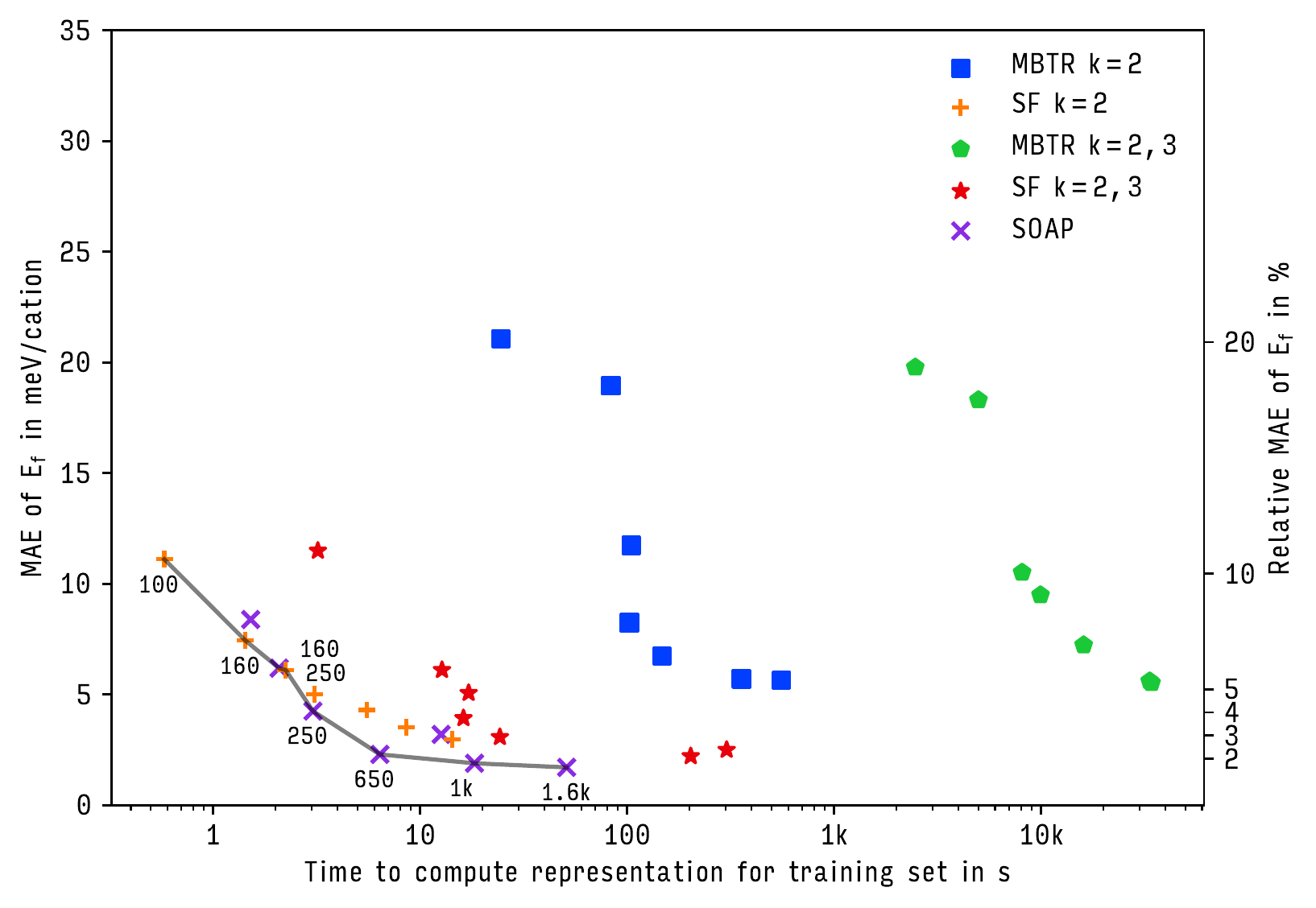}
		
		Dataset~\dstcor.
	\end{minipage}
	
	\medskip
	
	\begin{minipage}[t]{0.5\linewidth-1ex}
		\caption{\emph{Learning curves for mean absolute error} (MAE) of representations in \cref{sec:representations} on datasets \dsgdb{} (top), \dsba{} (middle), and \dstcor{} (bottom).
		Shown is MAE of energy predictions on out-of-sample-data as a function of training set size.
		Boxes, whiskers, bars, crosses show interquartile range, total range, median, mean, respectively.
		Lines are fits to theoretical asymptotic error.
		See Glossary for abbreviations.	
		\label{si:fig:learncurvesmae}}
	\end{minipage}%
	\hfill%
	\begin{minipage}[t]{0.5\linewidth-1ex}
		\caption{\emph{Compute times} for representations in \cref{sec:representations} on datasets \dsgdb{}~(top), \dsba{}~(middle), and \dstcor{}~(bottom).
		Shown is mean absolute error (MAE) of energy predictions on out-of-sample-data as a function of time needed to compute representations.
		Lines indicate Pareto frontiers, inset numbers show training set sizes.
		See Glossary for abbreviations.	
		\label{si:fig:timingsmae}}
	\end{minipage}
\end{figure*}

Let $y_i$, $f_i$, $e_i = f_i - y_i$ denote $i$-th observed label, prediction and residual.
Root mean squared error (RMSE) and mean absolute error (MAE) are given by 
\begin{equation*}
	\text{RMSE} = \sqrt{ \frac{1}{n} \sum_{i=1}^n e_i^2 }, 
	\quad
	\text{MAE} = \frac{1}{n} \sum_{i=1}^n |e_i| .
\end{equation*}
The canonical loss for least-squares regression is RMSE (as it is optimized by the regression).
We also provide MAE since it is often reported in the literature (\cref{si:fig:learncurvesmae,si:fig:timingsmae}).

RMSE and MAE are scale-dependent, and thus not suited for comparison across different datasets.
We therefore also report the scale-independent relative RMSE (rRMSE),
\begin{equation*}
	\text{rRMSE} = \frac{\text{RMSE}}{\sqrt{ \frac{1}{n} \sum_{i=1}^n (y_i - \bar{y})^2 }}
	= \frac{\text{RMSE}}{\sigma_y}
	= \frac{\text{RMSE}}{\text{RMSE}^{\ast}} ,
\end{equation*}
where $\bar{y} = \frac{1}{n} \sum_{i=1}^n y_i$ is the mean of the observed labels and $\sigma_y$ is their standard deviation.
The rRMSE can be seen as RMSE relative to the RMSE of a baseline model RMSE$^{\ast}$ that always predicts the mean of the labels.
While the latter is more naturally computed using training labels and the former using validation labels, 
as long as as the assumption of independent and identically distributed data holds, the number of samples is more important.

See \erefs{ctg2017x,b2019x} and references therein for an extended discussion of error metrics.

\subsection{Compute times}\label{si:computetimes}
\Cref{si:tab:compcostrepr,si:tab:compcostkernelm,si:tab:compcostoverview} present empirical computational costs, measured by processor wall-time, for calculating representations and kernel matrices, respectively.
Experiments were run on a single core of an Intel Xeon E5-2698v4 \SI{2.2}{\giga\hertz} processor.

For \cref{si:tab:compcostrepr}, representations of the 10\,k, 1\,k, 600 systems (datasets \dsgdb, \dsba, \dstco) in the first outer validation set were computed en bloc and the result divided by number of systems; this was repeated three times.

Similarly, for \cref{si:tab:compcostkernelm} kernel matrices between the representations of these systems were computed, also over three repetitions. 
The results were divided by the number of entries in the respective kernel matrices, yielding average kernel evaluation times.

\Cref{si:tab:compcostoverview} presents a summary overview of compute times for representations and kernel matrices.

\begin{table*}[hbtp]
	\caption{\emph{Computational cost of calculating representations} in milliseconds of processor wall-time on a single core.
		Shown are mean $\pm$ standard deviation over three repetitions of the time to compute a single system (molecule or unit cell).
		\label{si:tab:compcostrepr}}
		
	\centering

	\bigskip
	\medskip	
	
	(a) Dataset \dsgdb{}.
	
	\medskip
		
	\begin{tabular}{l rrrrrrr}
		\toprule
		& \multicolumn{6}{c}{Training set size} \\ \cmidrule(lr){2-7}
		Representation & 
		\multicolumn{1}{c}{\num{100}} & 
		\multicolumn{1}{c}{\num{250}} & 
		\multicolumn{1}{c}{\num{650}} & 
		\multicolumn{1}{c}{\num{1600}} & 
		\multicolumn{1}{c}{\num{4000}} & 
		\multicolumn{1}{c}{\num{10000}} \\ 
		\midrule
		MBTR $k=2$ & \num{1.3 \pm 0.1} & \num{0.8 \pm 0.1} & \num{1.1 \pm 0.1} & \num{0.6 \pm 0.1} & \num{0.4 \pm 0.1} & \num{0.4 \pm 0.1} \\
SF $k=2$ & \num{1.5 \pm 0.1} & \num{1.6 \pm 0.1} & \num{1.4 \pm 0.1} & \num{1.2 \pm 0.1} & \num{1.1 \pm 0.1} & \num{1.4 \pm 0.1} \\
MBTR $k=2,3$ & \num{9.6 \pm 0.1} & \num{26.5 \pm 0.1} & \num{7.3 \pm 0.1} & \num{12.1 \pm 0.1} & \num{7.5 \pm 0.1} & \num{7.0 \pm 0.1} \\
SF $k=2,3$ & \num{2.5 \pm 0.2} & \num{2.1 \pm 0.1} & \num{3.9 \pm 0.2} & \num{1.4 \pm 0.1} & \num{3.6 \pm 0.1} & \num{3.0 \pm 0.1} \\
SOAP & \num{1.2 \pm 0.1} & \num{1.4 \pm 0.1} & \num{2.1 \pm 0.1} & \num{1.6 \pm 0.1} & \num{2.6 \pm 0.1} & \num{2.4 \pm 0.1} \\
		\bottomrule
    \end{tabular}
    
    \bigskip
	\medskip	
    
    (b) Dataset~\dsba{}.
    
    \medskip
    
    \begin{tabular}{l rrrrrrr}
		\toprule
		& \multicolumn{6}{c}{Training set size} \\ \cmidrule(lr){2-7}
		Representation & 
		\multicolumn{1}{c}{\num{100}} & 
		\multicolumn{1}{c}{\num{250}} & 
		\multicolumn{1}{c}{\num{650}} & 
		\multicolumn{1}{c}{\num{1600}} & 
		\multicolumn{1}{c}{\num{4000}} & 
		\multicolumn{1}{c}{\num{10000}} \\ 
		\midrule
		MBTR $k=2$ & \num{6.3 \pm 0.3} & \num{15.6 \pm 0.1} & \num{5.4 \pm 0.1} & \num{15.6 \pm 0.1} & \num{18.4 \pm 0.1} & \num{16.1 \pm 0.1} \\
SF $k=2$ & \num{3.3 \pm 0.7} & \num{1.6 \pm 0.1} & \num{2.0 \pm 0.1} & \num{2.8 \pm 0.2} & \num{5.1 \pm 1.5} & \num{4.9 \pm 0.1} \\
MBTR $k=2,3$ & \num{30.9 \pm 0.4} & \num{302.0 \pm 0.1} & \num{440.8 \pm 0.1} & \num{269.6 \pm 0.1} & \num{428.5 \pm 0.2} & \num{282.5 \pm 0.1} \\
SF $k=2,3$ & \num{13.7 \pm 0.1} & \num{27.3 \pm 0.8} & \num{16.1 \pm 0.2} & \num{20.3 \pm 0.2} & \num{42.1 \pm 0.1} & \num{42.4 \pm 0.3} \\
SOAP & \num{6.4 \pm 0.1} & \num{4.8 \pm 0.1} & \num{5.1 \pm 0.1} & \num{7.8 \pm 0.1} & \num{12.0 \pm 0.1} & \num{18.3 \pm 0.1} \\
		\bottomrule
    \end{tabular}
    
    \bigskip
	\medskip	
    
    (c) Dataset~\dstcor{}.
    
    \medskip
    
    \begin{tabular}{l rrrrrrr}
		\toprule
		& \multicolumn{6}{c}{Training set size} \\ \cmidrule(lr){2-8}
		Representation & 
		\multicolumn{1}{c}{\num{100}} & 
		\multicolumn{1}{c}{\num{160}} & 
		\multicolumn{1}{c}{\num{250}} & 
		\multicolumn{1}{c}{\num{400}} & 
		\multicolumn{1}{c}{\num{650}} & 
		\multicolumn{1}{c}{\num{1000}} &
		\multicolumn{1}{c}{\num{1600}} \\ 
		\midrule
		MBTR $k=2$ & \num{246 \pm 1} & \num{522 \pm 1} & \num{419 \pm 1} & \num{256 \pm 1} & \num{227 \pm 1} & \num{356 \pm 1} & \num{348 \pm 1} \\
SF $k=2$ & \num{6 \pm 1} & \num{9 \pm 1} & \num{9 \pm 1} & \num{8 \pm 1} & \num{8 \pm 1} & \num{9 \pm 1} & \num{9 \pm 1} \\
MBTR $k=2,3$ & \num{24688 \pm 2} & \num{31168 \pm 3} & \num{32408 \pm 1} & \num{24864 \pm 2} & \num{24728 \pm 3} & \num{33518 \pm 1} & \num{21377 \pm 5} \\
SF $k=2,3$ & \num{32 \pm 1} & \num{80 \pm 1} & \num{69 \pm 1} & \num{40 \pm 1} & \num{37 \pm 1} & \num{303 \pm 2} & \num{127 \pm 1} \\
SOAP & \num{15 \pm 1} & \num{13 \pm 1} & \num{12 \pm 1} & \num{32 \pm 1} & \num{10 \pm 1} & \num{18 \pm 1} & \num{32 \pm 1} \\
		\bottomrule
    \end{tabular}

    \bigskip
	\medskip	
    
    (d) Dataset~\dstcou{}.
    
    \medskip

    \begin{tabular}{l rrrrrrr}
		\toprule
		& \multicolumn{6}{c}{Training set size} \\ \cmidrule(lr){2-8}
		Representation & 
		\multicolumn{1}{c}{\num{100}} & 
		\multicolumn{1}{c}{\num{160}} & 
		\multicolumn{1}{c}{\num{250}} & 
		\multicolumn{1}{c}{\num{400}} & 
		\multicolumn{1}{c}{\num{650}} & 
		\multicolumn{1}{c}{\num{1000}} &
		\multicolumn{1}{c}{\num{1600}} \\ 
		\midrule
		MBTR $k=2$ & \num{213 \pm 2} & \num{410 \pm 1} & \num{226 \pm 1} & \num{169 \pm 1} & \num{370 \pm 1} & \num{396 \pm 1} & \num{216 \pm 1} \\
SF $k=2$ & \num{7 \pm 1} & \num{8 \pm 1} & \num{30 \pm 1} & \num{14 \pm 1} & \num{6 \pm 1} & \num{6 \pm 1} & \num{7 \pm 1} \\
MBTR $k=2,3$ & \num{17757 \pm 8} & \num{18286 \pm 2} & \num{10959 \pm 1} & \num{10225 \pm 1} & \num{22921 \pm 1} & \num{22296 \pm 2} & \num{18878 \pm 1} \\
SF $k=2,3$ & \num{62 \pm 1} & \num{295 \pm 3} & \num{120 \pm 2} & \num{368 \pm 2} & \num{14 \pm 1} & \num{8 \pm 1} & \num{17 \pm 1} \\
SOAP & \num{53 \pm 1} & \num{38 \pm 1} & \num{76 \pm 1} & \num{65 \pm 1} & \num{46 \pm 1} & \num{48 \pm 1} & \num{12 \pm 1} \\
		\bottomrule
    \end{tabular}
\end{table*}

\begin{table*}[hbtp]
	\caption{\emph{Computational costs of calculating kernel matrices} in microseconds of processor wall-time on a single core.
		Shown are mean $\pm$ standard deviation over three repetitions of the time to compute a single kernel matrix entry.
		\label{si:tab:compcostkernelm}}

	\centering
		
	\bigskip
    \medskip
		
	(a) Dataset \dsgdb{}.
	
	\medskip
		
	\begin{tabular}{l rrrrrrr}
		\toprule
		& \multicolumn{6}{c}{Training set size} \\ \cmidrule(lr){2-7}
		Representation & 
		\multicolumn{1}{c}{\num{100}} & 
		\multicolumn{1}{c}{\num{250}} & 
		\multicolumn{1}{c}{\num{650}} & 
		\multicolumn{1}{c}{\num{1600}} & 
		\multicolumn{1}{c}{\num{4000}} & 
		\multicolumn{1}{c}{\num{10000}} \\ 
		\midrule
		MBTR $k=2$ & \num{0.16 \pm 0.01} & \num{0.16 \pm 0.01} & \num{0.16 \pm 0.01} & \num{0.16 \pm 0.01} & \num{0.16 \pm 0.01} & \num{0.16 \pm 0.01} \\
SF $k=2$ & \num{13.74 \pm 0.04} & \num{14.86 \pm 0.01} & \num{12.17 \pm 0.04} & \num{10.14 \pm 0.01} & \num{9.80 \pm 0.14} & \num{12.22 \pm 0.01} \\
MBTR $k=2,3$ & \num{0.90 \pm 0.01} & \num{0.90 \pm 0.01} & \num{0.90 \pm 0.01} & \num{0.90 \pm 0.01} & \num{0.90 \pm 0.01} & \num{0.90 \pm 0.01} \\
SF $k=2,3$ & \num{15.16 \pm 0.04} & \num{14.35 \pm 0.01} & \num{21.75 \pm 0.03} & \num{14.27 \pm 0.01} & \num{21.89 \pm 0.09} & \num{17.02 \pm 0.01} \\
SOAP & \num{14.97 \pm 0.03} & \num{28.80 \pm 0.01} & \num{71.30 \pm 0.03} & \num{31.81 \pm 0.01} & \num{120.09 \pm 0.10} & \num{91.74 \pm 0.01} \\
		\bottomrule
    \end{tabular}
    
    \bigskip
    \medskip
    
    (b) Dataset~\dsba{}.
    
    \medskip
    
    \begin{tabular}{l rrrrrrr}
		\toprule
		& \multicolumn{6}{c}{Training set size} \\ \cmidrule(lr){2-7}
		Representation & 
		\multicolumn{1}{c}{\num{100}} & 
		\multicolumn{1}{c}{\num{250}} & 
		\multicolumn{1}{c}{\num{650}} & 
		\multicolumn{1}{c}{\num{1600}} & 
		\multicolumn{1}{c}{\num{4000}} & 
		\multicolumn{1}{c}{\num{10000}} \\ 
		\midrule
		MBTR $k=2$ & \num{0.64 \pm 0.02} & \num{0.63 \pm 0.01} & \num{0.63 \pm 0.01} & \num{0.63 \pm 0.01} & \num{0.64 \pm 0.01} & \num{0.63 \pm 0.01} \\
SF $k=2$ & \num{9.73 \pm 0.11} & \num{7.19 \pm 0.01} & \num{8.22 \pm 0.01} & \num{11.38 \pm 0.02} & \num{9.13 \pm 0.01} & \num{7.74 \pm 0.01} \\
MBTR $k=2,3$ & \num{7.36 \pm 0.62} & \num{6.91 \pm 0.02} & \num{6.95 \pm 0.01} & \num{6.95 \pm 0.01} & \num{6.94 \pm 0.01} & \num{6.94 \pm 0.01} \\
SF $k=2,3$ & \num{15.14 \pm 0.22} & \num{15.09 \pm 0.02} & \num{20.92 \pm 0.01} & \num{22.50 \pm 0.14} & \num{20.03 \pm 0.08} & \num{26.49 \pm 0.03} \\
SOAP & \num{108.49 \pm 1.24} & \num{66.37 \pm 2.01} & \num{74.27 \pm 0.45} & \num{43.86 \pm 0.07} & \num{80.48 \pm 0.47} & \num{62.11 \pm 0.01} \\
		\bottomrule
    \end{tabular}
    
    \bigskip
    \medskip
    
    (c) Dataset~\dstcor{}.
    
    \medskip

    \begin{tabular}{l rrrrrrr}
		\toprule
		& \multicolumn{6}{c}{Training set size} \\ \cmidrule(lr){2-8}
		Representation & 
		\multicolumn{1}{c}{\num{100}} & 
		\multicolumn{1}{c}{\num{160}} & 
		\multicolumn{1}{c}{\num{250}} & 
		\multicolumn{1}{c}{\num{400}} & 
		\multicolumn{1}{c}{\num{650}} & 
		\multicolumn{1}{c}{\num{1000}} &
		\multicolumn{1}{c}{\num{1600}} \\ 
		\midrule
		MBTR $k=2$ & \num{0.2 \pm 0.2} & \num{0.1 \pm 0.1} & \num{0.1 \pm 0.1} & \num{0.1 \pm 0.1} & \num{0.1 \pm 0.1} & \num{0.1 \pm 0.1} & \num{0.1 \pm 0.1} \\
SF $k=2$ & \num{74.7 \pm 0.1} & \num{66.7 \pm 0.1} & \num{66.7 \pm 0.1} & \num{70.7 \pm 0.1} & \num{78.6 \pm 0.1} & \num{78.7 \pm 0.1} & \num{74.7 \pm 0.2} \\
MBTR $k=2,3$ & \num{3.3 \pm 3.8} & \num{0.5 \pm 0.1} & \num{0.5 \pm 0.1} & \num{0.6 \pm 0.1} & \num{0.6 \pm 0.1} & \num{0.5 \pm 0.1} & \num{0.5 \pm 0.1} \\
SF $k=2,3$ & \num{84.0 \pm 0.3} & \num{97.4 \pm 0.1} & \num{104.7 \pm 0.6} & \num{118.9 \pm 0.1} & \num{95.7 \pm 0.3} & \num{111.4 \pm 0.1} & \num{132.2 \pm 0.2} \\
SOAP & \num{89.7 \pm 0.3} & \num{142.9 \pm 0.1} & \num{308.6 \pm 0.2} & \num{907.9 \pm 2.5} & \num{174.2 \pm 0.1} & \num{198.0 \pm 0.1} & \num{252.3 \pm 0.1} \\
		\bottomrule
    \end{tabular}
		
    \bigskip
    \medskip
    
    (c) Dataset~\dstcou{}.
    
    \medskip

    \begin{tabular}{l rrrrrrr}
		\toprule
		& \multicolumn{6}{c}{Training set size} \\ \cmidrule(lr){2-8}
		Representation & 
		\multicolumn{1}{c}{\num{100}} & 
		\multicolumn{1}{c}{\num{160}} & 
		\multicolumn{1}{c}{\num{250}} & 
		\multicolumn{1}{c}{\num{400}} & 
		\multicolumn{1}{c}{\num{650}} & 
		\multicolumn{1}{c}{\num{1000}} &
		\multicolumn{1}{c}{\num{1600}} \\ 
		\midrule
		MBTR $k=2$ & \num{0.1 \pm 0.1} & \num{0.1 \pm 0.1} & \num{0.1 \pm 0.1} & \num{0.1 \pm 0.1} & \num{0.1 \pm 0.1} & \num{0.1 \pm 0.1} & \num{0.1 \pm 0.1} \\
SF $k=2$ & \num{78.9 \pm 0.4} & \num{70.8 \pm 0.4} & \num{66.3 \pm 0.1} & \num{83.8 \pm 0.1} & \num{70.2 \pm 0.2} & \num{66.4 \pm 0.1} & \num{87.7 \pm 0.1} \\
MBTR $k=2,3$ & \num{0.7 \pm 0.3} & \num{0.8 \pm 0.1} & \num{0.5 \pm 0.1} & \num{1.1 \pm 0.1} & \num{0.5 \pm 0.1} & \num{0.5 \pm 0.1} & \num{0.5 \pm 0.1} \\
SF $k=2,3$ & \num{117.2 \pm 0.9} & \num{91.7 \pm 0.2} & \num{128.2 \pm 1.2} & \num{124.1 \pm 0.1} & \num{105.5 \pm 0.1} & \num{88.4 \pm 0.1} & \num{99.3 \pm 0.1} \\
SOAP & \num{294.5 \pm 2.7} & \num{79.3 \pm 0.1} & \num{607.1 \pm 0.4} & \num{199.3 \pm 0.1} & \num{542.0 \pm 3.8} & \num{382.4 \pm 0.7} & \num{110.4 \pm 0.1} \\
		\bottomrule
    \end{tabular}
\end{table*}

\begin{table*}[hbtp]
	\caption{\emph{Overview of computational costs for calculating representations and kernel matrices.}
		Shown are computational cost estimates for (a) training on 10\,k training samples and (b) prediction of 10\,k validation samples.
		Based on mean observed compute times $t_{\text{rep}}$ for representations and $t_{\text{kernel}}$ for kernel matrices from \cref{si:tab:compcostrepr,si:tab:compcostkernelm}, we estimate total training times as $N_{\text{train}} \cdot t_{\text{rep}} + N_{\text{train}}^2 \cdot t_{\text{kernel}}/2$ and prediction times as $N_{\text{test}} \cdot t_{\text{rep}} + N_{\text{train}} \cdot N_{\text{test}} \cdot t_{\text{kernel}}$.
		Training times do not include time to calculate regression weights.
		All times are rounded to the nearest second, minute, or hour.
		\label{si:tab:compcostoverview}}
	
	\centering
	
	\bigskip
	\medskip
	
	(a) Training times.
	
	\medskip
	
	\begin{center}
		\begin{tabular}{l rll rll rll}
			\toprule
			& \multicolumn{9}{c}{Dataset} \\ \cmidrule(lr){2-10}
			& \multicolumn{3}{c}{\dsgdb} & \multicolumn{3}{c}{\dsba} & \multicolumn{3}{c}{\dstco} \\ \cmidrule(lr){2-4} \cmidrule(lr){5-7} \cmidrule(lr){8-10}
			Representation 
			& \multicolumn{1}{c}{$t_{\text{rep}}$} & \multicolumn{1}{c}{$t_{\text{kernel}}$} & \multicolumn{1}{c}{total} 
			& \multicolumn{1}{c}{$t_{\text{rep}}$} & \multicolumn{1}{c}{$t_{\text{kernel}}$} & \multicolumn{1}{c}{total} 
			& \multicolumn{1}{c}{$t_{\text{rep}}$} & \multicolumn{1}{c}{$t_{\text{kernel}}$} & \multicolumn{1}{c}{total}\\
		\midrule
		MBTR $k=2$ & 8s &+ 8s &= 15s & 2m &+ 32s &= 3m & 57m &+ 6s &= 57m \\
SF $k=2$ & 14s &+ 10m &= 10m & 33s &+ 7m &= 8m & 1m &+ 1h &= 1h \\
MBTR $k=2,3$ & 2m &+ 45s &= 3m & 49m &+ 6m &= 55m & 76h &+ 46s &= 77h \\
SF $k=2,3$ & 28s &+ 15m &= 15m & 4m &+ 17m &= 21m & 16m &+ 1h &= 2h \\
SOAP & 19s &+ 50m &= 50m & 2m &+ 1h &= 1h & 3m &+ 4h &= 4h \\
		\bottomrule
	\end{tabular}
	\end{center}
	
	\bigskip
	\medskip

	(b) Prediction times.
	
	\medskip
	
	\begin{center}
		\begin{tabular}{l rll rll rll}
			\toprule
			& \multicolumn{9}{c}{Dataset} \\ \cmidrule(lr){2-10}
			& \multicolumn{3}{c}{\dsgdb} & \multicolumn{3}{c}{\dsba} & \multicolumn{3}{c}{\dstco} \\ \cmidrule(lr){2-4} \cmidrule(lr){5-7} \cmidrule(lr){8-10}
			Representation 
			& \multicolumn{1}{c}{$t_{\text{rep}}$} & \multicolumn{1}{c}{$t_{\text{kernel}}$} & \multicolumn{1}{c}{total} 
			& \multicolumn{1}{c}{$t_{\text{rep}}$} & \multicolumn{1}{c}{$t_{\text{kernel}}$} & \multicolumn{1}{c}{total} 
			& \multicolumn{1}{c}{$t_{\text{rep}}$} & \multicolumn{1}{c}{$t_{\text{kernel}}$} & \multicolumn{1}{c}{total}\\
		\midrule
		MBTR $k=2$ & 8s &+ 16s &= 23s & 2m &+ 1m &= 3m & 57m &+ 13s &= 57m \\
SF $k=2$ & 14s &+ 20m &= 20m & 33s &+ 15m &= 15m & 1m &+ 2h &= 2h \\
MBTR $k=2,3$ & 2m &+ 1m &= 3m & 49m &+ 12m &= 1h & 76h &+ 2m &= 77h \\
SF $k=2,3$ & 28s &+ 29m &= 29m & 4m &+ 33m &= 38m & 16m &+ 3h &= 3h \\
SOAP & 19s &+ 2h &= 2h & 2m &+ 2h &= 2h & 3m &+ 8h &= 8h \\
		\bottomrule
	\end{tabular}
	\end{center}
\end{table*}

\subsection{Analysis details}\label{si:analysisdetails}
Predictive accuracy as measured by rRMSE is worse for solid-state datasets compared to the molecular \dsgdb{} one. 
This might indicate that periodic systems pose harder learning tasks than molecules.

MBTR performs worse for solid-state datasets than for the \dsgdb{} one, in particular for \dstcor{}. 
This might be due to increasing difficulty of the learning problem with system size (see discussion in \cref{sec:benchmark}) and 
lack of intrinsic scaling with number of atoms, impeding interpolation between unit cells of different size.
The high computational cost of MBTR with $k=3$ for large periodic systems also renders HP optimization more difficult.

For the \dsgdb{} dataset at $1\,600$ training samples, we observe an increase in RMSE standard deviation compared to neighbouring training set sizes for most methods. 
Comparing to MAE, which exhibits no such effect, and investigating errors individually, revealed that this is due to outliers, that is, few predictions with high error in some, but not all, outer splits. 
The problematic structures are ring molecules, and are not present in the outer training split used for HP optimization. 
This stresses the importance of carefully stratifying benchmark datasets.

\subsection{\dstcou{} dataset}\label{si:dstcounrel}
\begin{figure*}[hp]
	\begin{minipage}[t]{0.5\linewidth-1ex}\centering
		\includegraphics[width=\linewidth]{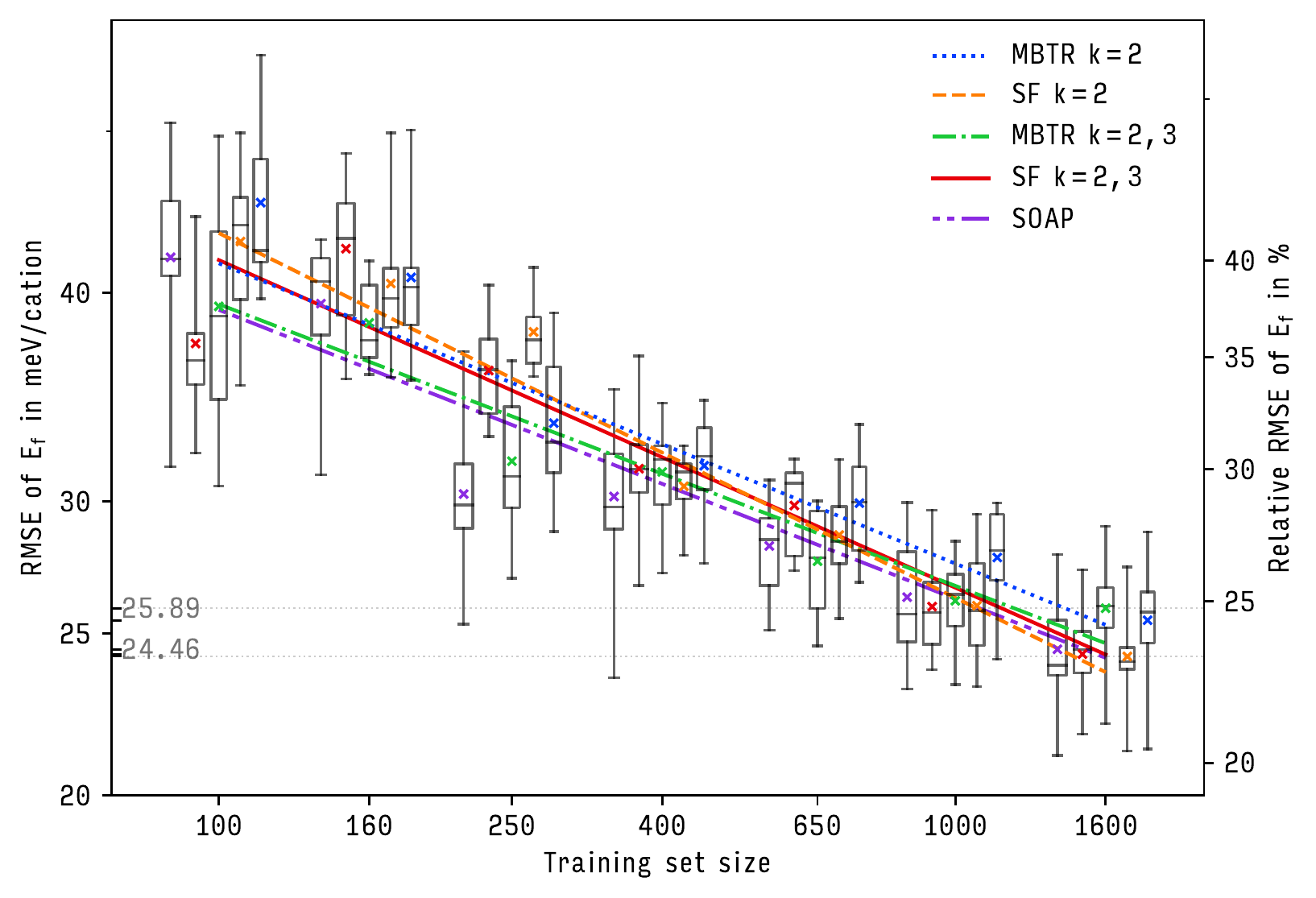}
		
		Learning curves (RMSE).
	\end{minipage}%
	\hfill%
	\begin{minipage}[t]{0.5\linewidth-1ex}\centering
		\includegraphics[width=\linewidth]{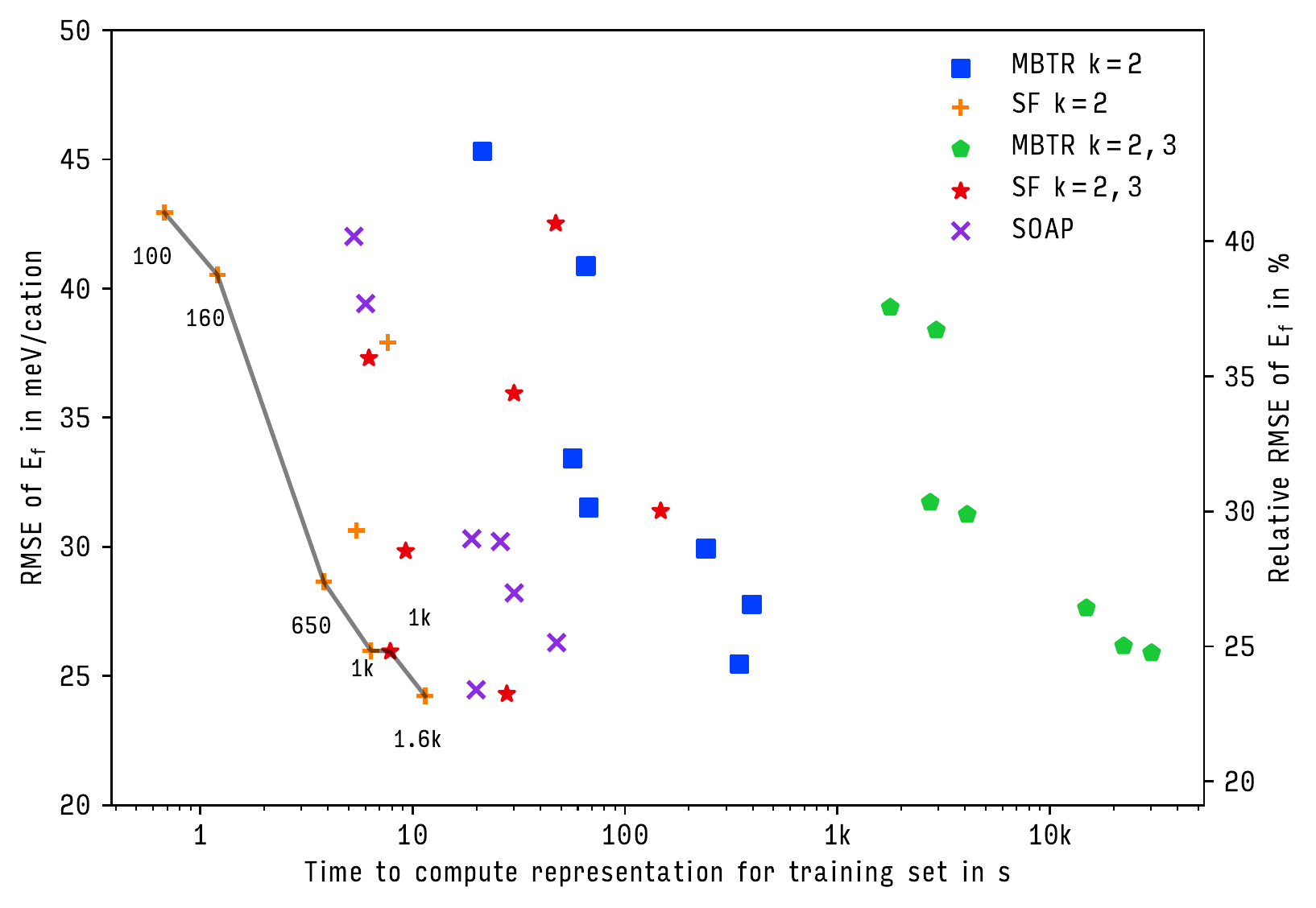}
		
		Prediction errors (RMSE) versus compute times.
	\end{minipage}
	
	\medskip
	
	\begin{minipage}[t]{0.5\linewidth-1ex}\centering
		\includegraphics[width=\linewidth]{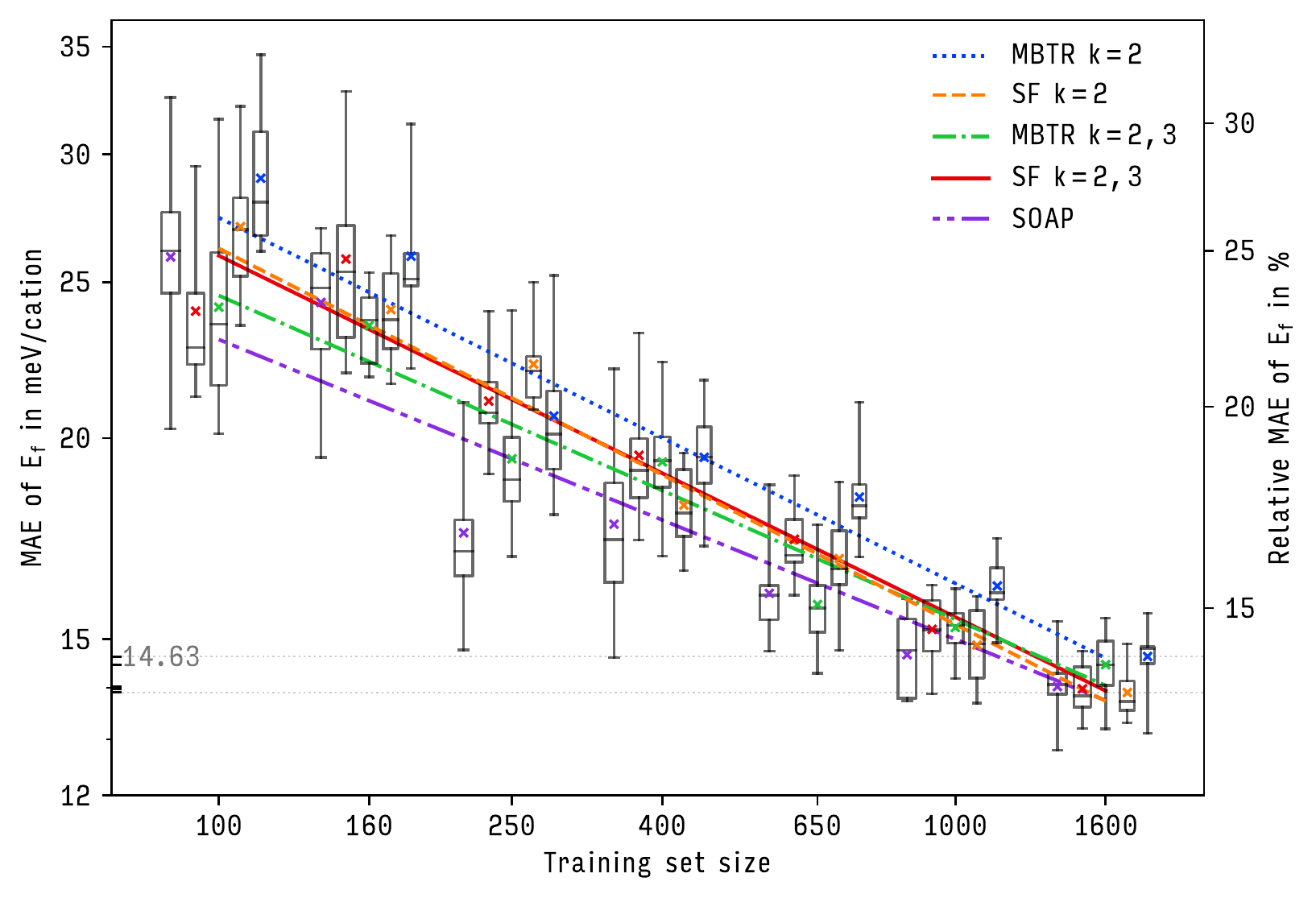}
		
		Learning curves (MAE).
	\end{minipage}%
	\hfill%
	\begin{minipage}[t]{0.5\linewidth-1ex}\centering
		\includegraphics[width=\linewidth]{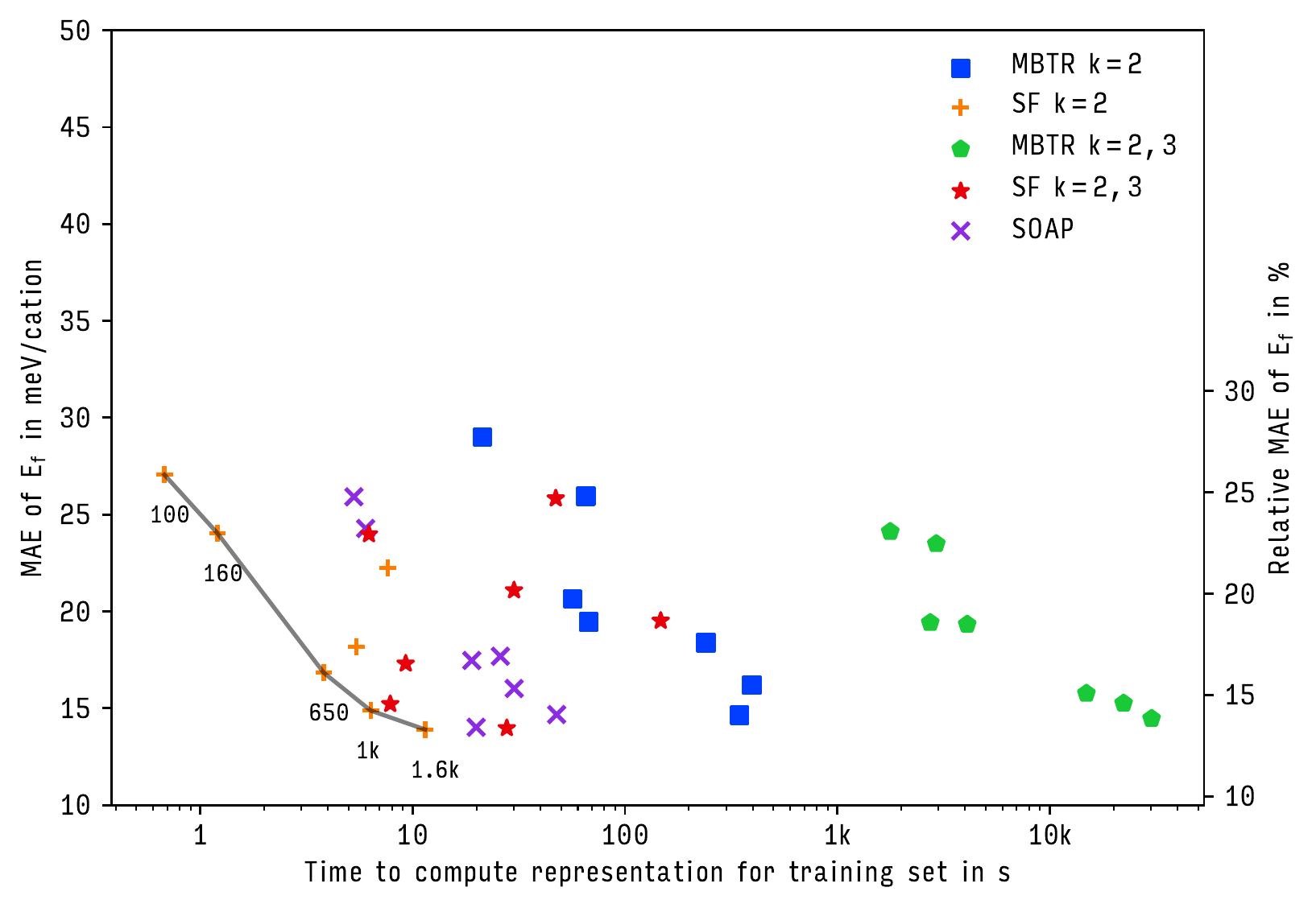}
		
		Prediction errors (MAE) versus compute times.
	\end{minipage}
	
	\medskip
	
	\begin{minipage}[t]{0.5\linewidth-1ex}
		\caption{\emph{Learning curves for dataset}~\dstcou{} of the representations in \cref{sec:representations}.
		Shown are root mean squared error (RMSE, top) and mean absolute error (MAE, bottom) of energy predictions on out-of-sample-data as a function of training set size.
		Boxes, whiskers, bars, crosses show interquartile range, total range, median, mean.
		Lines are fits to theoretical asymptotic error.
		See Glossary for abbreviations.	
		\label{si:fig:learncurvesnmd18u}}
	\end{minipage}%
	\hfill%
	\begin{minipage}[t]{0.5\linewidth-1ex}
		\caption{\emph{Compute times for dataset}~\dstcou{} of the representations in \cref{sec:representations}.
		Shown are root mean squared error (RMSE, top) and mean absolute error (MAE, bottom) of energy predictions on out-of-sample-data as a function of time needed to compute representations.
		Lines indicate Pareto frontiers, inset numbers show training set sizes.
		See Glossary for abbreviations.	
		\label{si:fig:timingsnmd18u}}
	\end{minipage}
\end{figure*}

\Cref{si:fig:learncurvesnmd18u,si:fig:timingsnmd18u}, and, \cref{si:tab:perfrmse,si:tab:perfmae,si:tab:compcostrepr,si:tab:compcostkernelm,si:tab:compcostoverview}
present results for energy predictions on the \dstcou{} dataset, that is, the \dstco{} dataset with approximate geometries obtained from Vegard's rule. 
In contrast to relaxed structures, such geometries can be obtained at almost no cost, and could be used in virtual screening campaigns.

We observe
(i) a strong increase in prediction errors (14--21\,\% for rRMSE),  
(ii) collapse of all representations to similar performance,
(iii) large differences between MAE and RMSE, indicating significant outliers.
From this, we conclude that 
the map from unrelaxed structures to ground-state energies is harder to learn than the map from relaxed structures to their energies, and,
that here the representation is not the limiting factor, and other sources of error dominate.

\subsection{Comparison with literature-reported errors}\label{si:complit}
Due to different conditions, such as sampling, regression and HP optimization methods, comparisons with performance estimates reported in the literature must remain qualitative.
Frequently, only MAE is reported, which tends to result in lower absolute values and to de-emphasize outliers.
\cref{si:tab:litperf} presents selected performance estimates from the literature.
Overall, errors in this work appear to be compatible with reported ones.

\subsection{Comparison with DFT and experimental errors}\label{si:compdft}
The error of DFT simulations against experimentally measured observations depends on system and property, as well as choice of density functional and other parameters, such as convergence thresholds and $k$-point density.
For heats of formation and the Becke 3-parameter Lee-Yang-Parr (B3LYP) functional used for the \dsgdb{} dataset, 
(systematic) MAEs relative to experiment of $\approx$\;\SI{2.6}{\kilo\cal\per\mol} have been reported for small organic molecules containing only C, H, N, O. \citesi{tj2008x}
For cohesive energies and the Perdew-Burke-Ernzerhof (PBE) functional used for the \dsba{} and \dstco{} datasets,
values of approximately 200 to 300\,\si{\milli\eV/\atom} have been reported. \citesi{l2008x,lvvc2014x,zrts2018x}

For the PBE functional, reported MAEs in computed energies between different parametrizations of DFT codes and \mbox{RMSEs} between 20 different DFT codes on 71 elements in bulk crystalline form were approximately 2\,meV/atom and 1.7\,meV/atom, respectively; \citesi{lbbbbbcccdetal2016x}
the latter reduces to 0.6\,meV/atom for all-electron codes only.
The best models for bulk crystal reported here have RMSEs of 4.6\,meV/atom and 3.3\,meV/cation on the \dsba{} and \dstco{} datasets.
However, the former benchmark values are integrated over a $\pm\,6\,\%$ interval around the equilibrium volume, 
whereas the values reported here are computed at the minima themselves and therefore measure related but distinct quantities.
This suggests that prediction errors are at least $\approx$\;2--6 times larger than DFT-intrinsic variations.

\begin{table}[htbp]
	\caption{\emph{Performance estimates from the literature.}
		\small
		Ref. = Reference, MAE = mean absolute error, RMSE = root mean square error, $N$ = training set size.
		\label{si:tab:litperf}}
	\mbox{}

	\mbox{}\hfill\mbox{(a) \dsgdb{} {dataset.}}\hfill\mbox{}
	
	\medskip

	\begin{center}		
	\begin{tabular}{@{}lcccl@{}}
		\toprule
		     & \multicolumn{2}{c}{\ Error / \si{\kilo\cal\per\mol}} \\ \cmidrule(lr){2-3}
		Ref. & MAE & RMSE & $N$ & Method \\
		\midrule
		\citenumsi{ptm2018qx}        & {1.5\; } & 2.8  &  5\,k  & IDMBR\,$^{\text{a}}$ \\
		\citenumsi{cbfgl2020x}       & {0.72} & ---  & 10\,k  & SOAP   \\ 
		\citenumsi{ssktm2018x}       & {1.27} & ---  & 10\,k  & SchNet \\  
		\citenumsi{fchl2018qx}       & {0.44} & ---  & 10\,k  & FCHL\,$^{\text{b}}$   \\  
		\citenumsi{cbfgl2020x}       & {0.66} & ---  & 10\,k  & FCHL\,$^{\text{c}}$ \\
		\citenumsi{wmc2018x}         & {0.14} & ---  & 100\,k & SOAP\,$^{\text{d}}$ \\
		\citenumsi{sgtm2019x}        & {0.35} & 0.94 & 100\,k & SchNet \\
		\citenumsi{fhhgsdvkrv2017qx} & {0.58} & ---  & 118\,k & HDAD   \\[1ex]
		here                         & {0.49} & 0.90 & 10\,k  & SOAP   \\
		\bottomrule
		\multicolumn{5}{@{}p{0.8\linewidth}@{}}{%
			\rule{0pt}{3ex}%
			$^{\text{a}}$ inverse-distance many-body representation\newline
			\,$^{\text{b}}$ original FCHL18 version \citesi{fchl2018qx}\newline
			\,$^{\text{c}}$ revised FCHL19 version \citesi{cbfgl2020x}\newline
			\,$^{\text{d}}$ radial-scaling modification
		}
	\end{tabular}
	\end{center}

	\medskip
	
	\mbox{}\hfill%
		\mbox{(b) \dsba{} {dataset.}}%
	\hfill\mbox{}

	\medskip

	\begin{center}		
	\begin{tabular}{@{}lcccl@{}}
		\toprule
		     & \multicolumn{2}{c}{\ Error / \si{\milli\eV\per\atom}} \\ \cmidrule(lr){2-3}
		Ref. & MAE & RMSE & $N$ & Method \\
		\midrule
		\citenumsi{nrbsmrcwh2019x} & 5.3 & --- & 10\,k & MBTR \\
		\citenumsi{nrbsmrcwh2019x} & 3.4 & --- & 10\,k & MTP \\[1ex]
		here                       & 2.8 & 4.6 & 10\,k & SOAP \\
		\bottomrule
	\end{tabular}
	\end{center}
	
	\medskip

	(c) \dstcou{} {dataset.}
		Here, all representations performed roughly equally.
		At the time of printing, no published results existed for the relaxed \dstcor{} version.

	\medskip

	\begin{center}		
	\begin{tabular}{@{}lcccl@{}}
		\toprule
		     & \multicolumn{2}{c}{\ Error / \si{\milli\eV\per\cation}} \\ \cmidrule(lr){2-3}
		Ref. & MAE & RMSE & $N$ & Method \\
		\midrule
		\citenumsi{sgylbhglzs2019x} & 13 & --- & 2\,400 & SOAP \\[1ex]
		here                        & 14--15 & 24--26 & 1\,600 & all \\
		\bottomrule
	\end{tabular}
	\end{center}
\end{table}


\stepcounter{section}


\setbiblabelwidth{99} 
\phantomsection\addcontentsline{toc}{section}{References}
\begingroup
\raggedright
\bibliographysi{abbreviations,supplement}
\hypersetup{allcolors=linkcolor} 
\endgroup

\end{document}